%% file: delvalle-romero-santoslima_v3.tex
\documentclass[useAMS,usenatbib,usegraphicx]{mn2e}
\usepackage{amssymb}
\usepackage{amsmath}
\usepackage{natbib}
\usepackage{graphics}
\usepackage{color}
\usepackage{ulem}
\newcommand{\bs}{{bowshock}}

\newcommand{\bss}{{bowshocks}}

\title[Runaway stars as cosmic ray injectors inside molecular cloud]{Runaway stars as cosmic ray injectors inside molecular clouds}
\author[M.V. del Valle, G.E. Romero and R. Santos-Lima]{M.V. del Valle$^{1,2}$\thanks{E-mail: maria@iar-conicet.gov.ar}, G.E. Romero $^{1,2}$\thanks{E-mail: romero@iar-conicet.gov.ar} and R. Santos-Lima$^{3}$\thanks{E-mail: rlima@astro.iag.usp.br}\\
$^{1}$Instituto Argentino de Radioastronom\'{\i}a, C.C.5, (1894) Villa Elisa, Buenos Aires, Argentina\\
$^{2}$Facultad de Ciencias Astron\'omicas y Geof\'{\i}sicas,
Universidad Nacional de La Plata, Paseo del Bosque, 1900 La Plata, Argentina\\
$^{3}$Instituto de Astronomia, Geof\'isica e Ci\^encias Atmosf\'ericas, Universidade de S\~ao Paulo, S\~ao Paulo, SP 05508-090, Brazil}

\begin{document}

% \date{Accepted 1988 December 15. Received 1988 December 14; in original form 1988 October 11}

% \pagerange{\pageref{firstpage}--\pageref{lastpage}} \pubyear{2002}

\maketitle

\label{firstpage}

\begin{abstract}
{
Giant molecular clouds (GMCs) are a new population of gamma-ray sources, being the target of cosmic rays (CRs) -- locally accelerated or not --. 
These clouds host very young stellar clusters where massive star formation takes place. Eventually, some of the stars are ejected from the clusters, becoming runaway stars. These stars move supersonically through the cloud and develop bowshocks
where particles can be accelerated up to relativistic energies. As a result, the bowshocks present non-thermal emission, and inject relativistic protons in the cloud. These protons diffuse in the GMC interacting with the matter.

We present a model for the non-thermal radiation generated by protons and secondary pairs accelerated in the bowshocks of massive runaways stars within young
GMCs.  We solve the transport equation for primary protons and secondary pairs as the stars move through the cloud. We  present non-thermal emissivity maps in radio and in gamma rays as a function of time. We obtain X-ray luminosities of the
order $\sim$  ${ 10^{32}}$ erg~s$^{-1}$ and gamma-ray luminosities $\sim$ $10^{34}$ erg~s$^{-1}$. We conclude that, under some assumptions, relativistic protons from massive runaway stars interacting with matter in GMCs give rise to extended non-thermal sources.

}
\end{abstract}


\section{Introduction}

Molecular clouds (MCs) are good targets for galactic cosmic rays (CRs) since 
they are extended regions with great amounts of mass. These systems contain particle accelerators such as supernova remnants (SNRs), OB massive stars and pulsars. The relativistic particles accelerated in these sources add to the galactic cosmic-ray population that illuminates the clouds producing gamma rays \citep[e.g.][]{ch1y2:casse80,ch5:aharonian_atoyan_1996,ch1y2:torres05}. 

GMCs are a new class of extended gamma-ray sources \citep[e.g.,][]{ch1y2:ackermann12,ch1y2:fernandez13}. However, their potential as  passive gamma-ray sources have been claimed since the beginning of gamma-ray astronomy 
and their emission was predicted in theoretical models long time ago \citep[e.g.][]{ch1y2:kraushaar72,aha91,ch1y2:combi95,ch1y2:hunter97}.
Not only passive GMCs emit gamma rays; recently the smaller and gas-rich  star forming regions Chamaeleon, R Coronae Australis, Cepheus and Polaris have been detected by {\it Fermi} \citep{ch7:ackermann_etal_2012}.

GMCs harbour young stellar clusters where massive stars form. Many stars with masses  
 $M$ $>$ 8 $M_{\odot}$ might be ejected from the clusters  \citep[e.g.,][]{ch3:perets_subr_2012}, becoming runaway stars travelling through the cloud. Runaway stars have spatial velocities $>$ 30 km s$^{-1}$  \citep[e.g.][]{ch3:gies_bolton_1986}. Two mechanism have been proposed for the origin of the high velocities in these stars. In the binary-supernova scenario one of the stars in a binary system is expelled during the supernova explosion of its companion  \citep{ch3:blaauw_1961}. In the other scenario, the  dynamical-ejection model, the star is expelled through close gravitational interactions between members of a cluster or association (\citealp{ch3:leonard_duncan_1988}). Currently the latter process is believe to be the most frequent \citep{ch3:fujii_zwart_2011}, but both mechanisms surely  operate \citep{ch3:hoogerwerf_etal_2000}.

 The supersonic interaction between the wind of fugitive stars with the interstellar medium (ISM) produces a {\bs} \citep[e.g.,][]{ch3:vanburen_mccray_1988,ch3:peri_etal_2012}. The wind-swept material, gas, and dust  is heated by the shock and radiated away in the infra-red (IR) band (e.g., \citealt{ch3:vanburen_mccray_1988, ch3:kobulnicky_etal_2010}). 
 
Both observational \citep{ch6:benaglia10,lopez2012,delvalle2013} and theoretical research \citep{delvalle2012,delvalle2014} support the idea that {\bss} from massive runaway stars  accelerate particles up to relativistic energies. Electrons lose their energy in the acceleration region, while protons escape, convected away by the shocked wind without losing much of their energy. The escaped protons then   diffuse in the MC.   
 
  We propose here  that  protons accelerated in  {\bss} from massive runaway stars embedded in MCs  contribute to the cloud's cosmic-ray population that produces the   observed gamma rays in these systems. The relativistic protons interact with the MC matter via $p-p$ inelastic collisions  \citep{ch5:aharonian_atoyan_1996,ch7:bosch-ramon_etal_2005}.
  
The favourite sources for accelerating  particles in the ISM are  SNRs, which can inject particles with total energies up to $10^{50}$ erg;
it is believed that these sources are   responsible  for the bulk of the galactic CRs  (\citealt{ginzburg64,ch7:hillas_2005}). MCs near SNRs produce gamma emission \citep[e.g.,][]{combi98}. Three  H.E.S.S. sources are firmly associated with a  MC-SNR system. However, in the star forming regions immersed in GMCs,  besides SNRs, there exist very energetic sources such as young stars with powerful winds. A very massive star during its life (1-10 Myr) can inject into the ISM an amount of kinetic energy from its wind comparable to the SNR values. {However,  a young cluster  has not lived  enough time for a  SNR to develop and to be able to inject relativistic particles into the cloud}.  Here we proposed that  runaway massive stars\footnote{No evidence exists so far of a strong shock or of non-thermal emission from the bowshock of a low-velocity (non-binary) massive star.  A strong shock might not form due to the catastrophic  adiabatic wind losses  (\citealt{voelk1982}). In a runaway star the stagnation point is much closer to the star.} could contribute to the local density of relativistic particles  inside  MCs, and produce significant non-thermal emission. { An instrument with both good enough sensitivity and angular resolution, such as the forthcoming Cherenkov Telescope Array (CTA, see \citealt{actis11})\footnote{ However, {\it Fermi}, in target mode, might have sufficient sensitivity to detect the emission.} can detect the produced emission and its expected morphology, that we estimate in this work.}

We organized this article as follows. In the next section  we briefly introduce MCs and their gamma-ray emission. Section~\ref{runa-mc} deals with runaway stars in MCs. There we present the scenario adopted in this work. In Sec.~\ref{calc} we describe  the model and in Sec.~\ref{methods} we describe the numerical method we use in the calculations. The results are given in Sec.~\ref{resu}. Finally, in Sec.~\ref{ch7:dyc} we discuss the results  and  offer our conclusions. 

\section{Molecular clouds}\label{MCs}

MCs are dense and cold regions that constitutes the most dense component of the ISM. They have  temperatures of the order of 10 to 20 K, and average densities  of the order of  10$^2$ cm$^{-3}$. In these systems most of the new stars of the Galaxy are formed. Young stars are associated with the densest regions of the clouds ($n>10^4$ cm$^{-3}$). In these cores of  GMCs (with total masses between 10$^3$ and $10^{6}$ $M_{\odot}$) the most massive stars are born. 

In the Galaxy the molecular gas is typically concentrated in big complexes or segments of spiral arms  with sizes of the order of $\sim$ 1 kpc and masses of $10^{7}$~$M_{\odot}$. These systems can contain many GMCs with sizes of $\sim$ 100 pc and  masses of $\sim$ $10^{6}$ $M_{\odot}$. These GMCs also contain substructures such as the cores with sizes of the order of $\sim$ $0.1$  pc. In our Galaxy  smaller clouds also exist with masses of $\sim$ 500 $M_{\odot}$ (e.g., \citealt{ch7:larson_2003}). 

The clouds have structure and turbulence at all scales.
The gas density in these objects varies many orders of magnitude, the densest regions having densities as high as  $\sim$ $10^{5}$~cm$^{-3}$. The density profile is not well  known,  with different  substructures present in the clouds (filaments, clumps, cores, etc.). Usually, the following profile is adopted for the density \citep[e.g.,][]{ch7:gabici_etal_2007}:

\begin{equation}
n(R) = \frac{n_{0}}{1+\left( \frac{R}{R_{\rm n}}\right) ^{\beta}},
\label{ch7:densi-pro}
\end{equation}
where $R$ is the distance from the cloud centre and $R_{\rm n}$  is the core radius.  The index $\beta$ is a free parameter. 

MCs are magnetized, the magnetic field being important in their evolution and dynamics. The magnetic field is closely related to the gas density (\citealt{ch5:crutcher_1999}), and it is described by the following profile:

\begin{equation}
B \sim 100 \left(\frac{n}{10^{4}\,{\rm cm}^{-3}}\right)^{\eta}\,\,\mu{\rm G}.
\label{B}
\end{equation}
Here $\eta = 0.5$. Although the correlation given in Eq.~(\ref{B}) between the gas density and the magnetic field has been found for the cores of MCs with densities greater than $10^{3}$ cm$^{-3}$, it provides reasonable values for regions of lower density and it is usually extrapolated to the whole range of densities (e.g., \citealt{ch7:gabici_etal_2007, ch7:pedaletti_etal_2013}), {although this correlation can be substantially reduced by turbulence (\citealt{santos-lima2012})}. 

The average age of MCs is of $\sim$  10 Myr (e.g., \citealt{ch6:Bodenheimer2011}). The clouds are eventually destroyed and disrupted by ionization, outflows and winds produced by the young stars.

\subsection{Gamma rays from molecular clouds}

As mentioned above, some GMCs are  gamma-ray sources. Studies of the gamma-ray emission of nearby  MCs   (at distances $\sim$ 1 kpc) dated since the  {\it COS-B} days (e.g., \citealt{ch7:bloemen_etal_1984,ch7:hunter_etal_1994}). Theoretical works on the cosmic-ray illumination of nearby sources are even older  (e.g., \citealt{ch7:black_fazio_1973, ch4:montmerle_1979}). Diffuse gamma-ray emission has been detected from the galactic centre region, being spatially correlated with a GMC complex (\citealt{ch7:aharonian_etal_2006});                                                  star formation regions inside MCs also have been detected: Monoceros R2 (\citealt{ch7:marti_etal_2013}), Westerlund 2 (\citealt{ch7:reimer_etal_2008}), Westerlund 1 (\citealt{ch7:ohm_etal_2013}), the region of Cygnus (\citealt{ch7:aharonian_etal_2005}), and the Orion region, which includes three dense young star clusters  (\citealt{ch7:hartmann_2009}). Additionally,  {\it Fermi} has been  detecting nearby clouds  (at distances $d$ $<300$ pc)  in the energy range  250 MeV-10 GeV; these clouds have masses between $10^{3}-10^{4}$ $M_{\odot}$ (\citealt{ch7:ackermann_etal_2012}).  The  gamma-ray luminosities observed in MCs vary between $\sim$ $10^{33}$ and $10^{35}$ erg~s$^{-1}$. 

Gamma-ray emission is of special interest because, when it is detectable, its study provides a good tool for investigating acceleration and propagation of CRs in the Galaxy (e.g., \citealt{ch7:aharonian_2001}). MCs embedded  in the  galatic cosmic-ray \emph{sea} are expected to emit gamma rays as passive sources. If particles can freely penetrate the clouds, the gamma-ray spectrum is expected to mimic the cosmic-ray spectrum and the total gamma-ray luminosity  depends only on the total mass of the cloud  (e.g., \citealt{ch7:gabici_2011}). However, cosmic-ray penetration on MCs is a subject of debate. In general, the penetration might depend on the diffusion coefficient, a key parameter very hard to estimate both theoretically and observationally.

\section{Runaway stars in Molecular clouds}\label{runa-mc}
 
Numerical simulations and theoretical predictions indicate that many  massive stars can be ejected with high velocities from their formation clusters by gravitational encounters. Runaway stars then move through their parental MC. The probability to eject a star from a massive cluster with velocity  $V_{\star}$ is a power law $\propto$ $V_{\star}^{-\nu}$, where $\nu = 3/2$ for slow runaways, and $8/3$ for fast ones (\citealt{ch3:perets_subr_2012}). Additionally, the ejection probability increases with mass. N-body simulations show that  during its life a cluster can eject $\sim$ six stars with masses $>$ 8$M_{\odot}$, independently of the cluster mass (\citealt{ch3:fujii_zwart_2011}). Observational evidence consistent with these results is found, for example, in the R136 cluster: six massive runaway stars are associated with it (\citealt{ch7:gvaramadze_etal_2010, ch7:bestenlehner_etal_2011}). 

\begin{figure}
\begin{center}
\includegraphics[trim=0cm 0cm 0cm 0cm, clip=true,width=.5\textwidth,angle=0]{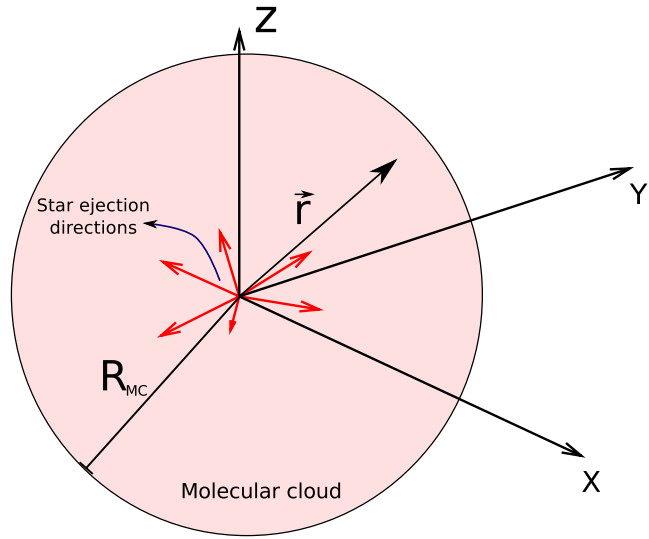}
\caption{Random ejection directions of the six runaway stars ejected at the centre of a spherical MC (not to scale).}
\label{ch7:estrellitas}
\end{center}
\end{figure}

Stellar  {\bss} inside MCs might not be detectable in the IR because the emission produced by the whole cloud overshines individual contributions. Nevertheless, the  presence of {\bss} can be inferred through the study of certain spectral lines and masers. Masers are  collisionally  excited by hydrogen which is heated  by  shock waves (e.g., \citealt{ch7:elitzur_1976}). 

A runaway star moving through an inhomogeneous medium produces variable non-thermal emission (\citealt{delvalle2014}). The electrons yield the bulk of the non-thermal radiation, while most of the accelerated protons escape without losing much of their energy and then diffuse in the environment. Gamma-ray emission and secondary electron-positron pairs are produced through $p-p$ interactions  with   matter. These pairs also diffuse in the cloud and  produce non-thermal radiation by synchrotron process. The injected power in protons by the  {\bs} is a fraction of the kinetic power of the wind: 
\begin{equation}
L_{\rm T} \sim  \frac{1}{2}{\dot M_{\rm w}}V_{\rm w}^2,
\end{equation}
where ${\dot M_{\rm w}}$  is the wind mass-loss rate and $V_{\rm w}^2$ is the wind terminal velocity. 

Here  we consider as a first approach a spherical young MC\footnote{MCs can be disrupted by the action of the winds and radiation of the new stars. Evolved MC are not expected to be spherical but annular; such a system  will be considered elsewhere.} of radius $R_{\rm MC} = 50$ pc and core radius  $R_{\rm n} = 0.5$ pc. This cloud hosts a massive young stellar cluster. The density profile of the cloud is given by Eq. (\ref{ch7:densi-pro}), with $\beta = 1$ ({\it boxy cloud}) and $n_{0}$ = $10^{4}$~cm$^{-3}$. The total mass of the cloud is  $M_{\rm  MC}$ $\sim$ $10^6$ $M_{\odot}$. We assume that the cluster ejects six massive stars in the last Myr. Additionally,  we adopt in our simulations that all stars have been ejected  at approximately the same time  in random spatial directions, as illustrated in  Fig.~\ref{ch7:estrellitas}.

The ejection probability of a star is  $\propto$ $V_{\star}^{-\nu}$, therefore it is more likely to eject stars of low velocity. We  consider then three stars with velocity  $V_{\star}$ $\sim$ $30$ km~s$^{-1}$, two stars of velocity $V_{\star}$ $\sim$ $65$ km~s$^{-1}$, and one  of $V_{\star}$ $\sim$ $100$ km~s$^{-1}$.  Also,  runaway stars of spectral types  O9 are more common than O4-type stars; then we consider one star  of type O4I, three of type O9I and  two intermediate cases;  the maximum energies we estimate for locally accelerated protons  are $10^{14}$ eV, $10^{12}$ eV and $10^{13}$ eV, respectively\footnote{These values are consistent with the maximum energies for protons obtained in \cite{delvalle2014}.}. These parameters are shown in Table~\ref{ch7:tab-star}. The stars inject protons continuously  during 1 Myr. The maximum energies that protons reach do not depend on the medium density (see, \citealt{delvalle2014})
 and neither does the injected power, so these quantities are constant during the motion of the stars through the cloud.

\begin{table*}
\begin{center}

\begin{tabular}{lllllll}
\hline\noalign{\smallskip}
 ${\star}$ \# & $V_{\star}$ [km~s$^{-1}$]& $\dot{M}_{\rm w}$ [$M_{\odot}$~yr$^{-1}$] & $V_{\rm w}$  [km~s$^{-1}$] & Power [erg~s$^{-1}$] & Max. energy [TeV] & $\tau$ [Myr] \\[0.001cm]
\hline\noalign{\smallskip}
1 & 30  & $10^{-4}$ &  2200  &  3$\times 10^{35}$ & 100 & 1.6 \\[0.001cm]
2 & 30  & $10^{-6}$ &  800  &  4$\times 10^{32}$  & 1  & 1.6 \\[0.001cm]
3 & 30  & $10^{-5}$ &  1000 &  7$\times 10^{33}$ & 10  &  1.6 \\[0.001cm]
4 & 65  & $10^{-5}$ &  1000 &  7$\times 10^{33}$  & 10 & 0.7 \\[0.001cm]
5 & 65  & $10^{-6}$ &  800  &  4$\times 10^{32}$  & 1  & 0.7 \\[0.001cm]
6 & 100 & $10^{-6}$ &  800  &  4$\times 10^{32}$  & 1  & 0.5 \\[0.001cm]
\hline\\[0.05cm]
\end{tabular}	
\caption[]{Wind velocity and mass-loss rate, stellar velocity,  injected power, proton maximum energies and crossing time ($\tau = R_{\rm MC}/V_{\star}$) for each runaway star considered (see text).}
\label{ch7:tab-star}
\end{center}
\end{table*}

\section{Physical model}\label{calc}

As mentioned, we do not take into account the physical details of the injectors (the stellar {\bss}); we consider them as punctual, moving with constant velocity  $V_{\star}$ through the cloud, without perturbing the density distribution. The relativistic protons diffuse into the cloud.  Because of the linear nature of the problem, we solve the equations for each star and sum all contributions to obtain the final result.

The spectral energy distribution $N_p$ of the protons follows the transport equation: 

\begin{eqnarray}
%\frac{\partial N_p(\vec{r}, E,t)}{\partial t} = \vec{\nabla} {\cdot} \left[D(\vec{r},E)\nabla N_p(\vec{r},E,t)\right] -  \frac{\partial}{\partial E} \left(P(\vec{r},E)\,N_p(\vec{r},E,t)  \right) +  
%+ \,\, Q_p(\vec{r},E,t)
&& \frac{\partial N_p(\vec{r}, E,t)}{\partial t} = \vec{\nabla} {\cdot} \left[D(\vec{r},E)\nabla N_p(\vec{r},E,t)\right] -  \nonumber\\ && \frac{\partial}{\partial E} \left(P(\vec{r},E)\,N_p(\vec{r},E,t)  \right) +  \,\, Q_p(\vec{r},E,t),
\label{ch7:geral}
\end{eqnarray}
where $D(\vec{r},E)$ is the diffusion coefficient of the particles, $P(\vec{r},E) \equiv -({\rm d}E/{\rm d}t)$ is the radiative energy loss rate and  $Q_p(\vec{r},E,t)$ is the injection function. We adopt a scalar diffusion coefficient\footnote{In its more general form, the diffusion coefficient is a tensor (e.g., \citealt{ch7:berezinskii_etal_1990}).} that depends only on the particle energy  $D(\vec{r},E) \equiv D(E)$; given the geometry of the scenario considered here, we adopt 
a spherical coordinate system $(R,\theta,\phi)$, with its origin at the cloud centre  (see Fig.~\ref{ch7:estrellitas}). 
{During the calculation for each  star (see details in \S5), we adopt a coordinate system in 
which the star moves along the polar axis. Thus the} 
proton density function $N_p$ depends spatially only on  $R$ and $\theta$, i.e. $N_p$ $\equiv$ $N_p(R,\theta,E,t)$. In this coordinate system Eq.~(\ref{ch7:geral}) becomes: 

\begin{eqnarray}
%\begin{aligned}
 \frac{\partial N_p}{\partial t}
= & D(E)\left[\frac{1}{R^2} \frac{\partial}{\partial R}
  \left( R^2 \frac{\partial N_p}{\partial R} \right) +  
\frac{1}{R^2{\sin}{\theta}} \frac{\partial}{\partial \theta} 
  \left( \sin \theta \frac{\partial}{\partial \theta} \right)\right] \nonumber\\ 
  & - \frac{\partial}{\partial E} \left(P(R,\theta,E)\,N_p  \right)  
+ Q_p(R,\theta,E,t).
\label{ch7:esfer}
%\end{aligned}
\end{eqnarray}

The dominant losses for protons  are  $p-p$ inelastic collisions, while the ionization losses are negligible in the range of energies considered (\citealt{ch5:aharonian_atoyan_1996}). The loss term is $P(R,\theta,E) = E\,(n(R)\,c\, \kappa_{pp}\sigma_{pp})$ where $\kappa_{pp}$ is the inelasticity ($\sim$ 0.45) and $\sigma_{pp}$ is the cross section for inelastic collisions. The  dependence on $R$ and $\theta$ of the radiative losses is given  through the density  -- see Eq.~(\ref{ch7:densi-pro}) --. 

The $\sigma_{pp}$ cross section can be approximated by  \citep{ch7:kelner_etal_2006}:
\begin{equation}
\sigma_{pp} = (34.3+1.88L+0.25L^{2})\biggl[1-\biggl(\frac{E_{\rm th}}{E_{p}}\biggr)^{4}\biggr]^{2} {\mbox mb},
\label{apa:sigma_pp}
\end{equation}
where $L = \ln(E_{p}/1{\rm TeV})$ {and $E_{\rm th} = 1.22$ GeV}.

The injection function $Q_p$ is: 
\begin{eqnarray}
Q_p(R,\theta,E,t) = N_{0}\,E^{-2}\,{\delta}^3(\vec{r} - \vec{r}_{\star}), 
\end{eqnarray}
where $\vec{r}_{\star} = \vec{V}_{\star}t$ is the position of each star with respect to the cloud centre; $N_{0}$ is the normalization constant  which depends on the injected power in relativistic particles for each star, as discussed above. We only consider proton injection when the stars are inside the cloud (i.e. $t \leq \tau$).

The spectral energy distribution of secondary pairs  $N_{e^{\pm}}$ obeys the same transport equation as protons (Eq.~(\ref{ch7:esfer})), but the radiative  term  $P(R,\theta,E)$ and the injection function $Q_{e^{\pm}}(R,\theta,E,t)$ are different. The pairs are injected through the decay of charged pions  produced in the $p-p$ collisions (e.g., \citealt{orellana2007}). The injection of leptons can be calculated from (\citealt{ch7:kelner_etal_2006}): 
 
\begin{eqnarray}
Q_{e^{\pm}}(R,\theta,E,t) = && \int^{\infty}_{E_{e^{\pm}}} \sigma_{pp}(E_p) \,n(R)\,c\,N_p(R,\theta,E_p,t)\times \nonumber\\ &&
F(E_{e^{\pm}},E_p)\,\frac{{\rm d}E_p}{E_p},
\end{eqnarray} 
where $F(E_{e^{\pm}},E_p)$ is the electron spectrum of the decay ${\pi} \rightarrow \mu + \nu_{\mu}$.

The relevant losses for pairs are synchrotron radiation and relativistic Bremsstrahlung (see, for example, \citealt{ch7:aharonian_2004} and references therein).

\subsection{Diffusion}

The diffusion coefficient  is a key parameter in the study  of the diffusion of energetic particles. It is a poorly determined quantity, from both the  observational and theoretical point of view. The theoretical determination of the diffusion coefficient is a very complex task  (e.g., \citealt{ch7:yan_lazarian_2004, ch7:yan_lazarian_2008}) and observations are necessary to constraint the models.  Cosmic-ray diffusion is a non-linear process in which the CRs generate the instabilities that produce the turbulence they interact with  (e.g., 
\citealt{ch7:nava_gabici_2013}, and references therein).
Near the particle accelerators, where the population of CRs is very high, the diffusion coefficient can significantly decrease through instabilities  (\citealt{ch7:ptuskin_etal_2008, ch7:malkov_etal_2013,yan2011}). Furthermore, slow  diffusion is expected   in dense regions  (e.g., \citealt{ch7:ormes_etal_1988}).

Through   gamma-ray observations of the SNR W28 -- a remnant on its radiative phase, localized in a region of dense molecular gas -- illuminating a MC, a  significant suppression of the diffusion coefficient with respect to the averaged galactic values was reported (e.g., 
\citealt{ch7:gabici_2011}). {Also, \citet{yan_lazarian_schlickeiser_2012} using a self-consistent model for the acceleration of CRs and the gamma-ray production in W28, required a high suppression of the ISM diffusion coefficient to match the gamma observations}. In these studies  isotropic diffusion was considered. If the isotropic assumption is dropped, the conclusions are different. Under some physical conditions diffusion becomes anisotropic, particles diffuse preferentially parallel to the magnetic field lines  (e.g., \citealt{ch7:casse_etal_2002}). An anisotropic treatment was developed in \citet{ch7:nava_gabici_2013}, where the diffusion coefficient is not suppressed to fit the observations. Both scenarios fit the observations of the SNR W28, showing that  a great uncertainty exists on the diffusion coefficient.

Here we adopt a phenomenological approach similar to the one used in 
\citet{ch7:gabici_etal_2007}. The diffusion coefficient is a power law in the particle energy:  
\begin{equation}
D(E) = {\chi} D_{10} \left(\frac{E}{10\,{\rm GeV}}\right)^{\delta},
\end{equation}  
where $D_{10}$ is the value of $D$ at  $E = 10$ GeV. The power $\delta$ varies  between $0.3-0.6$ depending on the power-law spectrum of the turbulence of the magnetic field. The parameter $\chi < 1$ takes into account the suppression of the diffusion coefficient inside the turbulent medium of the cloud.  For $\delta$ and $D_{10}$ we take  values considered as typical in the Galaxy  $0.5$, and  $10^{27}$ cm$^{2}$~s$^{-1}$, respectively  (\citealt{ch7:berezinskii_etal_1990}). We study two cases of $\chi$:  $10^{-1}$  -- expected in dense regions  -- and  $10^{-3}$ -- an extreme case  --. 
{ As we are considering an isotropic diffusion coefficient, no specification of the magnetic field 
direction is required in our calculations.}

\subsection{Cosmic-ray sea}\label{CR-sea}
In order to compare the contribution of the injected protons with the cosmic-ray background of the Galaxy we calculate the  cosmic-ray flux expected in the cloud, { following \citet{ch7:gabici_etal_2007}}. 
CRs do not freely penetrate the cloud; they diffuse slowly, especially in dense regions. To get the cosmic-ray distributions  $N_p^{\rm CR}$ consistently  with the parameters  adopted, we solve the transport equation in steady state. This means,  Eq.~(\ref{ch7:esfer}) with $\partial / \partial t = 0$, null injection function and  the condition that the distribution  $N_p^{\rm CR}$ at the edges of the cloud match the cosmic-ray sea distribution. We consider only protons because it is the dominant cosmic-ray component. We also consider the secondary pairs that the high-energy  protons produce in their collisions with the cold protons.
We take the galactic cosmic-ray-flux equal to the locally   observed one\footnote{However, this assumption is {\it ad hoc}, see the discussion in  Sec.~\ref{ch7:dyc}.} (e.g., 
\citealt{ch7:simpson_1983}):
\begin{equation}
J_{\rm CR}^{\rm gal}(E) = 2.2 \left(\frac{E}{\rm GeV}\right)^{-2.75}\,\,{\rm cm}^{-2}\,{\rm s}^{-1}\,{\rm sr}^{-1}\,{\rm GeV}^{-1}.
\end{equation}
We also consider the case in which the background cosmic-ray flux is one order of magnitude less than the  locally observed flux, given by the latter equation. 

\subsection{Emission}
We calculate the  $p-p$ emissivity for protons and the synchrotron emission produced by the secondary pairs. In the ISM the luminosity produced by inverse Compton scattering is in general negligible compared to the $p-p$ contribution (see, \citealt{ch7:bosch-ramon_etal_2005}). Relativistic Bremsstrahlung is significant only at energies smaller than  1 GeV (\citealt{ch7:aharonian_2004, ch7:gabici_etal_2007}) and here we neglect it. In what follows we describe the numerical methods used.

\section{Numerical methods}\label{methods}

The transport equation is solved for a single cosmic-ray injector  at a time. We consider the star moving along the 
polar axis of an spherical system of coordinates. In this way, the system has azimuthal symmetry, allowing a reduction of the dimensionality of the problem. 
The coordinates of the particle
distributions resulting from each injector star are then rotated { (by randomly chosen angles, shown in Fig.~\ref{ch7:estrellitas})}
and interpolated into a 3D spatial grid. 
In this way the resulting distribution of particles is obtained by summing the contribution 
coming from each star. Below, we describe the numerical methods employed for solving the problem for one injector.

We evolve the transport equations for protons and pairs simultaneously (throughout this section $N$  represents the distribution of protons or pairs, without distinction)
in a discrete grid of the phase space 
$(E, R, \theta) \in [ 1 \; {\rm MeV}, \; 100 \; {\rm TeV} ] \times [ 0, \; 50 \; {\rm pc} ] \times [ 0, \; \pi ]$, 
using the finite-volumes method.
The phase space is therefore divided in a grid of cells with central values 
$E_l, R_i, \theta_j$ ($1 \le l \le L$, $1 \le i \le M$, and $1 \le j \le K$). The lengths of the cells ($E_l, R_i, \theta_j$) are given by
$\Delta E_l = E_{l+1/2} - E_{l-1/2}$,  $\Delta R_i = R_{i+1/2} - R_{i-1/2}$,  $\Delta \theta_j = \theta_{j+1/2} - \theta_{j-1/2}$, 
where $\alpha \pm 1/2$ ($\alpha = l,i,j$) are the values at the left/right interface of the cell.
The energy grid ($E_l$, $1\le l \le L$) is logarithmically spaced, while the radial
($R_i$, $1 \le i \le M$) and polar ($\theta_j$, $1 \le j \le K$) grids are uniformly spaced.
In the simulations presented in this work, we employ the grid resolution $(L,M,K) = 
{ (128,64,64).}$~\footnote{When solving the distribution of particles for the galactic CR background, we extend the energy
range to $[1 \; {\rm MeV}, \; 10^{3} \; {\rm TeV}]$. Because of radial symmetry, the grid resolution employed 
is $(L,M,K) = {(144,64,1)}$.}

The density of particles at a given  time $t$ is represented inside each cell by the average
value $N_{l,i,j}(t) \equiv N(E_l, R_i, \theta_j, t)$,
i.e., the number of particles inside the
discrete volume of the phase space $\Delta E_l \Delta R \Delta \theta$ at a time $t$ is given by: $N_{l,i,j}(t) \Delta E_l \Delta V_{i,j}$, 
where $\Delta V_{i,j} \approx 2 \pi R_{i}^{2} \Delta R \sin\theta_j \Delta \theta$.

At $t=0$ we consider $N_{l,j,k}(t=0)=0$ for protons and pairs (i.e., no particles)~\footnote{{ In order to obtain numerically the steady state solution} for the  distribution of the background CRs, 
we consider ${N_{\rm CR}}_{l,j,k}(t=0)=0$
inside the molecular cloud, and we evolve the transport equation { during  enough time for the solution to become time independent.}}.

The energy boundary conditions we impose to $N$ are
\begin{equation}
	N(E < 1\;{\rm MeV},R,\theta,t) = N (E > 100\;{\rm TeV},R,\theta,t) = 0,
\end{equation}
i.e., no particles outside the energy bounds. { In fact, these limits do not influence the system evolution, 
because the upper limit is above the maximum energy of the injected protons, 
at the same time that the advection in the energy space is always directed to smaller energies. 
The lower bound is physically fixed because relativistic particles have kinetic energies greater or of the order of their rest mass. 
}The spatial boundary conditions are
\begin{equation}
	N(E,R > 50\;{\rm pc},\theta,t) = 0,
\end{equation}
i.e., no particles outside  the MC, and
\begin{equation}
	\frac{\partial N(E,R,\theta=0,t)}{\partial \theta} = \frac{\partial N(E,R,\theta=\pi,t)}{\partial \theta} = 0,
\end{equation}
due to the azimuthal symmetry. 

For the  calculation of the CR background distribution (see Sec.~\ref{CR-sea}) the following
boundary conditions are used instead:
\begin{equation}
	N_{CR}(E > 10^{3}\;{\rm TeV},R,\theta,t) = 0,
\end{equation}
and
\begin{equation}
	N_{CR}(E,R > 50\;{\rm pc},\theta,t) = \frac{4 \pi}{c} J_{CR} \left( \frac{E}{1\;{\rm GeV}} \right)^{-\alpha},
\end{equation}
where the last condition is valid only for protons, and $J_{CR}$ is the CR flux at $E = 1$~GeV. The other boundary conditions are identical to the ones described before. 
{Here we should remark that the upper boundary in the energy space affects the solution because,
differently from the protons injected from the stars,
there is no established cut-off at the cosmic-ray highest energies. 
We keep the energy range broader enough to minimize these effects.}

The numerical integration of the transport equation is performed {using the operator splitting method, in the way described below}. 
Each time-step integration evolves the particle density distribution on the grid $N^{n}_{l,i,j} \equiv N_{l,i,j}(t^{n})$
from  time $t^{n}$ to  time $t^{n+1} = t^{n} + \Delta t$, through three sub-steps described bellow.  

Firstly, we integrate only the losses
term of the transport equation,
\begin{equation}
	\frac{\partial N(E,R,\theta,t)}{\partial t} = {\color{red} -} \frac{\partial }{\partial E} \left[
	F(E,N(E,R,\theta,t)) \right],
\end{equation}
where the flux $F(E, N(E,R,\theta,t)) \equiv P(E) N(E,R,\theta,t)$ is an advection in the energy space. In the finite-volume 
formulation, we employed an upwind scheme of second order~\footnote{{ We use the 
Piecewise Linear Method (PLM) with the Monotonic Central limiter, which is second order accurate on a uniform grid.}} for calculating the fluxes at the interface of the cells.
The intermediate solution $N^{n+1/3}_{l,i,j}$ is then obtained from the solution $N^{n}_{l,i,j}$ at time $t^{n}$
through the explicit Euler method:
\begin{equation}
	\frac{N^{n+1/3}_{l,i,j} - N^{n}_{l,i,j}}{\Delta t} = - \frac{1}{\Delta E_l} \left( F^{n}_{l+1/2,i,j} - F^{n}_{l-1/2,i,j} \right).
\label{eq:adv}
\end{equation}
Here $F^{n}_{l \pm 1/2,i,j}$ represents the numerical fluxes at the cell interfaces.

Secondly, we integrate only the diffusion part of the transport equation,
\begin{equation}
\frac{\partial N}{\partial t}
=  D(E)\left[\frac{1}{R^2} \frac{\partial}{\partial R}
  \left( R^2 \frac{\partial N}{\partial R} \right) 
+ \frac{1}{R^2{\sin}{\theta}} \frac{\partial}{\partial \theta} 
  \left( \sin \theta \frac{\partial N}{\partial \theta} \right)\right].
 \label{eq:diff}
\end{equation}
For solving such integration, we use the semi-implicit Cranck-Nicolson method, with the gradients calculated 
at the cell interfaces, using central differences. This scheme is, therefore, second order accurate. We then get a second intermediate solution,
$N^{n+2/3}_{l,i,j}$ from the solution $N^{n+1/3}_{l,i,j}$:
\begin{equation}
	\frac{N^{n+2/3}_{l,j,k} - N^{n+1/3}_{l,j,k}}{\Delta t} = \frac{1}{2} D(E) \left[L(N^{n+2/3}_{l,i,j}) + L(N^{n+1/3}_{l,i,j}) \right],
\label{eq:diff_implicit}
\end{equation}
with 
\begin{flalign}
	& L(N_{l,j,k}) = \nonumber \\
& \frac{1}{\Delta V_{i,j}} \left\{ \Delta S_{i+1/2,j} \frac{N_{i+1,j} - N_{i,j}}{\Delta R} - {\Delta}S_{i-1/2,j} \frac{N_{i,j} - N_{i-1,j}}{\Delta R} \right. \nonumber \\
	& + \Delta S_{i,j+1/2} \frac{N_{i,j+1} - N_{i,j}}{R_{i} \Delta \theta}
- \Delta S_{i,j-1/2} \frac{N_{i,j} - N_{i,j-1}}{R_{i} \Delta \theta} \left. 
\vphantom{
\Delta S_{i+1/2,j} \frac{N_{i+1,j} - N_{i,j}}{\Delta R_{i}} - dS_{i-1/2,j} \frac{N_{i,j} - N_{i-1,j}}{\Delta R_{i}}
}
\right\} &&,
\label{eq:diff_laplacian}
\end{flalign}
where $\Delta S_{i\pm1/2,j}$ and $\Delta S_{i,j\pm1/2}$ are the cell surfaces at the interfaces indicated by the indices. 
{ Our algorithm first tries to solve the linear system implied by Eqs.~(\ref{eq:diff_implicit}) and ~(\ref{eq:diff_laplacian}) using the iterative 
Krylov Space scheme GMRESR  
(\citealt{VanderVorst1994}); when it fails to converge to the solution (with relative residue $< 10^{-7}$ in the 2-norm), it employs our implementation 
of a Multigrid solver. }

Finally, we add the contributions due to the injection using the Euler explicit method:
\begin{equation}
	\frac{N^{n+1}_{l,i,j} - N^{n+2/3}_{l,i,j}}{\Delta t} = Q^{n+2/3}_{l,i,j}.
\end{equation}
Then, the final solution $N^{n+1}_{l,i,j}$ at  $t^{n+1}$ is obtained.

The time-steps $\Delta t$ are chosen in accordance with the Courant-Friedrichs-Lewy stability criterion
for the minimum time step of the advection equation ($\delta t_{\rm adv}$, Eq.~\ref{eq:adv}) and of the diffusion 
equation ($\delta t_{\rm dif}$, Eq.~\ref{eq:diff}). We also impose the condition that the time step must be smaller than the time it takes the star to cross one cell ($\delta t_{\rm inj}$). These three time-steps constraints 
are calculated with the following formulae:
\begin{equation}
	\delta t_{\rm adv} = \min \left\{ \frac{\Delta E_l}{P(E_l,R_i,\theta_j)} \right\},
\end{equation}
\begin{equation}
	\delta t_{\rm dif} = \min \left\{ \frac{\Delta \theta (\Delta R)^{2}}{D(E_l)} \right\},
\end{equation}
\begin{equation}
	\delta t_{\rm inj} = \frac{\Delta R}{V_{\star}}.
\end{equation}
The minimum is taken over all the grid values. The time-step is then 
$\delta t = \min \left\{ \epsilon_{\rm adv} \delta {t}_{\rm adv}, \epsilon_{\rm dif} \delta {t}_{\rm dif}, \epsilon_{\rm inj} \delta {t}_{\rm inj} \right\}$. 
We use the safety factors $\epsilon_{\rm adv}=0.5$, $\epsilon_{\rm inj}=0.5$. As the semi-implicit method used for the diffusion equation is unconditionally stable, we  use $\epsilon_{\rm dif}=10$. 

 We checked the convergence of the solutions presented below performing additional runs (not shown) of some of the models using different resolutions 
(lower and higher). In addition, we also performed several tests changing the order of the operators sequence
(advection, diffusion, injection), and we have not found significant difference between the results.

\section{Results}\label{resu}

\subsection{Particle  distributions}
In what follows we show series of maps of the protons and $e^{\pm}$ pairs distributions at a fixed energy and different times, adding the contributions of the six stars. The 2D maps are constructed integrating the 3D data along an arbitrary line of sight, chosen to be on the $z$ direction (see Fig.~\ref{ch7:estrellitas}).

In  Figs.~\ref{ch7:map-pro} and ~\ref{ch7:map-pro2} we show the evolution maps of the proton distribution, for two energies: $10$ GeV and $10$ TeV, for  $\chi = 10^{-1}$ and $\chi = 10^{-3}$, respectively (i.e., fast and slow diffusion). The most energetic particles diffuse faster because of the  power-law  dependence of  $D(E)$ with energy. The different stars can be identified in the maps during the evolution; particularly the stars \#6 and \#4 are seen while they abandon the core region of the cloud.  The most energetic star, \#1,  produces an important anisotropy in the particle distribution. 

\begin{figure*}
\centering
\begin{tabular}{c}
\input{chi0p1_FE_EN1} \\
\input{chi0p1_FE_EN4}
\end{tabular}
\caption{Distribution of $e^{\pm}$ pairs
at fixed energy projected along the line of sight for the case  $\chi = 10^{-1}$. Top: $E_e = 10$ GeV; bottom: $E_e = 10$ TeV. 
Time evolves from left to right. The  level curves correspond to 0.1 (dotted), and 1 (dashed) times the  pairs background distribution.}
\label{ch7:map-par}
\end{figure*}
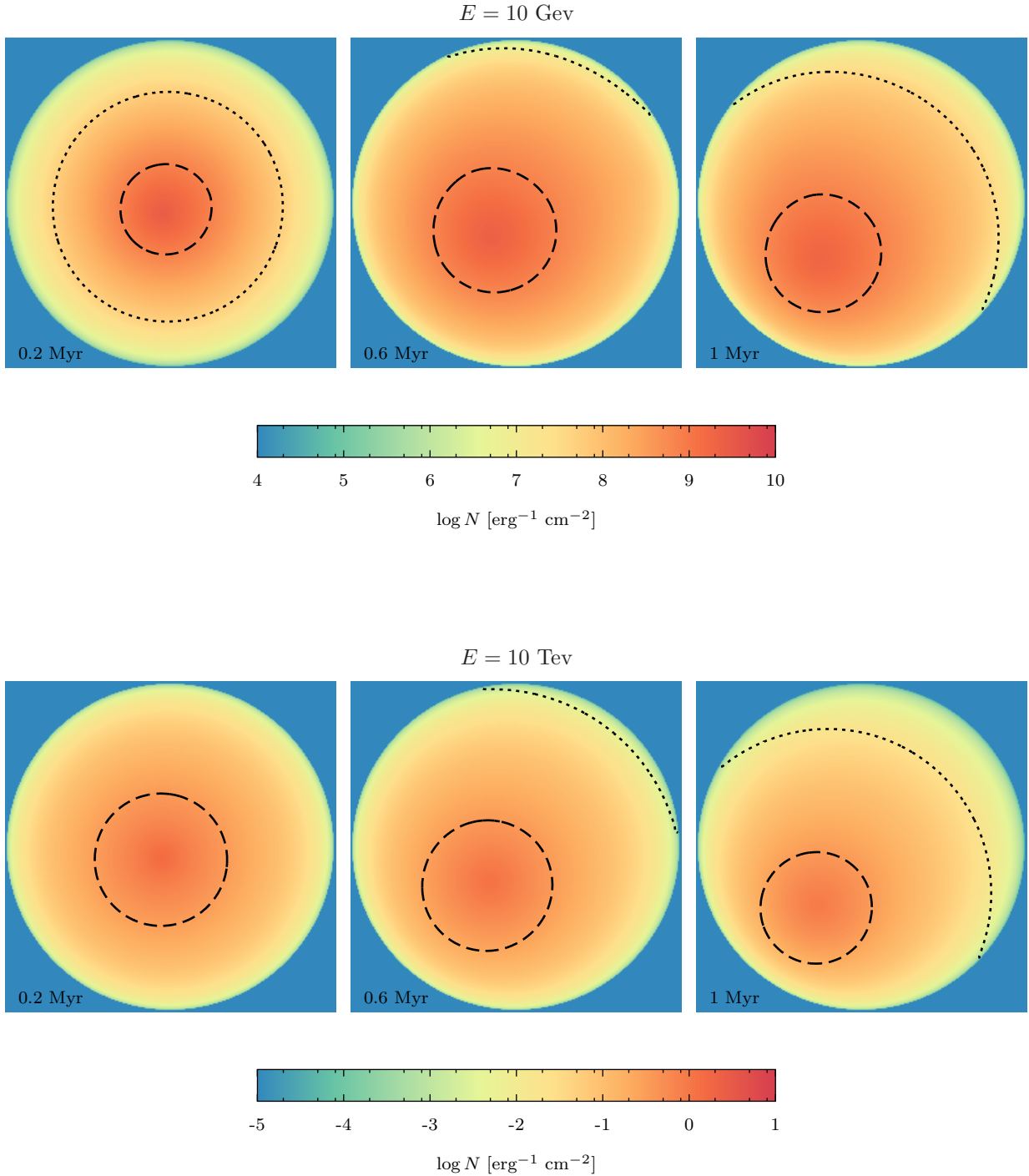

The Fig.~\ref{ch7:map-par} shows the evolution maps of the pairs  created in the $p-p$ interactions. Maps  correspond to two energies: $10$ GeV and $1$ TeV, for the case  $\chi = 10^{-1}$. The pair density is higher in the densest regions of the cloud; in the core also the synchrotron losses  are much more intense.

\begin{figure*}
\centering
\begin{tabular}{c}
\input{chi0p1_FP_EN1} \\
\input{chi0p1_FP_EN4}
\end{tabular}
\caption{Proton distribution at fixed energy projected along the line of sight for the case  $\chi = 10^{-1}$. Top: $E_p = 10$ GeV; bottom: $E_p = 10$ TeV. 
Time evolves from left to right. The level curves correspond to 0.1 (dotted), 1 (dashed), and 10 (continuous) times the CR background distribution.}
\label{ch7:map-pro}
\end{figure*}
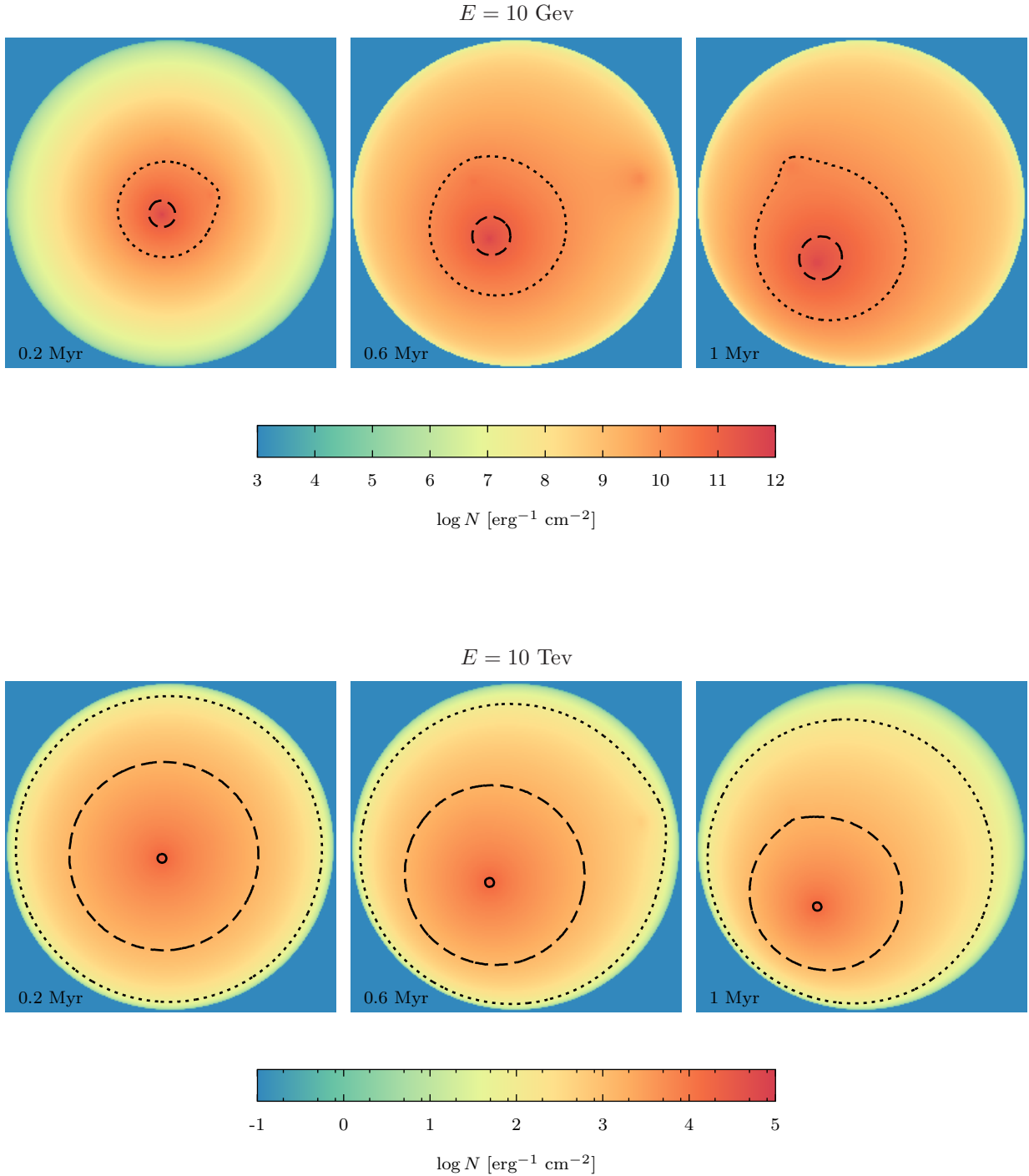

\begin{figure*}
\centering
\begin{tabular}{c}
\input{chi0p001_FP_EN1} \\
\input{chi0p001_FP_EN4}
\end{tabular}
\caption{Same as the last figure, for $\chi = 10^{-3}$.}
\label{ch7:map-pro2}
\end{figure*}
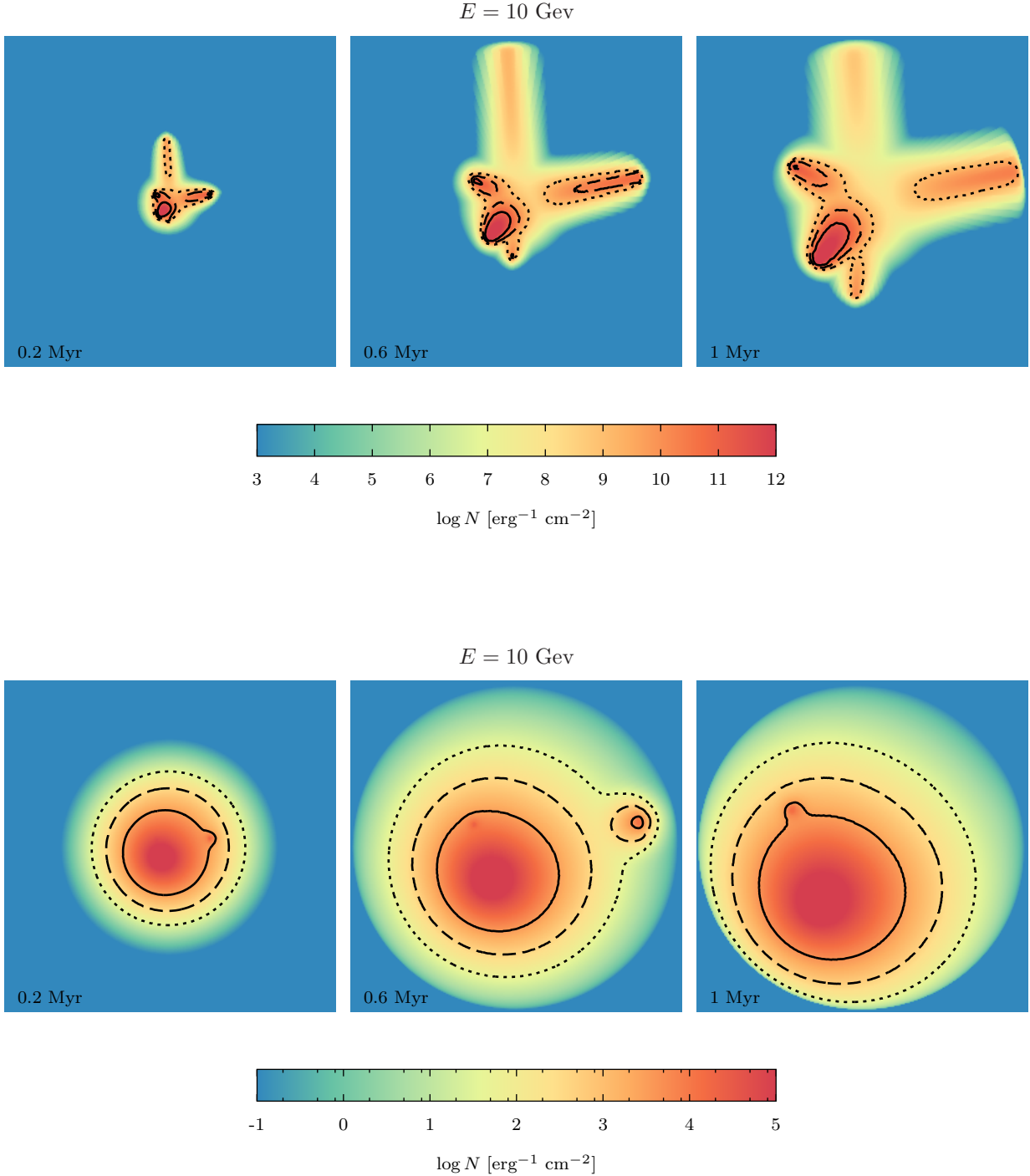

\subsection{Emissivity}

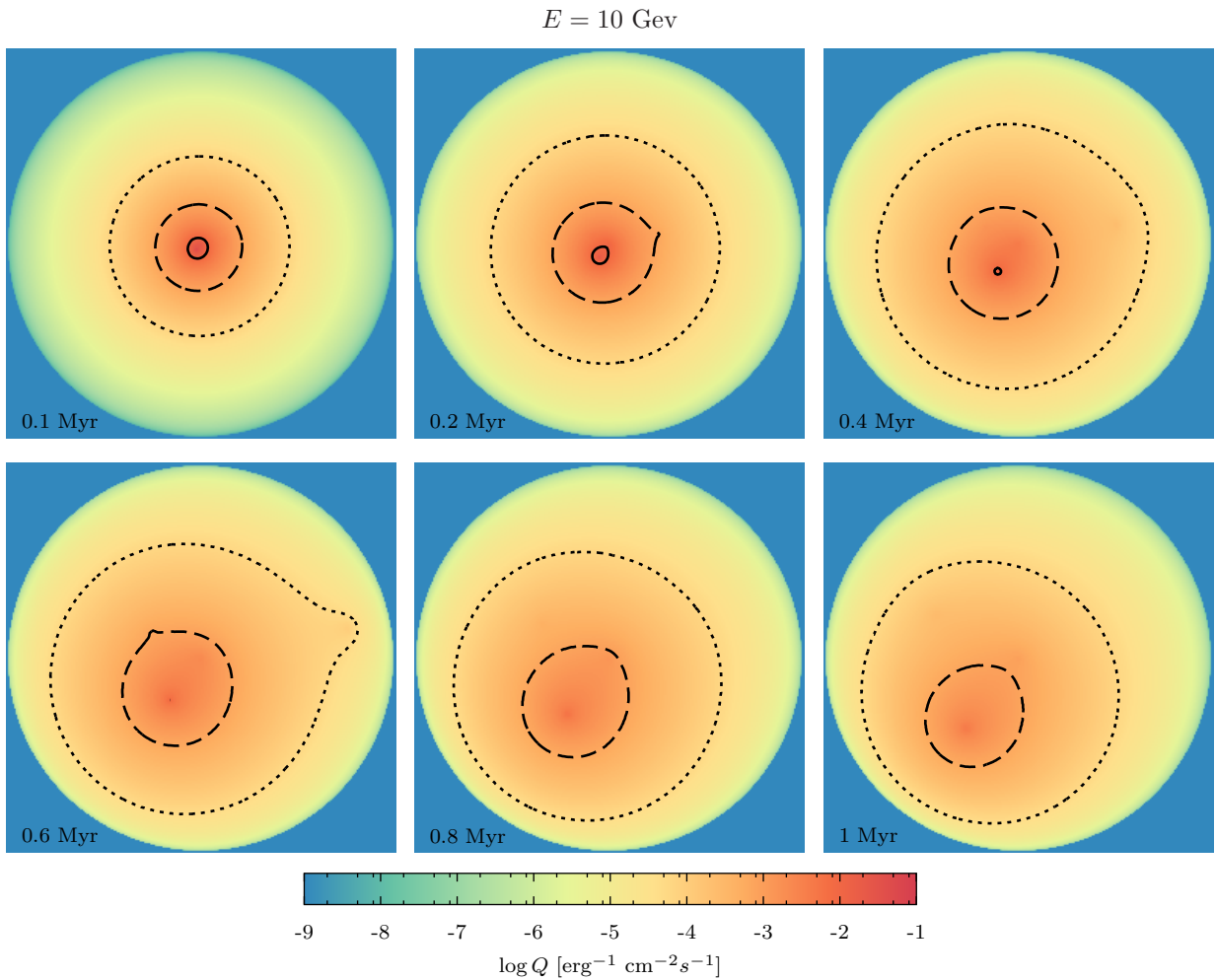
\begin{figure*}
\centering
\begin{tabular}{c}
\input{chi0p1_QGAMMA_EN2}
\end{tabular}
\caption{Gamma-ray emissivity evolution, projected along the line of sight for   $\chi = 10^{-1}$. The photon energy is $E = 10$ GeV. 
Time evolves from top left to bottom right.
The level curves represent  0.1 (dotted), 1 (dashed), and 10 (continuous) times the background emissivity.}
\label{ch7:mapa-pp}
\end{figure*}

\begin{figure*}
\centering
\begin{tabular}{c}
\input{chi0p001_QGAMMA_EN2}
\end{tabular}
\caption{Same as Fig.~\ref{ch7:mapa-pp} for slow diffusion ($\chi = 10^{-3}$).}
\label{ch7:mapa-pp2}
\end{figure*}
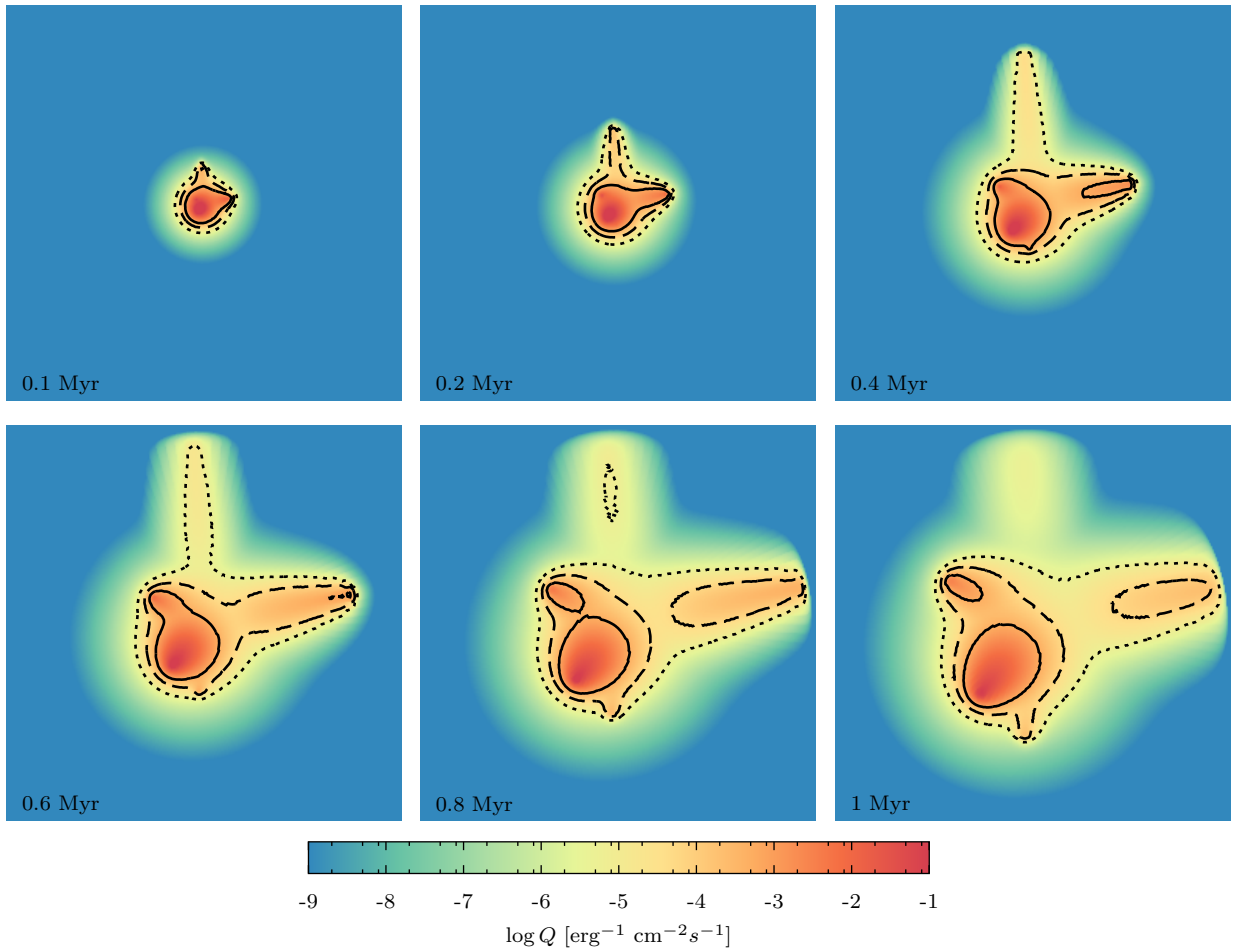

{We calculate the synchrotron emissivity produced by the interaction of the secondary pairs with the inhomogeneous magnetic field given by Eq.~(\ref{B}). We also  compute the emissivity produced in the $p-p$ collisions  by neutral pion decays; the target density is $n(R)$, the density of the MC mass, see Eq.~(\ref{ch7:densi-pro}).} The corresponding formulae can be found in, e.g., \cite{ch7:aharonian_2004} and references there in. Figures~\ref{ch7:mapa-pp} and \ref{ch7:mapa-pp2} show the evolution of the  gamma emissivity for $E = 10$ GeV, for both diffusion scenarios considered here respectively. The emission is highly anisotropic and its intensity follows the injectors motion. This is more clear in the case of slow diffusion. The maximum emissivity is reached immediately after the ejection of the stars, 
{while the high-energy particles are concentrated in the high density (and high magnetic field intensity) MC core.}

\begin{figure*}
\centering
\begin{tabular}{c}
\input{chi0p1_QSYNC_EN2}
\end{tabular}
\caption{Evolution of the synchrotron emissivity projected along the line of sight, for $\chi = 10^{-1}$. The photon energy is  $E = 1$ keV. 
Time evolves from top left to bottom right. The level curves represent  0.1 (dotted), 1 (dashed), and 10 (continuous) times the background emissivity.}
\label{ch7:mapa-sin}
\end{figure*}
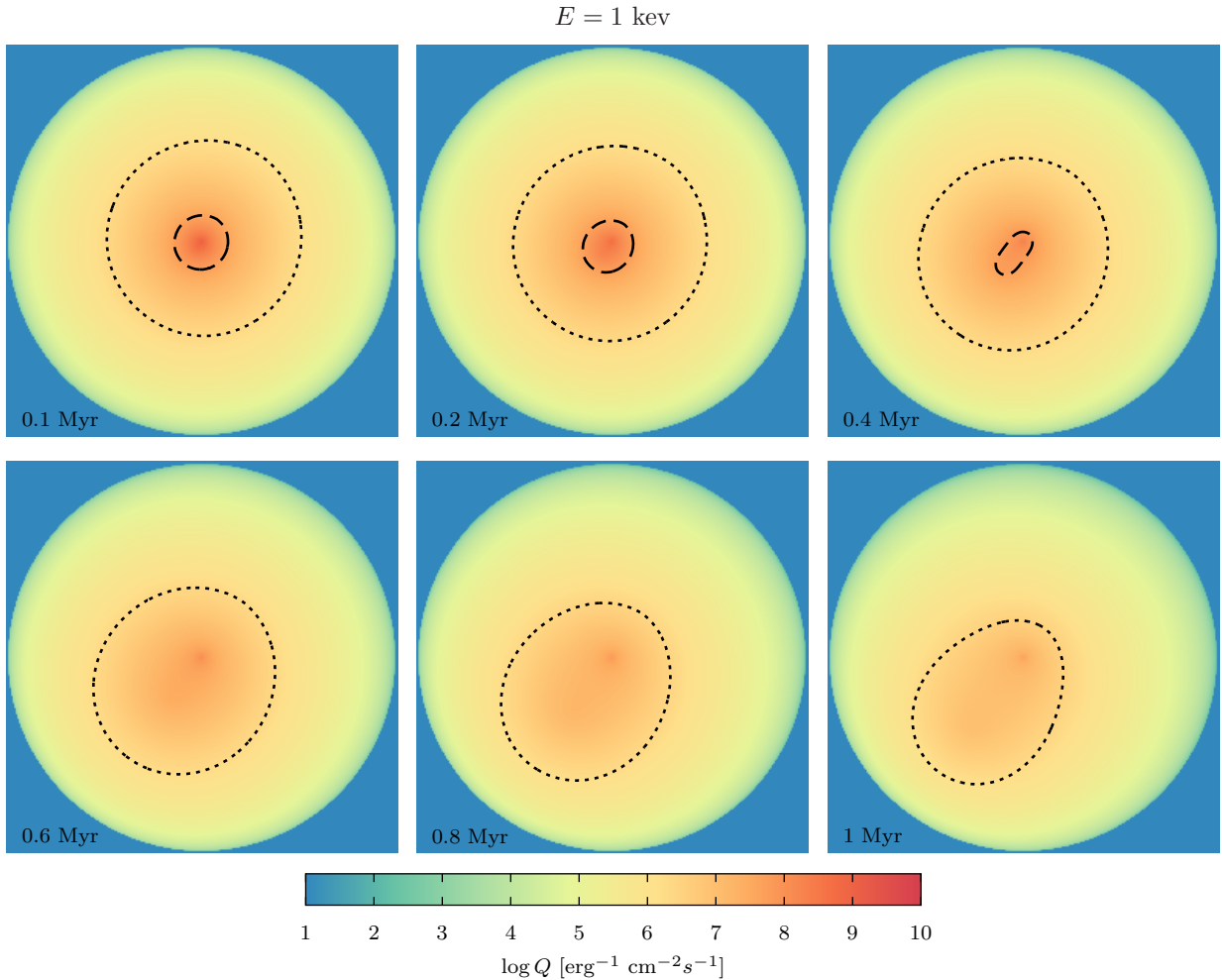

Pairs are created with high energies and they produce synchrotron radiation from radio to X-rays. In the maps displayed in  Fig.~\ref{ch7:mapa-sin} we show the evolution of the synchrotron  emissivity projected along the line of sight, for  $E = 1$ keV (soft X-rays) and $\chi = 10^{-1}$. Here we sum the contributions of all stars, but not all of them produce pairs that emit synchrotron radiation up to this energy. Because of  the dependence of this emission with the magnetic field ($\propto$ $B^{2}$), the radiation  is considerably more intense in the centre of the cloud, where $B$ is higher. 

\subsection{Spectral energy distributions}

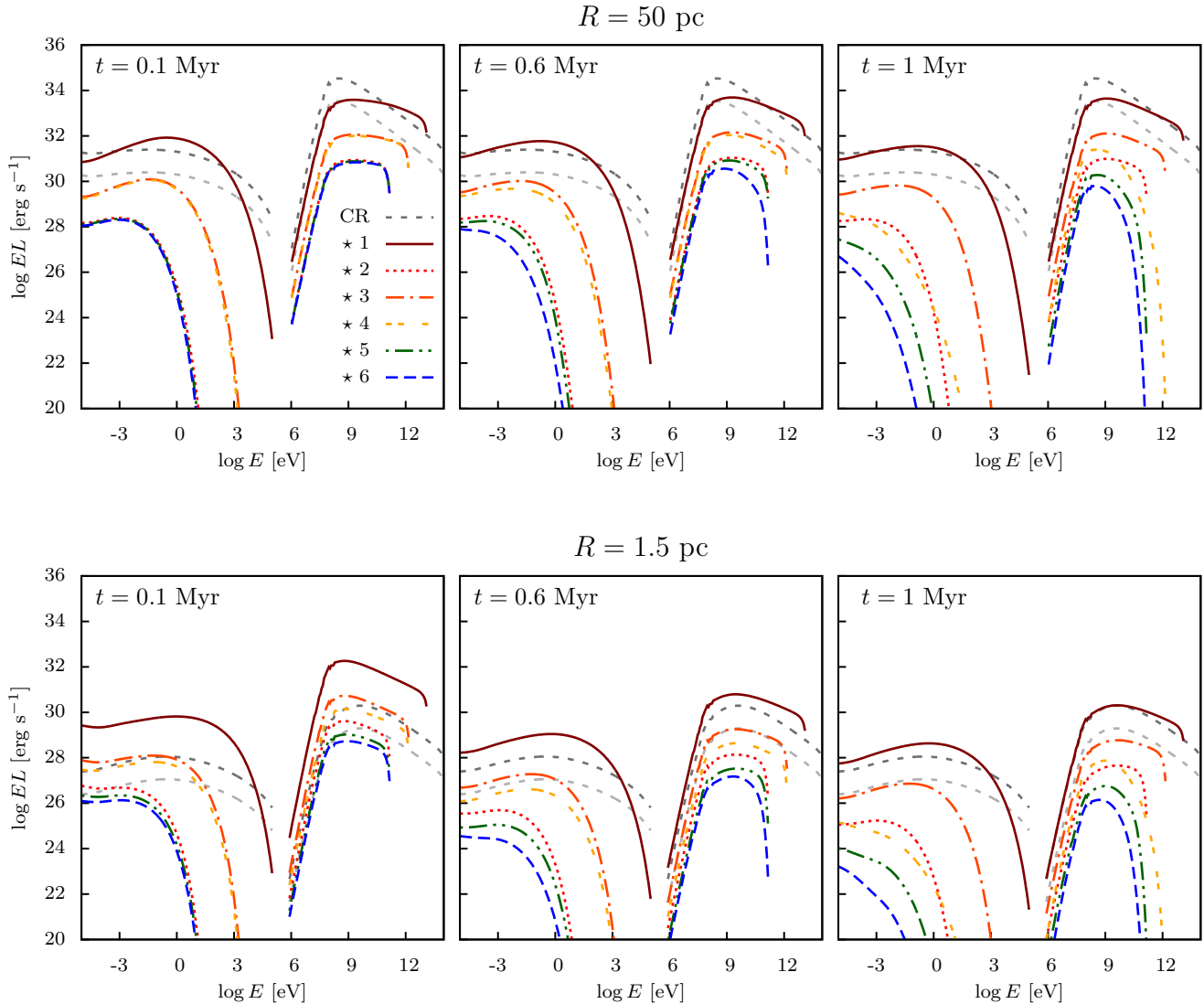
\begin{figure*}
\hfill
\begin{tabular}{c}
\input{chi0p1_SED} \\
\input{chi0p1_SED_CORE}
\end{tabular}
\caption{Evolution of the total SED  (top) and the nuclear SED  (bottom) in the  diffusion case $\chi = 10^{-1}$. The emission produced by the cosmic-ray background is indicated by ``CR''. The light-grey line corresponds to the case in which the flux of CRs is lower than the flux observed locally. Time evolves from left to right.}
\label{ch7:evol-sed}
\end{figure*}

Figure~\ref{ch7:evol-sed} shows the evolution of the total SEDs -- integrated over a sphere  of radius $R_{\rm MC} = 50$ pc -- and the nuclear  SED -- integrated inside a sphere of radius $r = 1.5$ pc --, for $\chi = 10^{-1}$. The luminosity produced by the cosmic-ray background  is also shown. 

The gamma emission and the synchrotron radiation produced by  stars  \#2-\#6 are negligible in the total SED, for both cosmic-ray backgrounds. The contribution of  star \#1 dominates  the spectrum over  the luminosity produced by the CRs (dark-grey line in the SEDs) {in the energy ranges $10^{-3}$~eV to $1$~keV and}  $100$~GeV to 10~TeV. If the background of CRs is the lowest case consider here  (light-grey line in the SEDs),  star~\#1 dominates the SED in the energy ranges  $10^{-5}$ eV - 10 keV and 1 MeV - 10 TeV.

In the case of the nuclear SED the non-thermal emission produced by  star \#1 dominates over the background emission. The emission from {the star~\#3} exceeds the background when the star is near the centre {in the energy ranges $10^{-3}$~eV - 10~eV and 1~MeV - 1~TeV}. If the cosmic-ray density is lower, the emission produced by the weaker stars (\#2, \#3 and \#4) exceeds the background at radio wavelengths and at  energies $\sim$ MeV, when the injection starts.  In this case the {contribution of the star \#3 is} greater than the background during a longer time, until  $\sim$ $0.7$ Myr.

\begin{figure*}
\hfill
\begin{tabular}{c}
\input{chi0p001_SED} \\
\input{chi0p001_SED_CORE}
\end{tabular}
\caption{Same as Fig.~\ref{ch7:evol-sed} in the case $\chi = 10^{-3}$.}
\label{ch7:evol-sed2}
\end{figure*}
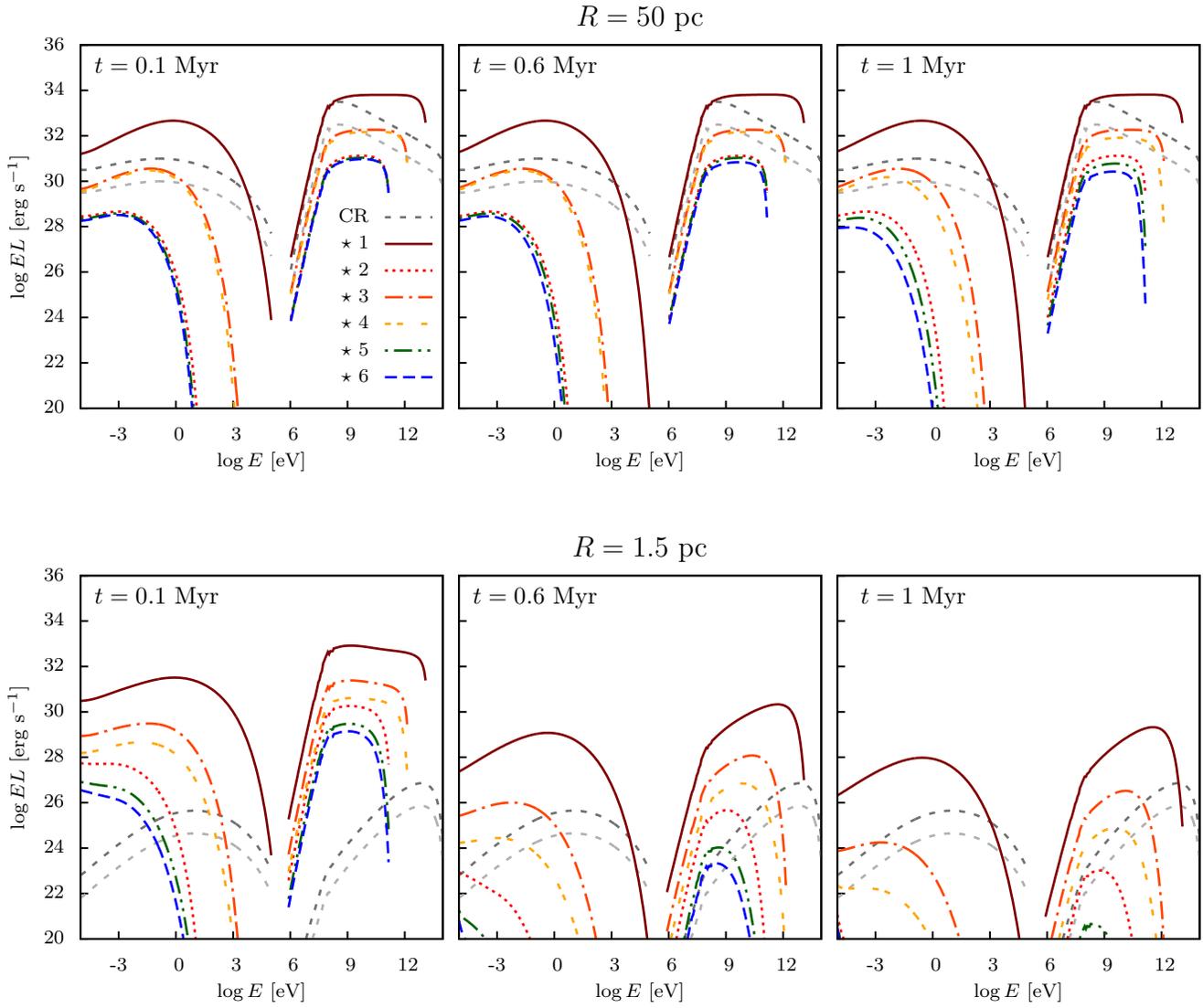

Figure~\ref{ch7:evol-sed2} shows the evolutions of the total and nuclear  SEDs for the case $\chi = 10^{-3}$ (slow diffusion). In this case the total SED  is dominated by the  contribution of  star \#1, from radio to hard X-rays, and  energies from $\sim$ MeV  to 10 TeV, during all the integration time. If the level of CRs is the lowest one considered here,  stars \#3 and \#4  also exceed the background during all the integration time {from the lowest energies up to $10^{2}$~eV and from 1~MeV to 1~TeV}. In the nuclear region the contribution of all stars exceeds the background (for both  values of the CR level considered), when the stars are near the core. The stars  \#3 and \#4 in the range of energies from  MeV to GeV and TeV, respectively, overcome the cosmic-ray contribution even at the final integration time (1 Myr). The slow diffusion makes the injected particles to stay longer inside the cloud, radiating in the denser regions; {the CRs penetrate less in the cloud because the larger diffusion time makes the energy losses more efficient.}

{It should be observed that the electron-positron pairs produced by the background CRs are limited in 
energy due to our energy grid limit of $10^{3}$ TeV for the CR protons. Changing the upper limit in the energy increases the number of high-energy pairs, and consequently increases the high-energy tail of the background  synchrotron emission. However, the increase on the emission is small and occurs at photon energies greater than 1~keV, where the synchrotron emission from the stars decreases.}

\begin{figure*}
\begin{tabular}{c c}
	\input{chi0p1_en_prot_time} &
	\input{chi0p1_en_elec_time} \\
	\input{chi0p1_L_gamm_time} &
	\input{chi0p1_L_sync_time}
\end{tabular}
\caption{Top: evolution of total energy in protons (left) and pairs 
(right). Bottom: evolution of the total gamma (left) and synchrotron 
(right) luminosity. Here $\chi = 10^{-1}$.}
\label{total}
\end{figure*}
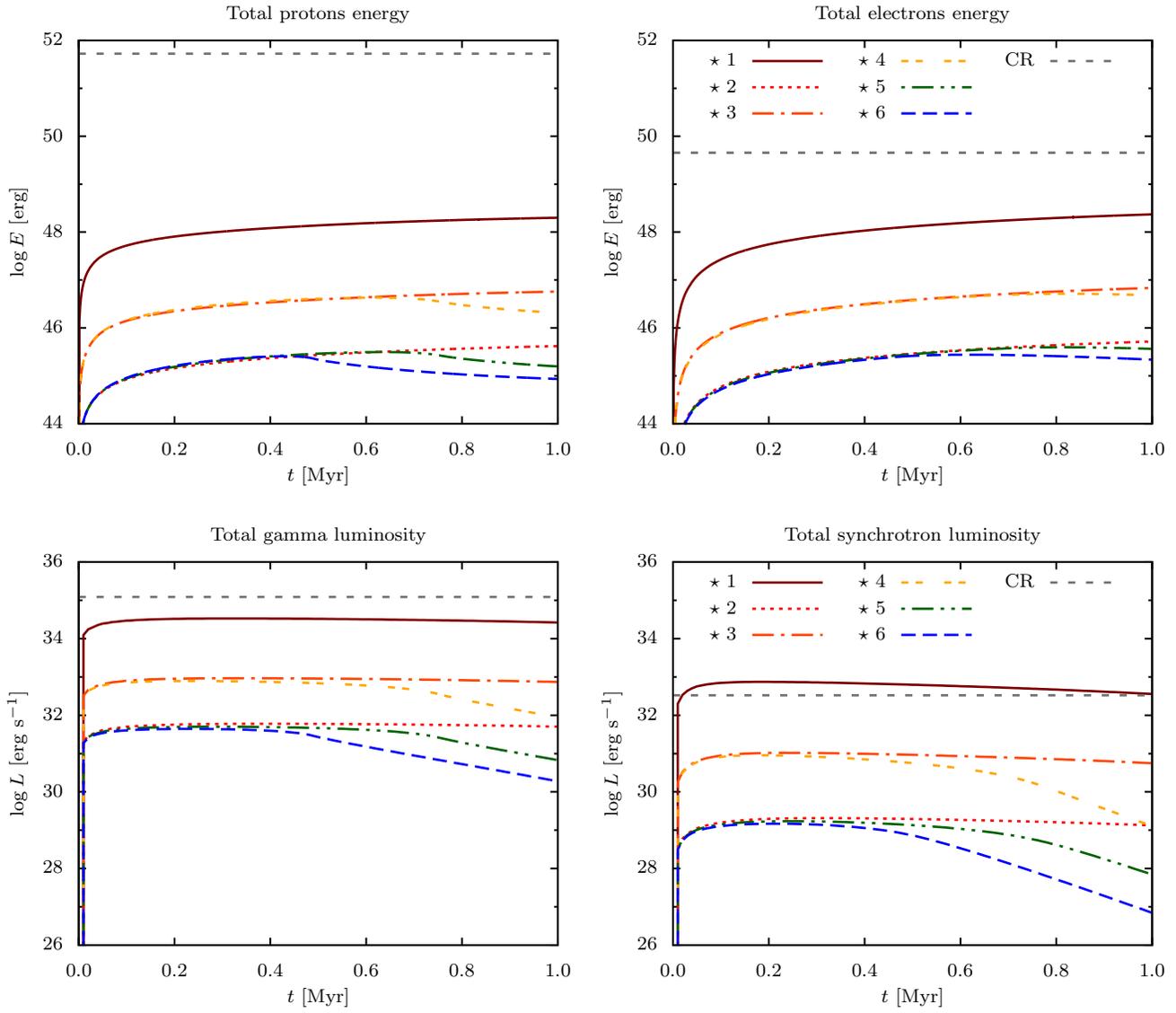

Finally,  the evolution of the total energy in protons (left) and pairs 
(right) for $\chi = 10^{-1}$ are shown in  Fig.~\ref{total} (top). The total energy of protons and pairs for stars \#1, \#2 and \#3 increases slowly with time. As the stars move away from the central region of the MC the energy losses diminish as $n$ and $B$ decrease. For stars \#4, \#5 and \#6 the total energy drops with time for $t \gtrsim \tau$ (see Table~\ref{ch7:tab-star}); this is because  the injection stops when the star leaves the MC.  The evolution of the total gamma (left) and synchrotron (right) luminosity, also for $\chi = 10^{-1}$, can be seen  in Fig.~\ref{total} (bottom). Initially, the total gamma luminosity increases very fast for all stars. As the stars move away from the centre, $n$ and $B$ decrease. However, the gamma luminosity does not decrease for stars \#1, \#2 and \#3. For the fastest stars (\#4, \#5 and \#6)  the gamma luminosity   decreases because they leave the MC during the integration time. The total synchrotron luminosity also increases very fast at the beginning, but after some time it starts to decrease for all stars; this might be because $B$ decreases faster than $n$ with $R$, and due to the quadratic dependence of the synchrotron radiation with $B$.

\section{Discussion and conclusions}\label{ch7:dyc}

The locally injected  protons and secondary pairs, under some assumptions,  dominate over the cosmic-ray sea that penetrates the cloud. The gamma-ray luminosity produced by the protons reaches values of the same order as some of the detected luminosities in  MCs, e.g., $\sim$ $10^{34}$~erg~s$^{-1}$.

A very energetic star, as the case of star  \#1, can inject a significant amount of protons in the medium and it is able to produce important gamma-ray emission over the whole cloud, overcoming the background emission produced by the cosmic-ray sea. This star dominates the gamma rays during most of the time in both cases considered here for the diffusion -- $\chi = 10^{-1}$ and  $\chi = 10^{-3}$ --. The contribution of the less powerful stars dominates locally over the cosmic-ray contribution. However, stars with weak winds, as  stars \#2, \#5 and \#6  do not  inject  enough  power to overcome globally the emission produced by the cosmic-ray sea. The injected power we adopt here depends on the acceleration model of particles in {\bss} of runaway stars (\citealt{delvalle2012,delvalle2014}), so the actual injected power in specific sources could differ. 

The non-thermal emission from radio to X-rays is significant, with luminosities of  {almost $10^{32}$} erg~s$^{-1}$. However, in a MC thermal radiation might dominate in many regions of this energy range. Additionally, the absorption produced by the MC matter is expected to be very intense from the IR to soft X-rays. Low radio frequencies are recommended, then, for the observationally study of large clouds.   

As can be inferred from the cases we present here, the results are very sensitive to particle diffusion and to the ambient cosmic-ray levels. Both quantities are not very well known. This fact makes the present research particularly valuable in the light of the inverse problem: detailed radio maps, including polarization, along with gamma-ray imaging might become a powerful tool to probe the physical conditions in the clouds.     

The hypothesis  that a uniform flux of CRs pervades the whole Galaxy can be inappropriate in some cases, especially at small spatial scales (\citealt{ch7:gabici_2013}). The locally observed cosmic-ray flux might not represent the cosmic-ray population of the whole Galaxy. The local cosmic-ray flux as a matter of fact, could be contaminated by a few local sources. This assumption requires an observational confirmation, that might come from gamma-ray  observations of nearby passive MCs (\citealt{ch7:aharonian_2004}). 

Regarding the diffusion coefficient, additional observations are necessary  to obtain solid constraints. The forthcoming gamma-ray observatory CTA might play a fundamental role in this subject because of its great angular resolution and sensitivity (e.g., \citealt{ch7:gabici_2013, ch7:pedaletti_etal_2013}).  

\section*{Acknowledgments}
This work is supported by AYA2010-21782-C03-01 (Spain) and PICT 2007-00848/2012-00878, Pr\'estamo BID (ANPCyT). R.S.L acknowledges support from the Brazilian agency FAPESP (2013/15115-8). This work has made use of the computing facilities of the Laboratory of Astroinformatics (IAG/USP, NAT/Unicsul), whose purchase was made possible by the Brazilian agency FAPESP (grant 2009/54006-4) and the INCT-A.
 
\bibliographystyle{mn2e}
\bibliography{refs_chap7}

\end{document}

%% file: chi0p1_FP_EN1.tex
% GNUPLOT: LaTeX picture with Postscript
\begingroup
  \makeatletter
  \providecommand\color[2][]{%
    \GenericError{(gnuplot) \space\space\space\@spaces}{%
      Package color not loaded in conjunction with
      terminal option `colourtext'%
    }{See the gnuplot documentation for explanation.%
    }{Either use 'blacktext' in gnuplot or load the package
      color.sty in LaTeX.}%
    \renewcommand\color[2][]{}%
  }%
  \providecommand\includegraphics[2][]{%
    \GenericError{(gnuplot) \space\space\space\@spaces}{%
      Package graphicx or graphics not loaded%
    }{See the gnuplot documentation for explanation.%
    }{The gnuplot epslatex terminal needs graphicx.sty or graphics.sty.}%
    \renewcommand\includegraphics[2][]{}%
  }%
  \providecommand\rotatebox[2]{#2}%
  \@ifundefined{ifGPcolor}{%
    \newif\ifGPcolor
    \GPcolortrue
  }{}%
  \@ifundefined{ifGPblacktext}{%
    \newif\ifGPblacktext
    \GPblacktextfalse
  }{}%
  % define a \g@addto@macro without @ in the name:
  \let\gplgaddtomacro\g@addto@macro
  % define empty templates for all commands taking text:
  \gdef\gplbacktext{}%
  \gdef\gplfronttext{}%
  \makeatother
  \ifGPblacktext
    % no textcolor at all
    \def\colorrgb#1{}%
    \def\colorgray#1{}%
  \else
    % gray or color?
    \ifGPcolor
      \def\colorrgb#1{\color[rgb]{#1}}%
      \def\colorgray#1{\color[gray]{#1}}%
      \expandafter\def\csname LTw\endcsname{\color{white}}%
      \expandafter\def\csname LTb\endcsname{\color{black}}%
      \expandafter\def\csname LTa\endcsname{\color{black}}%
      \expandafter\def\csname LT0\endcsname{\color[rgb]{1,0,0}}%
      \expandafter\def\csname LT1\endcsname{\color[rgb]{0,1,0}}%
      \expandafter\def\csname LT2\endcsname{\color[rgb]{0,0,1}}%
      \expandafter\def\csname LT3\endcsname{\color[rgb]{1,0,1}}%
      \expandafter\def\csname LT4\endcsname{\color[rgb]{0,1,1}}%
      \expandafter\def\csname LT5\endcsname{\color[rgb]{1,1,0}}%
      \expandafter\def\csname LT6\endcsname{\color[rgb]{0,0,0}}%
      \expandafter\def\csname LT7\endcsname{\color[rgb]{1,0.3,0}}%
      \expandafter\def\csname LT8\endcsname{\color[rgb]{0.5,0.5,0.5}}%
    \else
      % gray
      \def\colorrgb#1{\color{black}}%
      \def\colorgray#1{\color[gray]{#1}}%
      \expandafter\def\csname LTw\endcsname{\color{white}}%
      \expandafter\def\csname LTb\endcsname{\color{black}}%
      \expandafter\def\csname LTa\endcsname{\color{black}}%
      \expandafter\def\csname LT0\endcsname{\color{black}}%
      \expandafter\def\csname LT1\endcsname{\color{black}}%
      \expandafter\def\csname LT2\endcsname{\color{black}}%
      \expandafter\def\csname LT3\endcsname{\color{black}}%
      \expandafter\def\csname LT4\endcsname{\color{black}}%
      \expandafter\def\csname LT5\endcsname{\color{black}}%
      \expandafter\def\csname LT6\endcsname{\color{black}}%
      \expandafter\def\csname LT7\endcsname{\color{black}}%
      \expandafter\def\csname LT8\endcsname{\color{black}}%
    \fi
  \fi
  \setlength{\unitlength}{0.0500bp}%
  \begin{picture}(9360.00,5760.00)%
    \gplgaddtomacro\gplbacktext{%
    }%
    \gplgaddtomacro\gplfronttext{%
      \csname LTb\endcsname%
      \put(2340,356){\makebox(0,0){\strut{}3}}%
      \put(2859,356){\makebox(0,0){\strut{}4}}%
      \put(3379,356){\makebox(0,0){\strut{}5}}%
      \put(3899,356){\makebox(0,0){\strut{}6}}%
      \put(4419,356){\makebox(0,0){\strut{}7}}%
      \put(4939,356){\makebox(0,0){\strut{}8}}%
      \put(5459,356){\makebox(0,0){\strut{}9}}%
      \put(5979,356){\makebox(0,0){\strut{}10}}%
      \put(6499,356){\makebox(0,0){\strut{}11}}%
      \put(7019,356){\makebox(0,0){\strut{}12}}%
      \put(4679,0){\makebox(0,0){\strut{}$\log N$ [erg$^{-1}$ cm$^{-2}$]}}%
      \put(183,1503){\makebox(0,0)[l]{\strut{}$0.2$ Myr}}%
    }%
    \gplgaddtomacro\gplbacktext{%
      \put(4679,4593){\makebox(0,0){\large $E = 10$ Gev}}%
    }%
    \gplgaddtomacro\gplfronttext{%
      \csname LTb\endcsname%
      \put(3303,1503){\makebox(0,0)[l]{\strut{}$0.6$ Myr}}%
    }%
    \gplgaddtomacro\gplbacktext{%
    }%
    \gplgaddtomacro\gplfronttext{%
      \csname LTb\endcsname%
      \put(6423,1503){\makebox(0,0)[l]{\strut{}$1$ Myr}}%
    }%
    \gplbacktext
    \put(0,0){\includegraphics{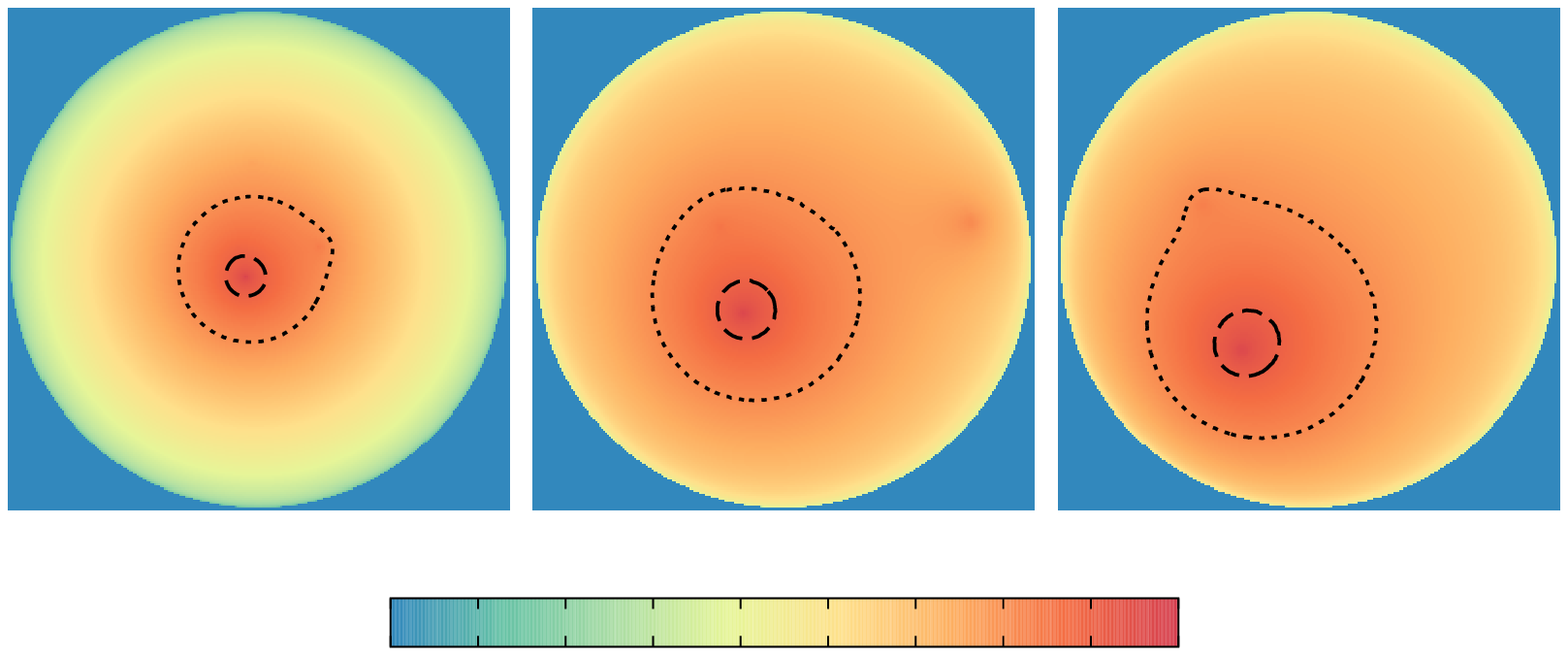}}%
    \gplfronttext
  \end{picture}%
\endgroup

%% file: chi0p1_FP_EN4.tex
% GNUPLOT: LaTeX picture with Postscript
\begingroup
  \makeatletter
  \providecommand\color[2][]{%
    \GenericError{(gnuplot) \space\space\space\@spaces}{%
      Package color not loaded in conjunction with
      terminal option `colourtext'%
    }{See the gnuplot documentation for explanation.%
    }{Either use 'blacktext' in gnuplot or load the package
      color.sty in LaTeX.}%
    \renewcommand\color[2][]{}%
  }%
  \providecommand\includegraphics[2][]{%
    \GenericError{(gnuplot) \space\space\space\@spaces}{%
      Package graphicx or graphics not loaded%
    }{See the gnuplot documentation for explanation.%
    }{The gnuplot epslatex terminal needs graphicx.sty or graphics.sty.}%
    \renewcommand\includegraphics[2][]{}%
  }%
  \providecommand\rotatebox[2]{#2}%
  \@ifundefined{ifGPcolor}{%
    \newif\ifGPcolor
    \GPcolortrue
  }{}%
  \@ifundefined{ifGPblacktext}{%
    \newif\ifGPblacktext
    \GPblacktextfalse
  }{}%
  % define a \g@addto@macro without @ in the name:
  \let\gplgaddtomacro\g@addto@macro
  % define empty templates for all commands taking text:
  \gdef\gplbacktext{}%
  \gdef\gplfronttext{}%
  \makeatother
  \ifGPblacktext
    % no textcolor at all
    \def\colorrgb#1{}%
    \def\colorgray#1{}%
  \else
    % gray or color?
    \ifGPcolor
      \def\colorrgb#1{\color[rgb]{#1}}%
      \def\colorgray#1{\color[gray]{#1}}%
      \expandafter\def\csname LTw\endcsname{\color{white}}%
      \expandafter\def\csname LTb\endcsname{\color{black}}%
      \expandafter\def\csname LTa\endcsname{\color{black}}%
      \expandafter\def\csname LT0\endcsname{\color[rgb]{1,0,0}}%
      \expandafter\def\csname LT1\endcsname{\color[rgb]{0,1,0}}%
      \expandafter\def\csname LT2\endcsname{\color[rgb]{0,0,1}}%
      \expandafter\def\csname LT3\endcsname{\color[rgb]{1,0,1}}%
      \expandafter\def\csname LT4\endcsname{\color[rgb]{0,1,1}}%
      \expandafter\def\csname LT5\endcsname{\color[rgb]{1,1,0}}%
      \expandafter\def\csname LT6\endcsname{\color[rgb]{0,0,0}}%
      \expandafter\def\csname LT7\endcsname{\color[rgb]{1,0.3,0}}%
      \expandafter\def\csname LT8\endcsname{\color[rgb]{0.5,0.5,0.5}}%
    \else
      % gray
      \def\colorrgb#1{\color{black}}%
      \def\colorgray#1{\color[gray]{#1}}%
      \expandafter\def\csname LTw\endcsname{\color{white}}%
      \expandafter\def\csname LTb\endcsname{\color{black}}%
      \expandafter\def\csname LTa\endcsname{\color{black}}%
      \expandafter\def\csname LT0\endcsname{\color{black}}%
      \expandafter\def\csname LT1\endcsname{\color{black}}%
      \expandafter\def\csname LT2\endcsname{\color{black}}%
      \expandafter\def\csname LT3\endcsname{\color{black}}%
      \expandafter\def\csname LT4\endcsname{\color{black}}%
      \expandafter\def\csname LT5\endcsname{\color{black}}%
      \expandafter\def\csname LT6\endcsname{\color{black}}%
      \expandafter\def\csname LT7\endcsname{\color{black}}%
      \expandafter\def\csname LT8\endcsname{\color{black}}%
    \fi
  \fi
  \setlength{\unitlength}{0.0500bp}%
  \begin{picture}(9360.00,5760.00)%
    \gplgaddtomacro\gplbacktext{%
    }%
    \gplgaddtomacro\gplfronttext{%
      \csname LTb\endcsname%
      \put(2339,356){\makebox(0,0){\strut{}-1}}%
      \put(3119,356){\makebox(0,0){\strut{}0}}%
      \put(3899,356){\makebox(0,0){\strut{}1}}%
      \put(4679,356){\makebox(0,0){\strut{}2}}%
      \put(5459,356){\makebox(0,0){\strut{}3}}%
      \put(6239,356){\makebox(0,0){\strut{}4}}%
      \put(7019,356){\makebox(0,0){\strut{}5}}%
      \put(4679,0){\makebox(0,0){\strut{}$\log N$ [erg$^{-1}$ cm$^{-2}$]}}%
      \put(183,1503){\makebox(0,0)[l]{\strut{}$0.2$ Myr}}%
    }%
    \gplgaddtomacro\gplbacktext{%
      \put(4679,4593){\makebox(0,0){\large $E = 10$ Tev}}%
    }%
    \gplgaddtomacro\gplfronttext{%
      \csname LTb\endcsname%
      \put(3303,1503){\makebox(0,0)[l]{\strut{}$0.6$ Myr}}%
    }%
    \gplgaddtomacro\gplbacktext{%
    }%
    \gplgaddtomacro\gplfronttext{%
      \csname LTb\endcsname%
      \put(6423,1503){\makebox(0,0)[l]{\strut{}$1$ Myr}}%
    }%
    \gplbacktext
    \put(0,0){\includegraphics{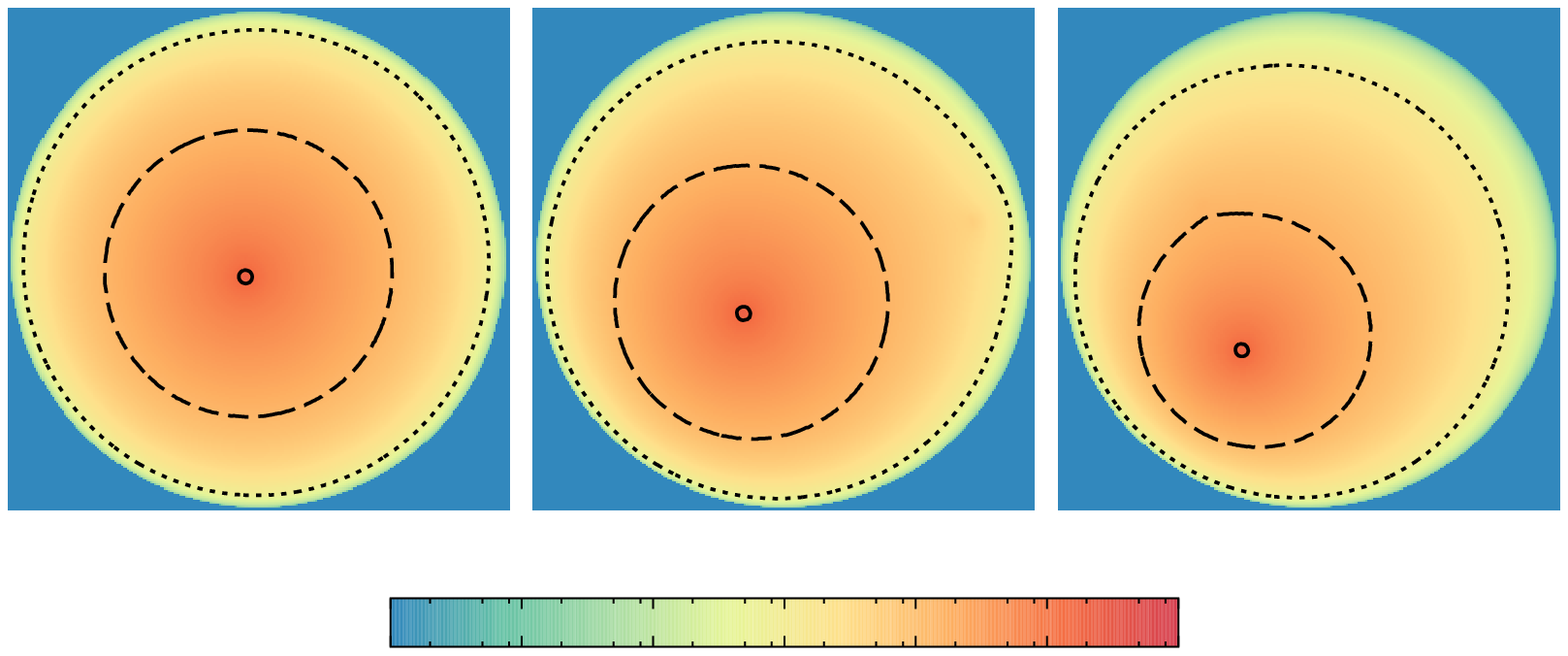}}%
    \gplfronttext
  \end{picture}%
\endgroup

%% file: chi0p001_FP_EN1.tex
% GNUPLOT: LaTeX picture with Postscript
\begingroup
  \makeatletter
  \providecommand\color[2][]{%
    \GenericError{(gnuplot) \space\space\space\@spaces}{%
      Package color not loaded in conjunction with
      terminal option `colourtext'%
    }{See the gnuplot documentation for explanation.%
    }{Either use 'blacktext' in gnuplot or load the package
      color.sty in LaTeX.}%
    \renewcommand\color[2][]{}%
  }%
  \providecommand\includegraphics[2][]{%
    \GenericError{(gnuplot) \space\space\space\@spaces}{%
      Package graphicx or graphics not loaded%
    }{See the gnuplot documentation for explanation.%
    }{The gnuplot epslatex terminal needs graphicx.sty or graphics.sty.}%
    \renewcommand\includegraphics[2][]{}%
  }%
  \providecommand\rotatebox[2]{#2}%
  \@ifundefined{ifGPcolor}{%
    \newif\ifGPcolor
    \GPcolortrue
  }{}%
  \@ifundefined{ifGPblacktext}{%
    \newif\ifGPblacktext
    \GPblacktextfalse
  }{}%
  % define a \g@addto@macro without @ in the name:
  \let\gplgaddtomacro\g@addto@macro
  % define empty templates for all commands taking text:
  \gdef\gplbacktext{}%
  \gdef\gplfronttext{}%
  \makeatother
  \ifGPblacktext
    % no textcolor at all
    \def\colorrgb#1{}%
    \def\colorgray#1{}%
  \else
    % gray or color?
    \ifGPcolor
      \def\colorrgb#1{\color[rgb]{#1}}%
      \def\colorgray#1{\color[gray]{#1}}%
      \expandafter\def\csname LTw\endcsname{\color{white}}%
      \expandafter\def\csname LTb\endcsname{\color{black}}%
      \expandafter\def\csname LTa\endcsname{\color{black}}%
      \expandafter\def\csname LT0\endcsname{\color[rgb]{1,0,0}}%
      \expandafter\def\csname LT1\endcsname{\color[rgb]{0,1,0}}%
      \expandafter\def\csname LT2\endcsname{\color[rgb]{0,0,1}}%
      \expandafter\def\csname LT3\endcsname{\color[rgb]{1,0,1}}%
      \expandafter\def\csname LT4\endcsname{\color[rgb]{0,1,1}}%
      \expandafter\def\csname LT5\endcsname{\color[rgb]{1,1,0}}%
      \expandafter\def\csname LT6\endcsname{\color[rgb]{0,0,0}}%
      \expandafter\def\csname LT7\endcsname{\color[rgb]{1,0.3,0}}%
      \expandafter\def\csname LT8\endcsname{\color[rgb]{0.5,0.5,0.5}}%
    \else
      % gray
      \def\colorrgb#1{\color{black}}%
      \def\colorgray#1{\color[gray]{#1}}%
      \expandafter\def\csname LTw\endcsname{\color{white}}%
      \expandafter\def\csname LTb\endcsname{\color{black}}%
      \expandafter\def\csname LTa\endcsname{\color{black}}%
      \expandafter\def\csname LT0\endcsname{\color{black}}%
      \expandafter\def\csname LT1\endcsname{\color{black}}%
      \expandafter\def\csname LT2\endcsname{\color{black}}%
      \expandafter\def\csname LT3\endcsname{\color{black}}%
      \expandafter\def\csname LT4\endcsname{\color{black}}%
      \expandafter\def\csname LT5\endcsname{\color{black}}%
      \expandafter\def\csname LT6\endcsname{\color{black}}%
      \expandafter\def\csname LT7\endcsname{\color{black}}%
      \expandafter\def\csname LT8\endcsname{\color{black}}%
    \fi
  \fi
  \setlength{\unitlength}{0.0500bp}%
  \begin{picture}(9360.00,5760.00)%
    \gplgaddtomacro\gplbacktext{%
    }%
    \gplgaddtomacro\gplfronttext{%
      \csname LTb\endcsname%
      \put(2340,356){\makebox(0,0){\strut{}3}}%
      \put(2859,356){\makebox(0,0){\strut{}4}}%
      \put(3379,356){\makebox(0,0){\strut{}5}}%
      \put(3899,356){\makebox(0,0){\strut{}6}}%
      \put(4419,356){\makebox(0,0){\strut{}7}}%
      \put(4939,356){\makebox(0,0){\strut{}8}}%
      \put(5459,356){\makebox(0,0){\strut{}9}}%
      \put(5979,356){\makebox(0,0){\strut{}10}}%
      \put(6499,356){\makebox(0,0){\strut{}11}}%
      \put(7019,356){\makebox(0,0){\strut{}12}}%
      \put(4679,0){\makebox(0,0){\strut{}$\log N$ [erg$^{-1}$ cm$^{-2}$]}}%
      \put(183,1503){\makebox(0,0)[l]{\strut{}$0.2$ Myr}}%
    }%
    \gplgaddtomacro\gplbacktext{%
      \put(4679,4593){\makebox(0,0){\large $E = 10$ Gev}}%
    }%
    \gplgaddtomacro\gplfronttext{%
      \csname LTb\endcsname%
      \put(3303,1503){\makebox(0,0)[l]{\strut{}$0.6$ Myr}}%
    }%
    \gplgaddtomacro\gplbacktext{%
    }%
    \gplgaddtomacro\gplfronttext{%
      \csname LTb\endcsname%
      \put(6423,1503){\makebox(0,0)[l]{\strut{}$1$ Myr}}%
    }%
    \gplbacktext
    \put(0,0){\includegraphics{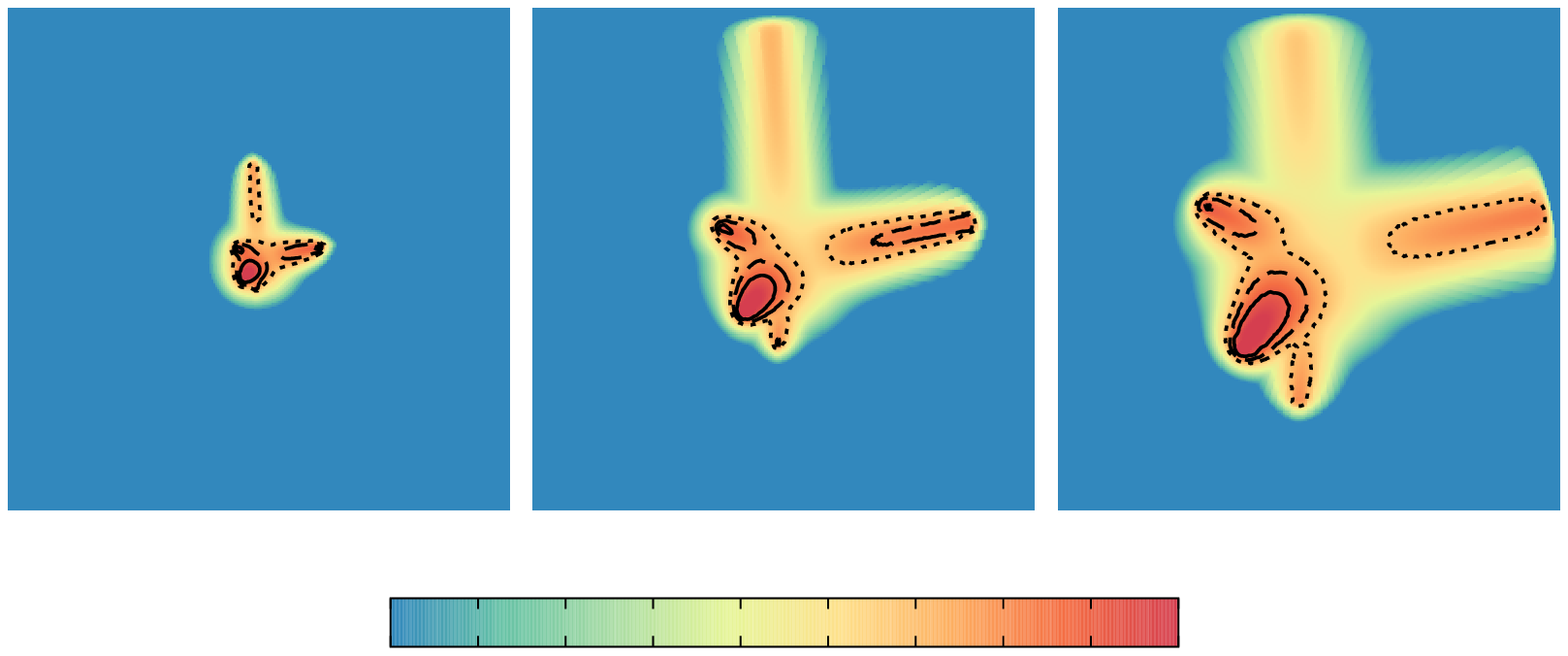}}%
    \gplfronttext
  \end{picture}%
\endgroup

%% file: chi0p001_FP_EN4.tex
% GNUPLOT: LaTeX picture with Postscript
\begingroup
  \makeatletter
  \providecommand\color[2][]{%
    \GenericError{(gnuplot) \space\space\space\@spaces}{%
      Package color not loaded in conjunction with
      terminal option `colourtext'%
    }{See the gnuplot documentation for explanation.%
    }{Either use 'blacktext' in gnuplot or load the package
      color.sty in LaTeX.}%
    \renewcommand\color[2][]{}%
  }%
  \providecommand\includegraphics[2][]{%
    \GenericError{(gnuplot) \space\space\space\@spaces}{%
      Package graphicx or graphics not loaded%
    }{See the gnuplot documentation for explanation.%
    }{The gnuplot epslatex terminal needs graphicx.sty or graphics.sty.}%
    \renewcommand\includegraphics[2][]{}%
  }%
  \providecommand\rotatebox[2]{#2}%
  \@ifundefined{ifGPcolor}{%
    \newif\ifGPcolor
    \GPcolortrue
  }{}%
  \@ifundefined{ifGPblacktext}{%
    \newif\ifGPblacktext
    \GPblacktextfalse
  }{}%
  % define a \g@addto@macro without @ in the name:
  \let\gplgaddtomacro\g@addto@macro
  % define empty templates for all commands taking text:
  \gdef\gplbacktext{}%
  \gdef\gplfronttext{}%
  \makeatother
  \ifGPblacktext
    % no textcolor at all
    \def\colorrgb#1{}%
    \def\colorgray#1{}%
  \else
    % gray or color?
    \ifGPcolor
      \def\colorrgb#1{\color[rgb]{#1}}%
      \def\colorgray#1{\color[gray]{#1}}%
      \expandafter\def\csname LTw\endcsname{\color{white}}%
      \expandafter\def\csname LTb\endcsname{\color{black}}%
      \expandafter\def\csname LTa\endcsname{\color{black}}%
      \expandafter\def\csname LT0\endcsname{\color[rgb]{1,0,0}}%
      \expandafter\def\csname LT1\endcsname{\color[rgb]{0,1,0}}%
      \expandafter\def\csname LT2\endcsname{\color[rgb]{0,0,1}}%
      \expandafter\def\csname LT3\endcsname{\color[rgb]{1,0,1}}%
      \expandafter\def\csname LT4\endcsname{\color[rgb]{0,1,1}}%
      \expandafter\def\csname LT5\endcsname{\color[rgb]{1,1,0}}%
      \expandafter\def\csname LT6\endcsname{\color[rgb]{0,0,0}}%
      \expandafter\def\csname LT7\endcsname{\color[rgb]{1,0.3,0}}%
      \expandafter\def\csname LT8\endcsname{\color[rgb]{0.5,0.5,0.5}}%
    \else
      % gray
      \def\colorrgb#1{\color{black}}%
      \def\colorgray#1{\color[gray]{#1}}%
      \expandafter\def\csname LTw\endcsname{\color{white}}%
      \expandafter\def\csname LTb\endcsname{\color{black}}%
      \expandafter\def\csname LTa\endcsname{\color{black}}%
      \expandafter\def\csname LT0\endcsname{\color{black}}%
      \expandafter\def\csname LT1\endcsname{\color{black}}%
      \expandafter\def\csname LT2\endcsname{\color{black}}%
      \expandafter\def\csname LT3\endcsname{\color{black}}%
      \expandafter\def\csname LT4\endcsname{\color{black}}%
      \expandafter\def\csname LT5\endcsname{\color{black}}%
      \expandafter\def\csname LT6\endcsname{\color{black}}%
      \expandafter\def\csname LT7\endcsname{\color{black}}%
      \expandafter\def\csname LT8\endcsname{\color{black}}%
    \fi
  \fi
  \setlength{\unitlength}{0.0500bp}%
  \begin{picture}(9360.00,5760.00)%
    \gplgaddtomacro\gplbacktext{%
    }%
    \gplgaddtomacro\gplfronttext{%
      \csname LTb\endcsname%
      \put(2339,356){\makebox(0,0){\strut{}-1}}%
      \put(3119,356){\makebox(0,0){\strut{}0}}%
      \put(3899,356){\makebox(0,0){\strut{}1}}%
      \put(4679,356){\makebox(0,0){\strut{}2}}%
      \put(5459,356){\makebox(0,0){\strut{}3}}%
      \put(6239,356){\makebox(0,0){\strut{}4}}%
      \put(7019,356){\makebox(0,0){\strut{}5}}%
      \put(4679,0){\makebox(0,0){\strut{}$\log N$ [erg$^{-1}$ cm$^{-2}$]}}%
      \put(183,1503){\makebox(0,0)[l]{\strut{}$0.2$ Myr}}%
    }%
    \gplgaddtomacro\gplbacktext{%
      \put(4679,4593){\makebox(0,0){\large $E = 10$ Gev}}%
    }%
    \gplgaddtomacro\gplfronttext{%
      \csname LTb\endcsname%
      \put(3303,1503){\makebox(0,0)[l]{\strut{}$0.6$ Myr}}%
    }%
    \gplgaddtomacro\gplbacktext{%
    }%
    \gplgaddtomacro\gplfronttext{%
      \csname LTb\endcsname%
      \put(6423,1503){\makebox(0,0)[l]{\strut{}$1$ Myr}}%
    }%
    \gplbacktext
    \put(0,0){\includegraphics{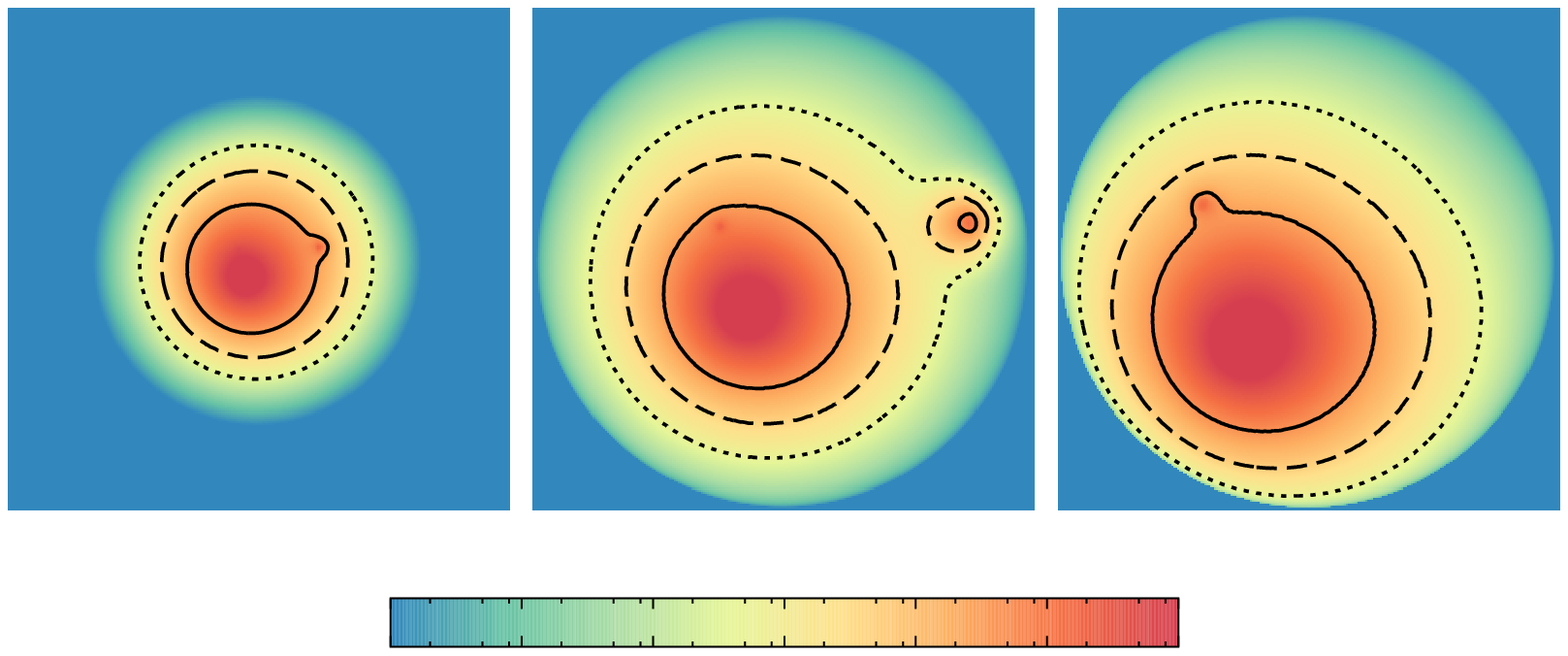}}%
    \gplfronttext
  \end{picture}%
\endgroup

%% file: chi0p1_FE_EN1.tex
% GNUPLOT: LaTeX picture with Postscript
\begingroup
  \makeatletter
  \providecommand\color[2][]{%
    \GenericError{(gnuplot) \space\space\space\@spaces}{%
      Package color not loaded in conjunction with
      terminal option `colourtext'%
    }{See the gnuplot documentation for explanation.%
    }{Either use 'blacktext' in gnuplot or load the package
      color.sty in LaTeX.}%
    \renewcommand\color[2][]{}%
  }%
  \providecommand\includegraphics[2][]{%
    \GenericError{(gnuplot) \space\space\space\@spaces}{%
      Package graphicx or graphics not loaded%
    }{See the gnuplot documentation for explanation.%
    }{The gnuplot epslatex terminal needs graphicx.sty or graphics.sty.}%
    \renewcommand\includegraphics[2][]{}%
  }%
  \providecommand\rotatebox[2]{#2}%
  \@ifundefined{ifGPcolor}{%
    \newif\ifGPcolor
    \GPcolortrue
  }{}%
  \@ifundefined{ifGPblacktext}{%
    \newif\ifGPblacktext
    \GPblacktextfalse
  }{}%
  % define a \g@addto@macro without @ in the name:
  \let\gplgaddtomacro\g@addto@macro
  % define empty templates for all commands taking text:
  \gdef\gplbacktext{}%
  \gdef\gplfronttext{}%
  \makeatother
  \ifGPblacktext
    % no textcolor at all
    \def\colorrgb#1{}%
    \def\colorgray#1{}%
  \else
    % gray or color?
    \ifGPcolor
      \def\colorrgb#1{\color[rgb]{#1}}%
      \def\colorgray#1{\color[gray]{#1}}%
      \expandafter\def\csname LTw\endcsname{\color{white}}%
      \expandafter\def\csname LTb\endcsname{\color{black}}%
      \expandafter\def\csname LTa\endcsname{\color{black}}%
      \expandafter\def\csname LT0\endcsname{\color[rgb]{1,0,0}}%
      \expandafter\def\csname LT1\endcsname{\color[rgb]{0,1,0}}%
      \expandafter\def\csname LT2\endcsname{\color[rgb]{0,0,1}}%
      \expandafter\def\csname LT3\endcsname{\color[rgb]{1,0,1}}%
      \expandafter\def\csname LT4\endcsname{\color[rgb]{0,1,1}}%
      \expandafter\def\csname LT5\endcsname{\color[rgb]{1,1,0}}%
      \expandafter\def\csname LT6\endcsname{\color[rgb]{0,0,0}}%
      \expandafter\def\csname LT7\endcsname{\color[rgb]{1,0.3,0}}%
      \expandafter\def\csname LT8\endcsname{\color[rgb]{0.5,0.5,0.5}}%
    \else
      % gray
      \def\colorrgb#1{\color{black}}%
      \def\colorgray#1{\color[gray]{#1}}%
      \expandafter\def\csname LTw\endcsname{\color{white}}%
      \expandafter\def\csname LTb\endcsname{\color{black}}%
      \expandafter\def\csname LTa\endcsname{\color{black}}%
      \expandafter\def\csname LT0\endcsname{\color{black}}%
      \expandafter\def\csname LT1\endcsname{\color{black}}%
      \expandafter\def\csname LT2\endcsname{\color{black}}%
      \expandafter\def\csname LT3\endcsname{\color{black}}%
      \expandafter\def\csname LT4\endcsname{\color{black}}%
      \expandafter\def\csname LT5\endcsname{\color{black}}%
      \expandafter\def\csname LT6\endcsname{\color{black}}%
      \expandafter\def\csname LT7\endcsname{\color{black}}%
      \expandafter\def\csname LT8\endcsname{\color{black}}%
    \fi
  \fi
  \setlength{\unitlength}{0.0500bp}%
  \begin{picture}(9360.00,5760.00)%
    \gplgaddtomacro\gplbacktext{%
    }%
    \gplgaddtomacro\gplfronttext{%
      \csname LTb\endcsname%
      \put(2340,356){\makebox(0,0){\strut{}4}}%
      \put(3119,356){\makebox(0,0){\strut{}5}}%
      \put(3899,356){\makebox(0,0){\strut{}6}}%
      \put(4679,356){\makebox(0,0){\strut{}7}}%
      \put(5459,356){\makebox(0,0){\strut{}8}}%
      \put(6239,356){\makebox(0,0){\strut{}9}}%
      \put(7019,356){\makebox(0,0){\strut{}10}}%
      \put(4679,0){\makebox(0,0){\strut{}$\log N$ [erg$^{-1}$ cm$^{-2}$]}}%
      \put(183,1503){\makebox(0,0)[l]{\strut{}$0.2$ Myr}}%
    }%
    \gplgaddtomacro\gplbacktext{%
      \put(4679,4593){\makebox(0,0){\large $E = 10$ Gev}}%
    }%
    \gplgaddtomacro\gplfronttext{%
      \csname LTb\endcsname%
      \put(3303,1503){\makebox(0,0)[l]{\strut{}$0.6$ Myr}}%
    }%
    \gplgaddtomacro\gplbacktext{%
    }%
    \gplgaddtomacro\gplfronttext{%
      \csname LTb\endcsname%
      \put(6423,1503){\makebox(0,0)[l]{\strut{}$1$ Myr}}%
    }%
    \gplbacktext
    \put(0,0){\includegraphics{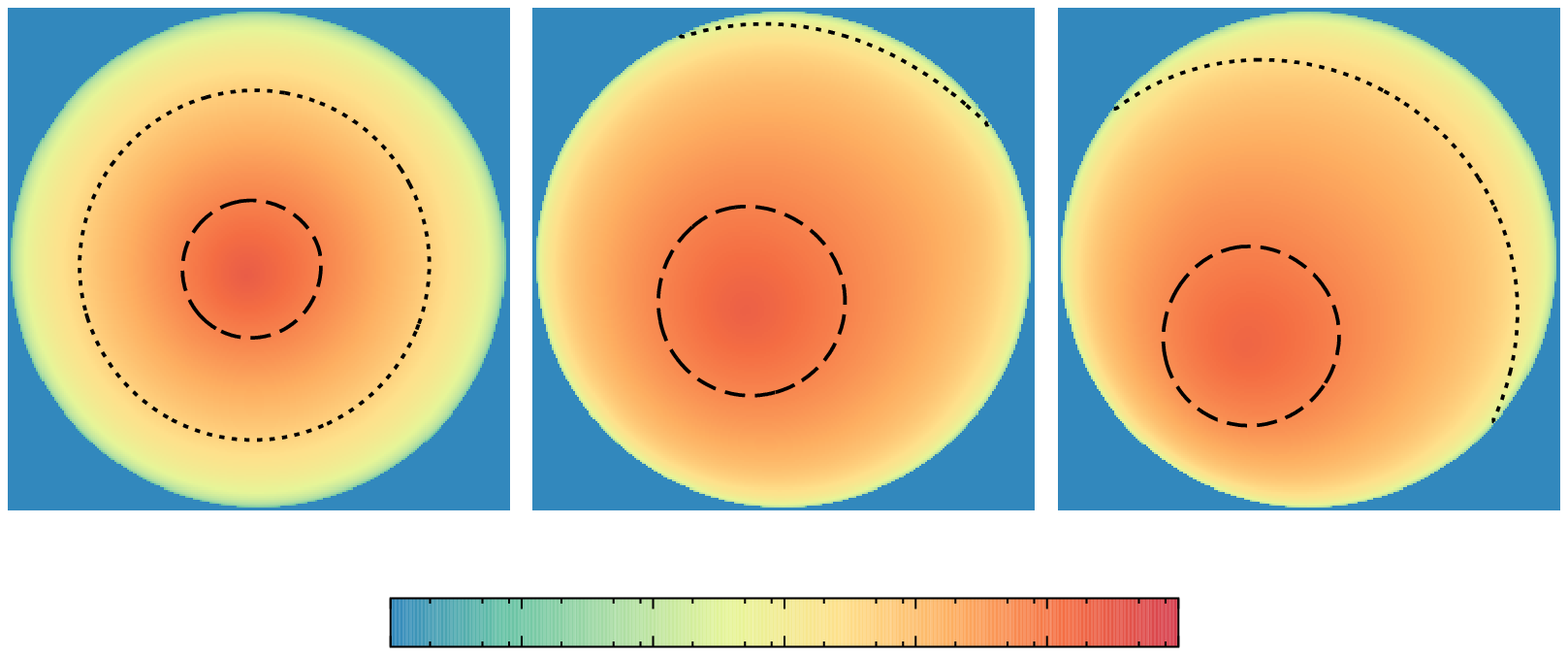}}%
    \gplfronttext
  \end{picture}%
\endgroup

%% file: chi0p1_FE_EN4.tex
% GNUPLOT: LaTeX picture with Postscript
\begingroup
  \makeatletter
  \providecommand\color[2][]{%
    \GenericError{(gnuplot) \space\space\space\@spaces}{%
      Package color not loaded in conjunction with
      terminal option `colourtext'%
    }{See the gnuplot documentation for explanation.%
    }{Either use 'blacktext' in gnuplot or load the package
      color.sty in LaTeX.}%
    \renewcommand\color[2][]{}%
  }%
  \providecommand\includegraphics[2][]{%
    \GenericError{(gnuplot) \space\space\space\@spaces}{%
      Package graphicx or graphics not loaded%
    }{See the gnuplot documentation for explanation.%
    }{The gnuplot epslatex terminal needs graphicx.sty or graphics.sty.}%
    \renewcommand\includegraphics[2][]{}%
  }%
  \providecommand\rotatebox[2]{#2}%
  \@ifundefined{ifGPcolor}{%
    \newif\ifGPcolor
    \GPcolortrue
  }{}%
  \@ifundefined{ifGPblacktext}{%
    \newif\ifGPblacktext
    \GPblacktextfalse
  }{}%
  % define a \g@addto@macro without @ in the name:
  \let\gplgaddtomacro\g@addto@macro
  % define empty templates for all commands taking text:
  \gdef\gplbacktext{}%
  \gdef\gplfronttext{}%
  \makeatother
  \ifGPblacktext
    % no textcolor at all
    \def\colorrgb#1{}%
    \def\colorgray#1{}%
  \else
    % gray or color?
    \ifGPcolor
      \def\colorrgb#1{\color[rgb]{#1}}%
      \def\colorgray#1{\color[gray]{#1}}%
      \expandafter\def\csname LTw\endcsname{\color{white}}%
      \expandafter\def\csname LTb\endcsname{\color{black}}%
      \expandafter\def\csname LTa\endcsname{\color{black}}%
      \expandafter\def\csname LT0\endcsname{\color[rgb]{1,0,0}}%
      \expandafter\def\csname LT1\endcsname{\color[rgb]{0,1,0}}%
      \expandafter\def\csname LT2\endcsname{\color[rgb]{0,0,1}}%
      \expandafter\def\csname LT3\endcsname{\color[rgb]{1,0,1}}%
      \expandafter\def\csname LT4\endcsname{\color[rgb]{0,1,1}}%
      \expandafter\def\csname LT5\endcsname{\color[rgb]{1,1,0}}%
      \expandafter\def\csname LT6\endcsname{\color[rgb]{0,0,0}}%
      \expandafter\def\csname LT7\endcsname{\color[rgb]{1,0.3,0}}%
      \expandafter\def\csname LT8\endcsname{\color[rgb]{0.5,0.5,0.5}}%
    \else
      % gray
      \def\colorrgb#1{\color{black}}%
      \def\colorgray#1{\color[gray]{#1}}%
      \expandafter\def\csname LTw\endcsname{\color{white}}%
      \expandafter\def\csname LTb\endcsname{\color{black}}%
      \expandafter\def\csname LTa\endcsname{\color{black}}%
      \expandafter\def\csname LT0\endcsname{\color{black}}%
      \expandafter\def\csname LT1\endcsname{\color{black}}%
      \expandafter\def\csname LT2\endcsname{\color{black}}%
      \expandafter\def\csname LT3\endcsname{\color{black}}%
      \expandafter\def\csname LT4\endcsname{\color{black}}%
      \expandafter\def\csname LT5\endcsname{\color{black}}%
      \expandafter\def\csname LT6\endcsname{\color{black}}%
      \expandafter\def\csname LT7\endcsname{\color{black}}%
      \expandafter\def\csname LT8\endcsname{\color{black}}%
    \fi
  \fi
  \setlength{\unitlength}{0.0500bp}%
  \begin{picture}(9360.00,5760.00)%
    \gplgaddtomacro\gplbacktext{%
    }%
    \gplgaddtomacro\gplfronttext{%
      \csname LTb\endcsname%
      \put(2339,356){\makebox(0,0){\strut{}-5}}%
      \put(3119,356){\makebox(0,0){\strut{}-4}}%
      \put(3899,356){\makebox(0,0){\strut{}-3}}%
      \put(4679,356){\makebox(0,0){\strut{}-2}}%
      \put(5459,356){\makebox(0,0){\strut{}-1}}%
      \put(6239,356){\makebox(0,0){\strut{}0}}%
      \put(7019,356){\makebox(0,0){\strut{}1}}%
      \put(4679,0){\makebox(0,0){\strut{}$\log N$ [erg$^{-1}$ cm$^{-2}$]}}%
      \put(183,1503){\makebox(0,0)[l]{\strut{}$0.2$ Myr}}%
    }%
    \gplgaddtomacro\gplbacktext{%
      \put(4679,4593){\makebox(0,0){\large $E = 10$ Tev}}%
    }%
    \gplgaddtomacro\gplfronttext{%
      \csname LTb\endcsname%
      \put(3303,1503){\makebox(0,0)[l]{\strut{}$0.6$ Myr}}%
    }%
    \gplgaddtomacro\gplbacktext{%
    }%
    \gplgaddtomacro\gplfronttext{%
      \csname LTb\endcsname%
      \put(6423,1503){\makebox(0,0)[l]{\strut{}$1$ Myr}}%
    }%
    \gplbacktext
    \put(0,0){\includegraphics{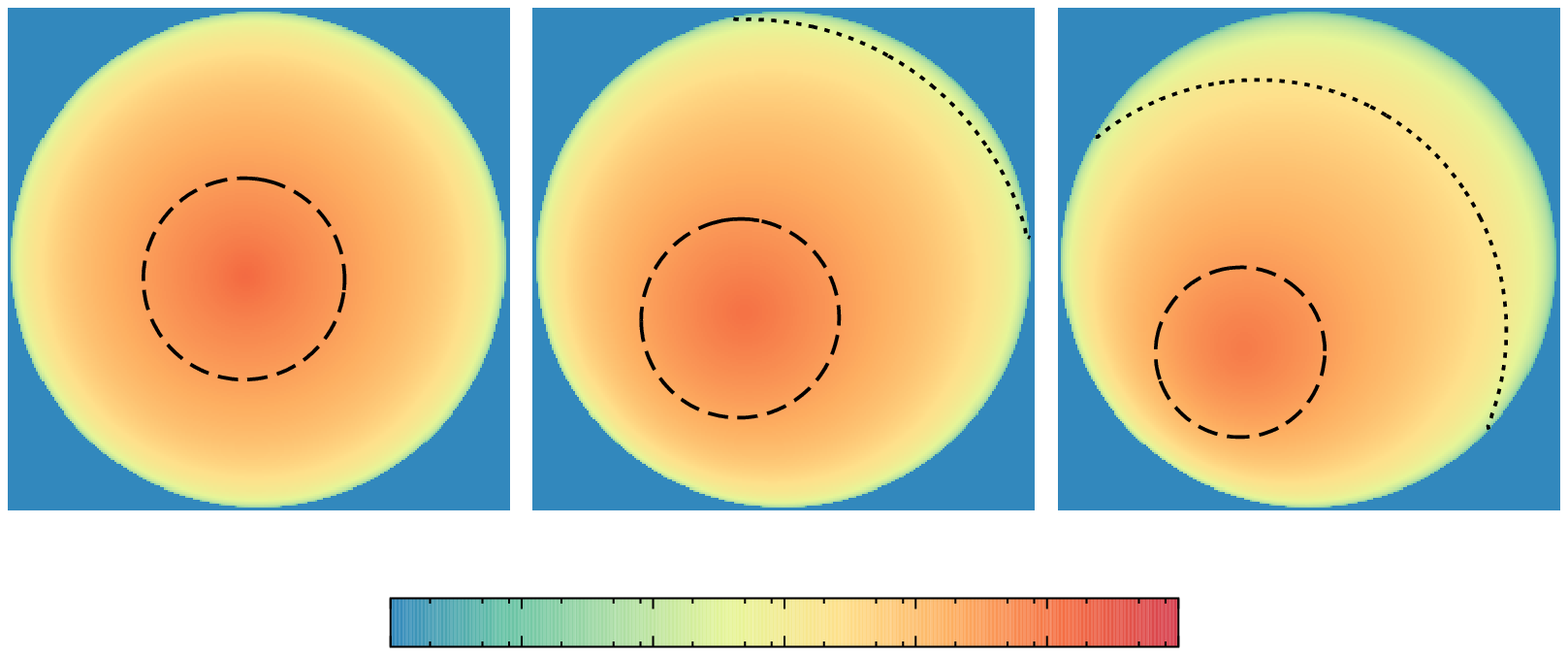}}%
    \gplfronttext
  \end{picture}%
\endgroup

%% file: chi0p1_QGAMMA_EN2.tex
% GNUPLOT: LaTeX picture with Postscript
\begingroup
  \makeatletter
  \providecommand\color[2][]{%
    \GenericError{(gnuplot) \space\space\space\@spaces}{%
      Package color not loaded in conjunction with
      terminal option `colourtext'%
    }{See the gnuplot documentation for explanation.%
    }{Either use 'blacktext' in gnuplot or load the package
      color.sty in LaTeX.}%
    \renewcommand\color[2][]{}%
  }%
  \providecommand\includegraphics[2][]{%
    \GenericError{(gnuplot) \space\space\space\@spaces}{%
      Package graphicx or graphics not loaded%
    }{See the gnuplot documentation for explanation.%
    }{The gnuplot epslatex terminal needs graphicx.sty or graphics.sty.}%
    \renewcommand\includegraphics[2][]{}%
  }%
  \providecommand\rotatebox[2]{#2}%
  \@ifundefined{ifGPcolor}{%
    \newif\ifGPcolor
    \GPcolortrue
  }{}%
  \@ifundefined{ifGPblacktext}{%
    \newif\ifGPblacktext
    \GPblacktextfalse
  }{}%
  % define a \g@addto@macro without @ in the name:
  \let\gplgaddtomacro\g@addto@macro
  % define empty templates for all commands taking text:
  \gdef\gplbacktext{}%
  \gdef\gplfronttext{}%
  \makeatother
  \ifGPblacktext
    % no textcolor at all
    \def\colorrgb#1{}%
    \def\colorgray#1{}%
  \else
    % gray or color?
    \ifGPcolor
      \def\colorrgb#1{\color[rgb]{#1}}%
      \def\colorgray#1{\color[gray]{#1}}%
      \expandafter\def\csname LTw\endcsname{\color{white}}%
      \expandafter\def\csname LTb\endcsname{\color{black}}%
      \expandafter\def\csname LTa\endcsname{\color{black}}%
      \expandafter\def\csname LT0\endcsname{\color[rgb]{1,0,0}}%
      \expandafter\def\csname LT1\endcsname{\color[rgb]{0,1,0}}%
      \expandafter\def\csname LT2\endcsname{\color[rgb]{0,0,1}}%
      \expandafter\def\csname LT3\endcsname{\color[rgb]{1,0,1}}%
      \expandafter\def\csname LT4\endcsname{\color[rgb]{0,1,1}}%
      \expandafter\def\csname LT5\endcsname{\color[rgb]{1,1,0}}%
      \expandafter\def\csname LT6\endcsname{\color[rgb]{0,0,0}}%
      \expandafter\def\csname LT7\endcsname{\color[rgb]{1,0.3,0}}%
      \expandafter\def\csname LT8\endcsname{\color[rgb]{0.5,0.5,0.5}}%
    \else
      % gray
      \def\colorrgb#1{\color{black}}%
      \def\colorgray#1{\color[gray]{#1}}%
      \expandafter\def\csname LTw\endcsname{\color{white}}%
      \expandafter\def\csname LTb\endcsname{\color{black}}%
      \expandafter\def\csname LTa\endcsname{\color{black}}%
      \expandafter\def\csname LT0\endcsname{\color{black}}%
      \expandafter\def\csname LT1\endcsname{\color{black}}%
      \expandafter\def\csname LT2\endcsname{\color{black}}%
      \expandafter\def\csname LT3\endcsname{\color{black}}%
      \expandafter\def\csname LT4\endcsname{\color{black}}%
      \expandafter\def\csname LT5\endcsname{\color{black}}%
      \expandafter\def\csname LT6\endcsname{\color{black}}%
      \expandafter\def\csname LT7\endcsname{\color{black}}%
      \expandafter\def\csname LT8\endcsname{\color{black}}%
    \fi
  \fi
  \setlength{\unitlength}{0.0500bp}%
  \begin{picture}(9360.00,7200.00)%
    \gplgaddtomacro\gplbacktext{%
    }%
    \gplgaddtomacro\gplfronttext{%
      \csname LTb\endcsname%
      \put(183,4164){\makebox(0,0)[l]{\strut{}$0.1$ Myr}}%
    }%
    \gplgaddtomacro\gplbacktext{%
      \put(4674,7252){\makebox(0,0){\large $E = 10$ Gev}}%
    }%
    \gplgaddtomacro\gplfronttext{%
      \csname LTb\endcsname%
      \put(3299,4164){\makebox(0,0)[l]{\strut{}$0.2$ Myr}}%
    }%
    \gplgaddtomacro\gplbacktext{%
    }%
    \gplgaddtomacro\gplfronttext{%
      \csname LTb\endcsname%
      \put(6426,4164){\makebox(0,0)[l]{\strut{}$0.4$ Myr}}%
    }%
    \gplgaddtomacro\gplbacktext{%
    }%
    \gplgaddtomacro\gplfronttext{%
      \csname LTb\endcsname%
      \put(2340,260){\makebox(0,0){\strut{}-9}}%
      \put(2924,260){\makebox(0,0){\strut{}-8}}%
      \put(3509,260){\makebox(0,0){\strut{}-7}}%
      \put(4094,260){\makebox(0,0){\strut{}-6}}%
      \put(4679,260){\makebox(0,0){\strut{}-5}}%
      \put(5264,260){\makebox(0,0){\strut{}-4}}%
      \put(5849,260){\makebox(0,0){\strut{}-3}}%
      \put(6434,260){\makebox(0,0){\strut{}-2}}%
      \put(7018,260){\makebox(0,0){\strut{}-1}}%
      \put(4679,0){\makebox(0,0){\strut{}$\log Q$ [erg$^{-1}$ cm$^{-2} s^{-1}$]}}%
      \put(183,996){\makebox(0,0)[l]{\strut{}$0.6$ Myr}}%
    }%
    \gplgaddtomacro\gplbacktext{%
    }%
    \gplgaddtomacro\gplfronttext{%
      \csname LTb\endcsname%
      \put(3299,995){\makebox(0,0)[l]{\strut{}$0.8$ Myr}}%
    }%
    \gplgaddtomacro\gplbacktext{%
    }%
    \gplgaddtomacro\gplfronttext{%
      \csname LTb\endcsname%
      \put(6426,996){\makebox(0,0)[l]{\strut{}$1$ Myr}}%
    }%
    \gplbacktext
    \put(0,0){\includegraphics{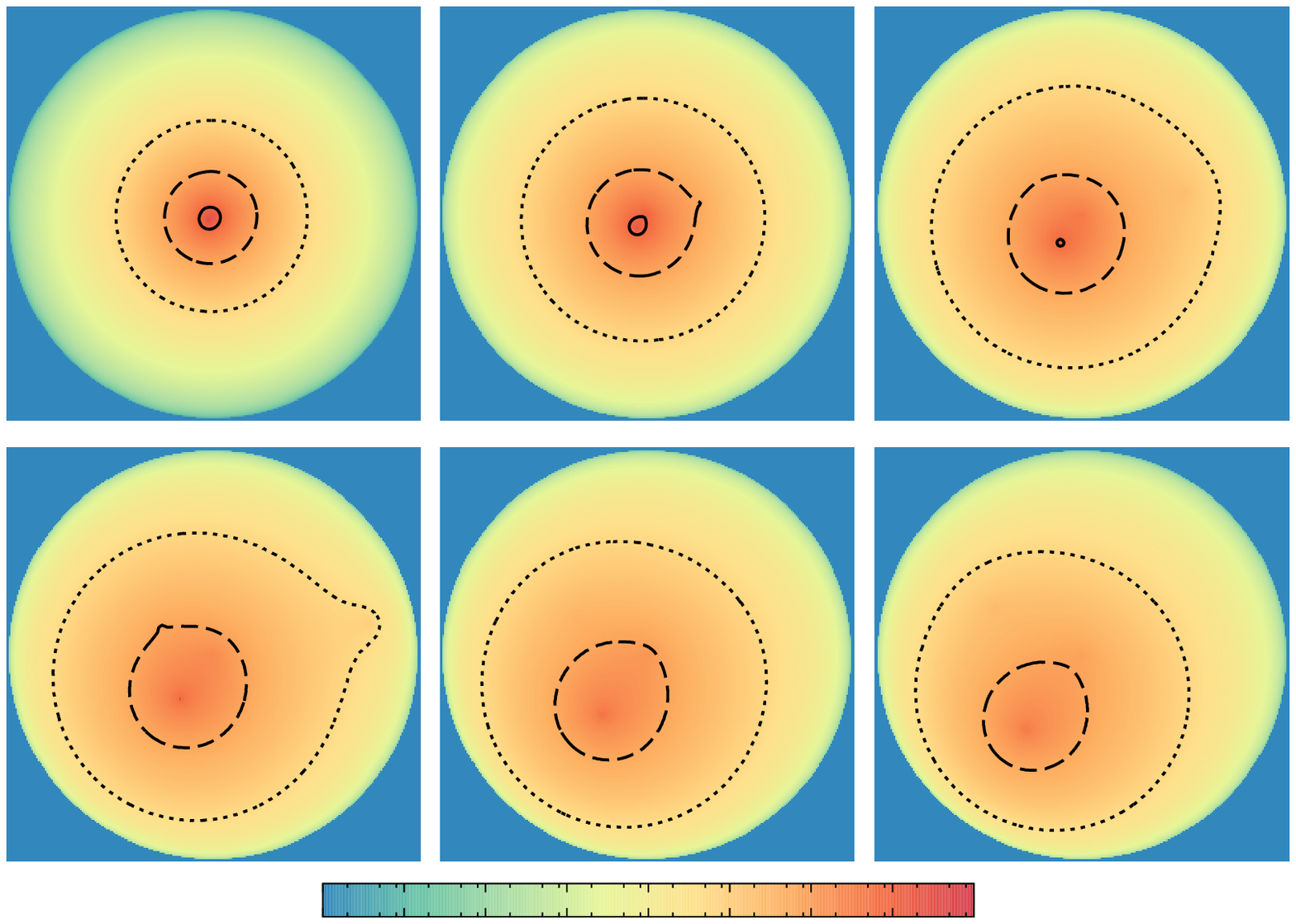}}%
    \gplfronttext
  \end{picture}%
\endgroup

%% file: chi0p001_QGAMMA_EN2.tex
% GNUPLOT: LaTeX picture with Postscript
\begingroup
  \makeatletter
  \providecommand\color[2][]{%
    \GenericError{(gnuplot) \space\space\space\@spaces}{%
      Package color not loaded in conjunction with
      terminal option `colourtext'%
    }{See the gnuplot documentation for explanation.%
    }{Either use 'blacktext' in gnuplot or load the package
      color.sty in LaTeX.}%
    \renewcommand\color[2][]{}%
  }%
  \providecommand\includegraphics[2][]{%
    \GenericError{(gnuplot) \space\space\space\@spaces}{%
      Package graphicx or graphics not loaded%
    }{See the gnuplot documentation for explanation.%
    }{The gnuplot epslatex terminal needs graphicx.sty or graphics.sty.}%
    \renewcommand\includegraphics[2][]{}%
  }%
  \providecommand\rotatebox[2]{#2}%
  \@ifundefined{ifGPcolor}{%
    \newif\ifGPcolor
    \GPcolortrue
  }{}%
  \@ifundefined{ifGPblacktext}{%
    \newif\ifGPblacktext
    \GPblacktextfalse
  }{}%
  % define a \g@addto@macro without @ in the name:
  \let\gplgaddtomacro\g@addto@macro
  % define empty templates for all commands taking text:
  \gdef\gplbacktext{}%
  \gdef\gplfronttext{}%
  \makeatother
  \ifGPblacktext
    % no textcolor at all
    \def\colorrgb#1{}%
    \def\colorgray#1{}%
  \else
    % gray or color?
    \ifGPcolor
      \def\colorrgb#1{\color[rgb]{#1}}%
      \def\colorgray#1{\color[gray]{#1}}%
      \expandafter\def\csname LTw\endcsname{\color{white}}%
      \expandafter\def\csname LTb\endcsname{\color{black}}%
      \expandafter\def\csname LTa\endcsname{\color{black}}%
      \expandafter\def\csname LT0\endcsname{\color[rgb]{1,0,0}}%
      \expandafter\def\csname LT1\endcsname{\color[rgb]{0,1,0}}%
      \expandafter\def\csname LT2\endcsname{\color[rgb]{0,0,1}}%
      \expandafter\def\csname LT3\endcsname{\color[rgb]{1,0,1}}%
      \expandafter\def\csname LT4\endcsname{\color[rgb]{0,1,1}}%
      \expandafter\def\csname LT5\endcsname{\color[rgb]{1,1,0}}%
      \expandafter\def\csname LT6\endcsname{\color[rgb]{0,0,0}}%
      \expandafter\def\csname LT7\endcsname{\color[rgb]{1,0.3,0}}%
      \expandafter\def\csname LT8\endcsname{\color[rgb]{0.5,0.5,0.5}}%
    \else
      % gray
      \def\colorrgb#1{\color{black}}%
      \def\colorgray#1{\color[gray]{#1}}%
      \expandafter\def\csname LTw\endcsname{\color{white}}%
      \expandafter\def\csname LTb\endcsname{\color{black}}%
      \expandafter\def\csname LTa\endcsname{\color{black}}%
      \expandafter\def\csname LT0\endcsname{\color{black}}%
      \expandafter\def\csname LT1\endcsname{\color{black}}%
      \expandafter\def\csname LT2\endcsname{\color{black}}%
      \expandafter\def\csname LT3\endcsname{\color{black}}%
      \expandafter\def\csname LT4\endcsname{\color{black}}%
      \expandafter\def\csname LT5\endcsname{\color{black}}%
      \expandafter\def\csname LT6\endcsname{\color{black}}%
      \expandafter\def\csname LT7\endcsname{\color{black}}%
      \expandafter\def\csname LT8\endcsname{\color{black}}%
    \fi
  \fi
  \setlength{\unitlength}{0.0500bp}%
  \begin{picture}(9360.00,7200.00)%
    \gplgaddtomacro\gplbacktext{%
    }%
    \gplgaddtomacro\gplfronttext{%
      \csname LTb\endcsname%
      \put(183,4164){\makebox(0,0)[l]{\strut{}$0.1$ Myr}}%
    }%
    \gplgaddtomacro\gplbacktext{%
      \put(4674,7252){\makebox(0,0){\large $E = 10$ Gev}}%
    }%
    \gplgaddtomacro\gplfronttext{%
      \csname LTb\endcsname%
      \put(3299,4164){\makebox(0,0)[l]{\strut{}$0.2$ Myr}}%
    }%
    \gplgaddtomacro\gplbacktext{%
    }%
    \gplgaddtomacro\gplfronttext{%
      \csname LTb\endcsname%
      \put(6426,4164){\makebox(0,0)[l]{\strut{}$0.4$ Myr}}%
    }%
    \gplgaddtomacro\gplbacktext{%
    }%
    \gplgaddtomacro\gplfronttext{%
      \csname LTb\endcsname%
      \put(2340,260){\makebox(0,0){\strut{}-9}}%
      \put(2924,260){\makebox(0,0){\strut{}-8}}%
      \put(3509,260){\makebox(0,0){\strut{}-7}}%
      \put(4094,260){\makebox(0,0){\strut{}-6}}%
      \put(4679,260){\makebox(0,0){\strut{}-5}}%
      \put(5264,260){\makebox(0,0){\strut{}-4}}%
      \put(5849,260){\makebox(0,0){\strut{}-3}}%
      \put(6434,260){\makebox(0,0){\strut{}-2}}%
      \put(7018,260){\makebox(0,0){\strut{}-1}}%
      \put(4679,0){\makebox(0,0){\strut{}$\log Q$ [erg$^{-1}$ cm$^{-2} s^{-1}$]}}%
      \put(183,996){\makebox(0,0)[l]{\strut{}$0.6$ Myr}}%
    }%
    \gplgaddtomacro\gplbacktext{%
    }%
    \gplgaddtomacro\gplfronttext{%
      \csname LTb\endcsname%
      \put(3299,995){\makebox(0,0)[l]{\strut{}$0.8$ Myr}}%
    }%
    \gplgaddtomacro\gplbacktext{%
    }%
    \gplgaddtomacro\gplfronttext{%
      \csname LTb\endcsname%
      \put(6426,996){\makebox(0,0)[l]{\strut{}$1$ Myr}}%
    }%
    \gplbacktext
    \put(0,0){\includegraphics{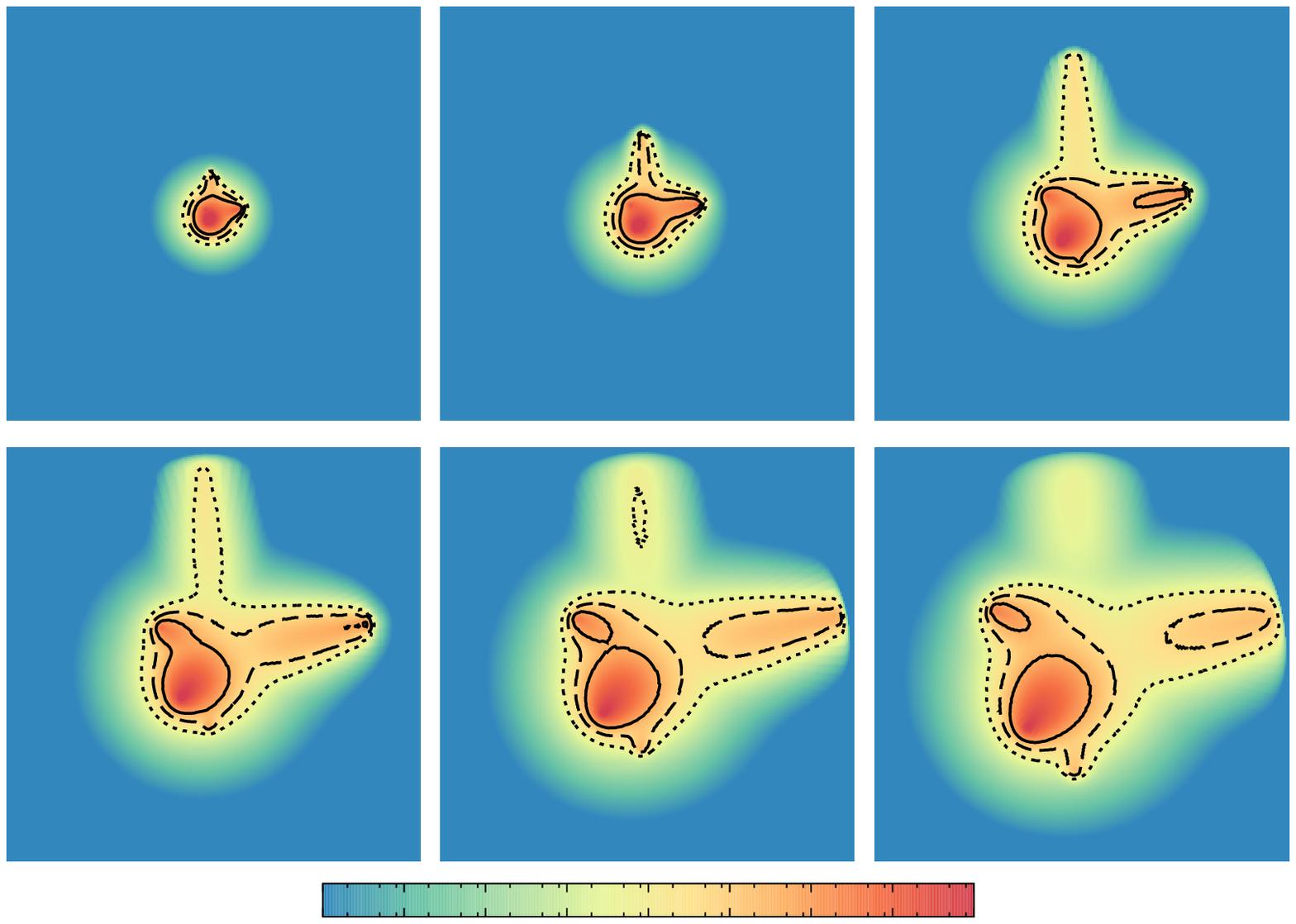}}%
    \gplfronttext
  \end{picture}%
\endgroup

%% file: chi0p1_QSYNC_EN2.tex
% GNUPLOT: LaTeX picture with Postscript
\begingroup
  \makeatletter
  \providecommand\color[2][]{%
    \GenericError{(gnuplot) \space\space\space\@spaces}{%
      Package color not loaded in conjunction with
      terminal option `colourtext'%
    }{See the gnuplot documentation for explanation.%
    }{Either use 'blacktext' in gnuplot or load the package
      color.sty in LaTeX.}%
    \renewcommand\color[2][]{}%
  }%
  \providecommand\includegraphics[2][]{%
    \GenericError{(gnuplot) \space\space\space\@spaces}{%
      Package graphicx or graphics not loaded%
    }{See the gnuplot documentation for explanation.%
    }{The gnuplot epslatex terminal needs graphicx.sty or graphics.sty.}%
    \renewcommand\includegraphics[2][]{}%
  }%
  \providecommand\rotatebox[2]{#2}%
  \@ifundefined{ifGPcolor}{%
    \newif\ifGPcolor
    \GPcolortrue
  }{}%
  \@ifundefined{ifGPblacktext}{%
    \newif\ifGPblacktext
    \GPblacktextfalse
  }{}%
  % define a \g@addto@macro without @ in the name:
  \let\gplgaddtomacro\g@addto@macro
  % define empty templates for all commands taking text:
  \gdef\gplbacktext{}%
  \gdef\gplfronttext{}%
  \makeatother
  \ifGPblacktext
    % no textcolor at all
    \def\colorrgb#1{}%
    \def\colorgray#1{}%
  \else
    % gray or color?
    \ifGPcolor
      \def\colorrgb#1{\color[rgb]{#1}}%
      \def\colorgray#1{\color[gray]{#1}}%
      \expandafter\def\csname LTw\endcsname{\color{white}}%
      \expandafter\def\csname LTb\endcsname{\color{black}}%
      \expandafter\def\csname LTa\endcsname{\color{black}}%
      \expandafter\def\csname LT0\endcsname{\color[rgb]{1,0,0}}%
      \expandafter\def\csname LT1\endcsname{\color[rgb]{0,1,0}}%
      \expandafter\def\csname LT2\endcsname{\color[rgb]{0,0,1}}%
      \expandafter\def\csname LT3\endcsname{\color[rgb]{1,0,1}}%
      \expandafter\def\csname LT4\endcsname{\color[rgb]{0,1,1}}%
      \expandafter\def\csname LT5\endcsname{\color[rgb]{1,1,0}}%
      \expandafter\def\csname LT6\endcsname{\color[rgb]{0,0,0}}%
      \expandafter\def\csname LT7\endcsname{\color[rgb]{1,0.3,0}}%
      \expandafter\def\csname LT8\endcsname{\color[rgb]{0.5,0.5,0.5}}%
    \else
      % gray
      \def\colorrgb#1{\color{black}}%
      \def\colorgray#1{\color[gray]{#1}}%
      \expandafter\def\csname LTw\endcsname{\color{white}}%
      \expandafter\def\csname LTb\endcsname{\color{black}}%
      \expandafter\def\csname LTa\endcsname{\color{black}}%
      \expandafter\def\csname LT0\endcsname{\color{black}}%
      \expandafter\def\csname LT1\endcsname{\color{black}}%
      \expandafter\def\csname LT2\endcsname{\color{black}}%
      \expandafter\def\csname LT3\endcsname{\color{black}}%
      \expandafter\def\csname LT4\endcsname{\color{black}}%
      \expandafter\def\csname LT5\endcsname{\color{black}}%
      \expandafter\def\csname LT6\endcsname{\color{black}}%
      \expandafter\def\csname LT7\endcsname{\color{black}}%
      \expandafter\def\csname LT8\endcsname{\color{black}}%
    \fi
  \fi
  \setlength{\unitlength}{0.0500bp}%
  \begin{picture}(9360.00,7200.00)%
    \gplgaddtomacro\gplbacktext{%
    }%
    \gplgaddtomacro\gplfronttext{%
      \csname LTb\endcsname%
      \put(183,4164){\makebox(0,0)[l]{\strut{}$0.1$ Myr}}%
    }%
    \gplgaddtomacro\gplbacktext{%
      \put(4674,7252){\makebox(0,0){\large $E = 1$ kev}}%
    }%
    \gplgaddtomacro\gplfronttext{%
      \csname LTb\endcsname%
      \put(3299,4164){\makebox(0,0)[l]{\strut{}$0.2$ Myr}}%
    }%
    \gplgaddtomacro\gplbacktext{%
    }%
    \gplgaddtomacro\gplfronttext{%
      \csname LTb\endcsname%
      \put(6426,4164){\makebox(0,0)[l]{\strut{}$0.4$ Myr}}%
    }%
    \gplgaddtomacro\gplbacktext{%
    }%
    \gplgaddtomacro\gplfronttext{%
      \csname LTb\endcsname%
      \put(2340,260){\makebox(0,0){\strut{}1}}%
      \put(2859,260){\makebox(0,0){\strut{}2}}%
      \put(3379,260){\makebox(0,0){\strut{}3}}%
      \put(3899,260){\makebox(0,0){\strut{}4}}%
      \put(4419,260){\makebox(0,0){\strut{}5}}%
      \put(4939,260){\makebox(0,0){\strut{}6}}%
      \put(5459,260){\makebox(0,0){\strut{}7}}%
      \put(5979,260){\makebox(0,0){\strut{}8}}%
      \put(6499,260){\makebox(0,0){\strut{}9}}%
      \put(7019,260){\makebox(0,0){\strut{}10}}%
      \put(4679,0){\makebox(0,0){\strut{}$\log Q$ [erg$^{-1}$ cm$^{-2} s^{-1}$]}}%
      \put(183,996){\makebox(0,0)[l]{\strut{}$0.6$ Myr}}%
    }%
    \gplgaddtomacro\gplbacktext{%
    }%
    \gplgaddtomacro\gplfronttext{%
      \csname LTb\endcsname%
      \put(3299,995){\makebox(0,0)[l]{\strut{}$0.8$ Myr}}%
    }%
    \gplgaddtomacro\gplbacktext{%
    }%
    \gplgaddtomacro\gplfronttext{%
      \csname LTb\endcsname%
      \put(6426,996){\makebox(0,0)[l]{\strut{}$1$ Myr}}%
    }%
    \gplbacktext
    \put(0,0){\includegraphics{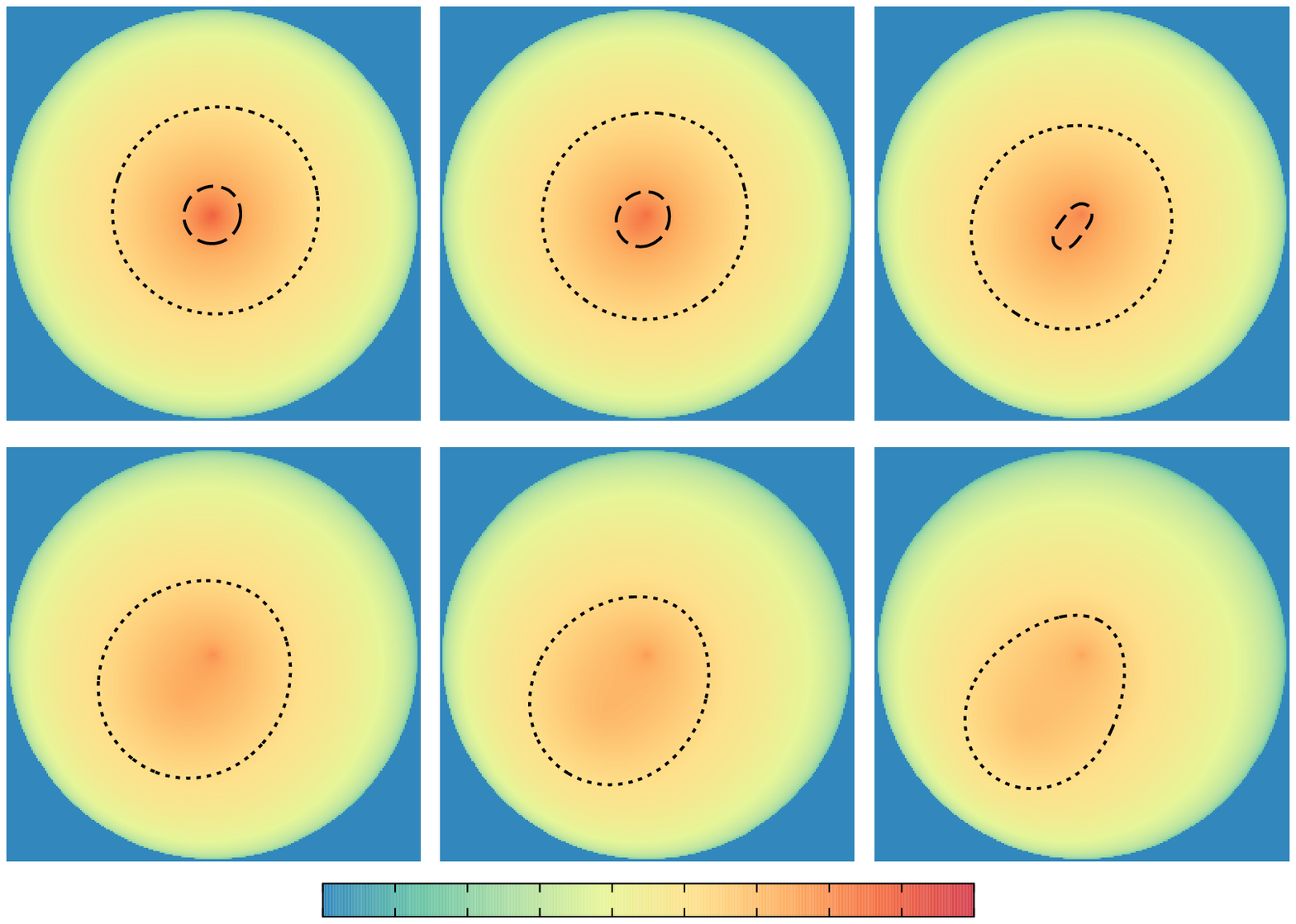}}%
    \gplfronttext
  \end{picture}%
\endgroup

%% file: chi0p1_SED.tex
% GNUPLOT: LaTeX picture with Postscript
\begingroup
  \makeatletter
  \providecommand\color[2][]{%
    \GenericError{(gnuplot) \space\space\space\@spaces}{%
      Package color not loaded in conjunction with
      terminal option `colourtext'%
    }{See the gnuplot documentation for explanation.%
    }{Either use 'blacktext' in gnuplot or load the package
      color.sty in LaTeX.}%
    \renewcommand\color[2][]{}%
  }%
  \providecommand\includegraphics[2][]{%
    \GenericError{(gnuplot) \space\space\space\@spaces}{%
      Package graphicx or graphics not loaded%
    }{See the gnuplot documentation for explanation.%
    }{The gnuplot epslatex terminal needs graphicx.sty or graphics.sty.}%
    \renewcommand\includegraphics[2][]{}%
  }%
  \providecommand\rotatebox[2]{#2}%
  \@ifundefined{ifGPcolor}{%
    \newif\ifGPcolor
    \GPcolortrue
  }{}%
  \@ifundefined{ifGPblacktext}{%
    \newif\ifGPblacktext
    \GPblacktextfalse
  }{}%
  % define a \g@addto@macro without @ in the name:
  \let\gplgaddtomacro\g@addto@macro
  % define empty templates for all commands taking text:
  \gdef\gplbacktext{}%
  \gdef\gplfronttext{}%
  \makeatother
  \ifGPblacktext
    % no textcolor at all
    \def\colorrgb#1{}%
    \def\colorgray#1{}%
  \else
    % gray or color?
    \ifGPcolor
      \def\colorrgb#1{\color[rgb]{#1}}%
      \def\colorgray#1{\color[gray]{#1}}%
      \expandafter\def\csname LTw\endcsname{\color{white}}%
      \expandafter\def\csname LTb\endcsname{\color{black}}%
      \expandafter\def\csname LTa\endcsname{\color{black}}%
      \expandafter\def\csname LT0\endcsname{\color[rgb]{1,0,0}}%
      \expandafter\def\csname LT1\endcsname{\color[rgb]{0,1,0}}%
      \expandafter\def\csname LT2\endcsname{\color[rgb]{0,0,1}}%
      \expandafter\def\csname LT3\endcsname{\color[rgb]{1,0,1}}%
      \expandafter\def\csname LT4\endcsname{\color[rgb]{0,1,1}}%
      \expandafter\def\csname LT5\endcsname{\color[rgb]{1,1,0}}%
      \expandafter\def\csname LT6\endcsname{\color[rgb]{0,0,0}}%
      \expandafter\def\csname LT7\endcsname{\color[rgb]{1,0.3,0}}%
      \expandafter\def\csname LT8\endcsname{\color[rgb]{0.5,0.5,0.5}}%
    \else
      % gray
      \def\colorrgb#1{\color{black}}%
      \def\colorgray#1{\color[gray]{#1}}%
      \expandafter\def\csname LTw\endcsname{\color{white}}%
      \expandafter\def\csname LTb\endcsname{\color{black}}%
      \expandafter\def\csname LTa\endcsname{\color{black}}%
      \expandafter\def\csname LT0\endcsname{\color{black}}%
      \expandafter\def\csname LT1\endcsname{\color{black}}%
      \expandafter\def\csname LT2\endcsname{\color{black}}%
      \expandafter\def\csname LT3\endcsname{\color{black}}%
      \expandafter\def\csname LT4\endcsname{\color{black}}%
      \expandafter\def\csname LT5\endcsname{\color{black}}%
      \expandafter\def\csname LT6\endcsname{\color{black}}%
      \expandafter\def\csname LT7\endcsname{\color{black}}%
      \expandafter\def\csname LT8\endcsname{\color{black}}%
    \fi
  \fi
  \setlength{\unitlength}{0.0500bp}%
  \begin{picture}(9360.00,4320.00)%
    \gplgaddtomacro\gplbacktext{%
      \csname LTb\endcsname%
      \put(0,660){\makebox(0,0)[r]{\strut{}20}}%
      \put(0,1035){\makebox(0,0)[r]{\strut{}22}}%
      \put(0,1410){\makebox(0,0)[r]{\strut{}24}}%
      \put(0,1785){\makebox(0,0)[r]{\strut{}26}}%
      \put(0,2160){\makebox(0,0)[r]{\strut{}28}}%
      \put(0,2534){\makebox(0,0)[r]{\strut{}30}}%
      \put(0,2909){\makebox(0,0)[r]{\strut{}32}}%
      \put(0,3284){\makebox(0,0)[r]{\strut{}34}}%
      \put(0,3659){\makebox(0,0)[r]{\strut{}36}}%
      \put(446,440){\makebox(0,0){\strut{}-3}}%
      \put(918,440){\makebox(0,0){\strut{} 0}}%
      \put(1390,440){\makebox(0,0){\strut{} 3}}%
      \put(1861,440){\makebox(0,0){\strut{} 6}}%
      \put(2333,440){\makebox(0,0){\strut{} 9}}%
      \put(2805,440){\makebox(0,0){\strut{} 12}}%
      \put(-374,2159){\rotatebox{-270}{\makebox(0,0){\strut{}$\log E L$ [erg s$^{-1}$]}}}%
      \put(1625,220){\makebox(0,0){\strut{}$\log E$ [eV]}}%
      \put(761,3472){\makebox(0,0){\large $t = 0.1$ Myr}}%
    }%
    \gplgaddtomacro\gplfronttext{%
      \csname LTb\endcsname%
      \put(2509,2237){\makebox(0,0)[r]{\strut{}CR}}%
      \csname LTb\endcsname%
      \put(2509,2017){\makebox(0,0)[r]{\strut{}$\star$ 1}}%
      \csname LTb\endcsname%
      \put(2509,1797){\makebox(0,0)[r]{\strut{}$\star$ 2}}%
      \csname LTb\endcsname%
      \put(2509,1577){\makebox(0,0)[r]{\strut{}$\star$ 3}}%
      \csname LTb\endcsname%
      \put(2509,1357){\makebox(0,0)[r]{\strut{}$\star$ 4}}%
      \csname LTb\endcsname%
      \put(2509,1137){\makebox(0,0)[r]{\strut{}$\star$ 5}}%
      \csname LTb\endcsname%
      \put(2509,917){\makebox(0,0)[r]{\strut{}$\star$ 6}}%
    }%
    \gplgaddtomacro\gplbacktext{%
      \csname LTb\endcsname%
      \put(3120,660){\makebox(0,0)[r]{\strut{}}}%
      \put(3120,1035){\makebox(0,0)[r]{\strut{}}}%
      \put(3120,1410){\makebox(0,0)[r]{\strut{}}}%
      \put(3120,1785){\makebox(0,0)[r]{\strut{}}}%
      \put(3120,2160){\makebox(0,0)[r]{\strut{}}}%
      \put(3120,2534){\makebox(0,0)[r]{\strut{}}}%
      \put(3120,2909){\makebox(0,0)[r]{\strut{}}}%
      \put(3120,3284){\makebox(0,0)[r]{\strut{}}}%
      \put(3120,3659){\makebox(0,0)[r]{\strut{}}}%
      \put(3566,440){\makebox(0,0){\strut{}-3}}%
      \put(4038,440){\makebox(0,0){\strut{} 0}}%
      \put(4510,440){\makebox(0,0){\strut{} 3}}%
      \put(4981,440){\makebox(0,0){\strut{} 6}}%
      \put(5453,440){\makebox(0,0){\strut{} 9}}%
      \put(5925,440){\makebox(0,0){\strut{} 12}}%
      \put(4745,220){\makebox(0,0){\strut{}$\log E$ [eV]}}%
      \put(4745,3879){\makebox(0,0){\Large $R = 50$ pc}}%
      \put(3881,3472){\makebox(0,0){\large $t = 0.6$ Myr}}%
    }%
    \gplgaddtomacro\gplfronttext{%
    }%
    \gplgaddtomacro\gplbacktext{%
      \csname LTb\endcsname%
      \put(6240,660){\makebox(0,0)[r]{\strut{}}}%
      \put(6240,1035){\makebox(0,0)[r]{\strut{}}}%
      \put(6240,1410){\makebox(0,0)[r]{\strut{}}}%
      \put(6240,1785){\makebox(0,0)[r]{\strut{}}}%
      \put(6240,2160){\makebox(0,0)[r]{\strut{}}}%
      \put(6240,2534){\makebox(0,0)[r]{\strut{}}}%
      \put(6240,2909){\makebox(0,0)[r]{\strut{}}}%
      \put(6240,3284){\makebox(0,0)[r]{\strut{}}}%
      \put(6240,3659){\makebox(0,0)[r]{\strut{}}}%
      \put(6686,440){\makebox(0,0){\strut{}-3}}%
      \put(7158,440){\makebox(0,0){\strut{} 0}}%
      \put(7630,440){\makebox(0,0){\strut{} 3}}%
      \put(8101,440){\makebox(0,0){\strut{} 6}}%
      \put(8573,440){\makebox(0,0){\strut{} 9}}%
      \put(9045,440){\makebox(0,0){\strut{} 12}}%
      \put(7865,220){\makebox(0,0){\strut{}$\log E$ [eV]}}%
      \put(7001,3472){\makebox(0,0){\large $t = 1$ Myr}}%
    }%
    \gplgaddtomacro\gplfronttext{%
    }%
    \gplbacktext
    \put(0,0){\includegraphics{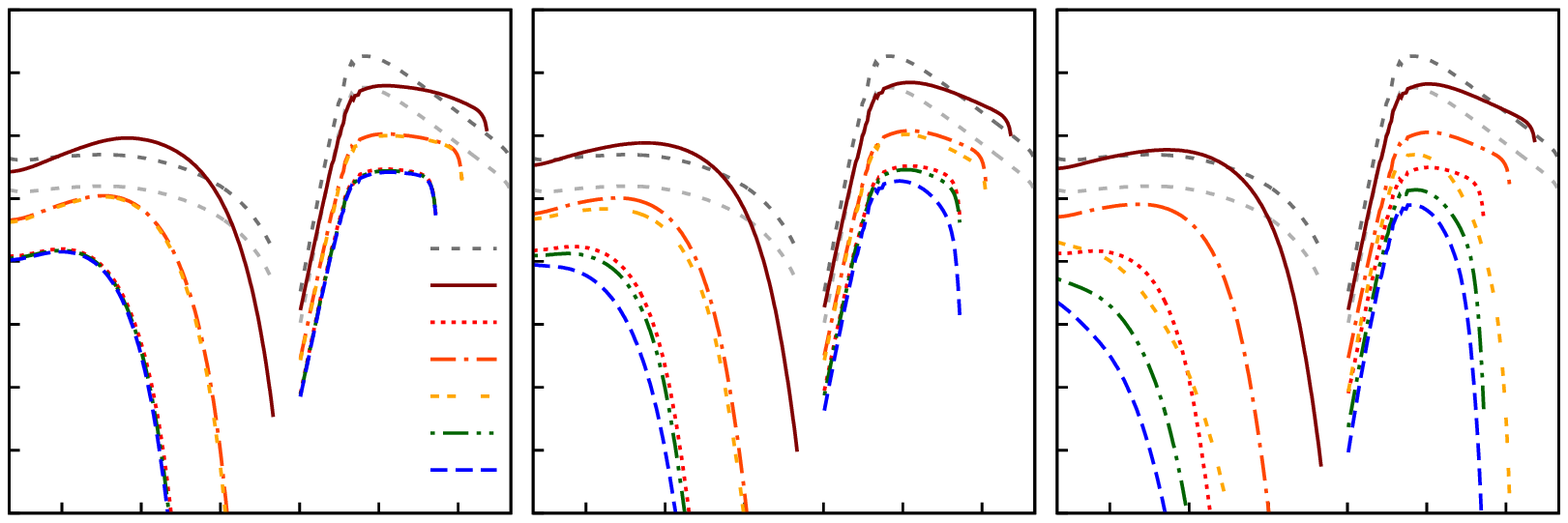}}%
    \gplfronttext
  \end{picture}%
\endgroup

%% file: chi0p1_SED_CORE.tex
% GNUPLOT: LaTeX picture with Postscript
\begingroup
  \makeatletter
  \providecommand\color[2][]{%
    \GenericError{(gnuplot) \space\space\space\@spaces}{%
      Package color not loaded in conjunction with
      terminal option `colourtext'%
    }{See the gnuplot documentation for explanation.%
    }{Either use 'blacktext' in gnuplot or load the package
      color.sty in LaTeX.}%
    \renewcommand\color[2][]{}%
  }%
  \providecommand\includegraphics[2][]{%
    \GenericError{(gnuplot) \space\space\space\@spaces}{%
      Package graphicx or graphics not loaded%
    }{See the gnuplot documentation for explanation.%
    }{The gnuplot epslatex terminal needs graphicx.sty or graphics.sty.}%
    \renewcommand\includegraphics[2][]{}%
  }%
  \providecommand\rotatebox[2]{#2}%
  \@ifundefined{ifGPcolor}{%
    \newif\ifGPcolor
    \GPcolortrue
  }{}%
  \@ifundefined{ifGPblacktext}{%
    \newif\ifGPblacktext
    \GPblacktextfalse
  }{}%
  % define a \g@addto@macro without @ in the name:
  \let\gplgaddtomacro\g@addto@macro
  % define empty templates for all commands taking text:
  \gdef\gplbacktext{}%
  \gdef\gplfronttext{}%
  \makeatother
  \ifGPblacktext
    % no textcolor at all
    \def\colorrgb#1{}%
    \def\colorgray#1{}%
  \else
    % gray or color?
    \ifGPcolor
      \def\colorrgb#1{\color[rgb]{#1}}%
      \def\colorgray#1{\color[gray]{#1}}%
      \expandafter\def\csname LTw\endcsname{\color{white}}%
      \expandafter\def\csname LTb\endcsname{\color{black}}%
      \expandafter\def\csname LTa\endcsname{\color{black}}%
      \expandafter\def\csname LT0\endcsname{\color[rgb]{1,0,0}}%
      \expandafter\def\csname LT1\endcsname{\color[rgb]{0,1,0}}%
      \expandafter\def\csname LT2\endcsname{\color[rgb]{0,0,1}}%
      \expandafter\def\csname LT3\endcsname{\color[rgb]{1,0,1}}%
      \expandafter\def\csname LT4\endcsname{\color[rgb]{0,1,1}}%
      \expandafter\def\csname LT5\endcsname{\color[rgb]{1,1,0}}%
      \expandafter\def\csname LT6\endcsname{\color[rgb]{0,0,0}}%
      \expandafter\def\csname LT7\endcsname{\color[rgb]{1,0.3,0}}%
      \expandafter\def\csname LT8\endcsname{\color[rgb]{0.5,0.5,0.5}}%
    \else
      % gray
      \def\colorrgb#1{\color{black}}%
      \def\colorgray#1{\color[gray]{#1}}%
      \expandafter\def\csname LTw\endcsname{\color{white}}%
      \expandafter\def\csname LTb\endcsname{\color{black}}%
      \expandafter\def\csname LTa\endcsname{\color{black}}%
      \expandafter\def\csname LT0\endcsname{\color{black}}%
      \expandafter\def\csname LT1\endcsname{\color{black}}%
      \expandafter\def\csname LT2\endcsname{\color{black}}%
      \expandafter\def\csname LT3\endcsname{\color{black}}%
      \expandafter\def\csname LT4\endcsname{\color{black}}%
      \expandafter\def\csname LT5\endcsname{\color{black}}%
      \expandafter\def\csname LT6\endcsname{\color{black}}%
      \expandafter\def\csname LT7\endcsname{\color{black}}%
      \expandafter\def\csname LT8\endcsname{\color{black}}%
    \fi
  \fi
  \setlength{\unitlength}{0.0500bp}%
  \begin{picture}(9360.00,4320.00)%
    \gplgaddtomacro\gplbacktext{%
      \csname LTb\endcsname%
      \put(0,660){\makebox(0,0)[r]{\strut{}20}}%
      \put(0,1035){\makebox(0,0)[r]{\strut{}22}}%
      \put(0,1410){\makebox(0,0)[r]{\strut{}24}}%
      \put(0,1785){\makebox(0,0)[r]{\strut{}26}}%
      \put(0,2160){\makebox(0,0)[r]{\strut{}28}}%
      \put(0,2534){\makebox(0,0)[r]{\strut{}30}}%
      \put(0,2909){\makebox(0,0)[r]{\strut{}32}}%
      \put(0,3284){\makebox(0,0)[r]{\strut{}34}}%
      \put(0,3659){\makebox(0,0)[r]{\strut{}36}}%
      \put(446,440){\makebox(0,0){\strut{}-3}}%
      \put(918,440){\makebox(0,0){\strut{} 0}}%
      \put(1390,440){\makebox(0,0){\strut{} 3}}%
      \put(1861,440){\makebox(0,0){\strut{} 6}}%
      \put(2333,440){\makebox(0,0){\strut{} 9}}%
      \put(2805,440){\makebox(0,0){\strut{} 12}}%
      \put(-374,2159){\rotatebox{-270}{\makebox(0,0){\strut{}$\log E L$ [erg s$^{-1}$]}}}%
      \put(1625,220){\makebox(0,0){\strut{}$\log E$ [eV]}}%
      \put(761,3472){\makebox(0,0){\large $t = 0.1$ Myr}}%
    }%
    \gplgaddtomacro\gplfronttext{%
    }%
    \gplgaddtomacro\gplbacktext{%
      \csname LTb\endcsname%
      \put(3120,660){\makebox(0,0)[r]{\strut{}}}%
      \put(3120,1035){\makebox(0,0)[r]{\strut{}}}%
      \put(3120,1410){\makebox(0,0)[r]{\strut{}}}%
      \put(3120,1785){\makebox(0,0)[r]{\strut{}}}%
      \put(3120,2160){\makebox(0,0)[r]{\strut{}}}%
      \put(3120,2534){\makebox(0,0)[r]{\strut{}}}%
      \put(3120,2909){\makebox(0,0)[r]{\strut{}}}%
      \put(3120,3284){\makebox(0,0)[r]{\strut{}}}%
      \put(3120,3659){\makebox(0,0)[r]{\strut{}}}%
      \put(3566,440){\makebox(0,0){\strut{}-3}}%
      \put(4038,440){\makebox(0,0){\strut{} 0}}%
      \put(4510,440){\makebox(0,0){\strut{} 3}}%
      \put(4981,440){\makebox(0,0){\strut{} 6}}%
      \put(5453,440){\makebox(0,0){\strut{} 9}}%
      \put(5925,440){\makebox(0,0){\strut{} 12}}%
      \put(4745,220){\makebox(0,0){\strut{}$\log E$ [eV]}}%
      \put(4745,3879){\makebox(0,0){\Large $R = 1.5$ pc}}%
      \put(3881,3472){\makebox(0,0){\large $t = 0.6$ Myr}}%
    }%
    \gplgaddtomacro\gplfronttext{%
    }%
    \gplgaddtomacro\gplbacktext{%
      \csname LTb\endcsname%
      \put(6240,660){\makebox(0,0)[r]{\strut{}}}%
      \put(6240,1035){\makebox(0,0)[r]{\strut{}}}%
      \put(6240,1410){\makebox(0,0)[r]{\strut{}}}%
      \put(6240,1785){\makebox(0,0)[r]{\strut{}}}%
      \put(6240,2160){\makebox(0,0)[r]{\strut{}}}%
      \put(6240,2534){\makebox(0,0)[r]{\strut{}}}%
      \put(6240,2909){\makebox(0,0)[r]{\strut{}}}%
      \put(6240,3284){\makebox(0,0)[r]{\strut{}}}%
      \put(6240,3659){\makebox(0,0)[r]{\strut{}}}%
      \put(6686,440){\makebox(0,0){\strut{}-3}}%
      \put(7158,440){\makebox(0,0){\strut{} 0}}%
      \put(7630,440){\makebox(0,0){\strut{} 3}}%
      \put(8101,440){\makebox(0,0){\strut{} 6}}%
      \put(8573,440){\makebox(0,0){\strut{} 9}}%
      \put(9045,440){\makebox(0,0){\strut{} 12}}%
      \put(7865,220){\makebox(0,0){\strut{}$\log E$ [eV]}}%
      \put(7001,3472){\makebox(0,0){\large $t = 1$ Myr}}%
    }%
    \gplgaddtomacro\gplfronttext{%
    }%
    \gplbacktext
    \put(0,0){\includegraphics{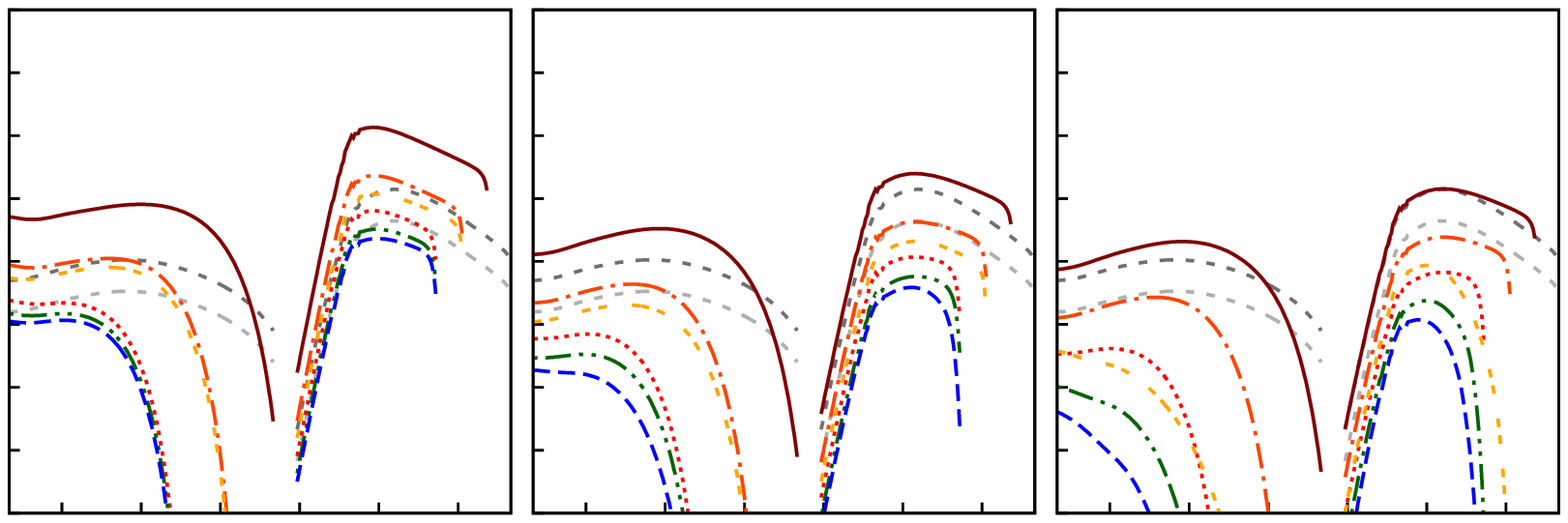}}%
    \gplfronttext
  \end{picture}%
\endgroup

%% file: chi0p001_SED.tex
% GNUPLOT: LaTeX picture with Postscript
\begingroup
  \makeatletter
  \providecommand\color[2][]{%
    \GenericError{(gnuplot) \space\space\space\@spaces}{%
      Package color not loaded in conjunction with
      terminal option `colourtext'%
    }{See the gnuplot documentation for explanation.%
    }{Either use 'blacktext' in gnuplot or load the package
      color.sty in LaTeX.}%
    \renewcommand\color[2][]{}%
  }%
  \providecommand\includegraphics[2][]{%
    \GenericError{(gnuplot) \space\space\space\@spaces}{%
      Package graphicx or graphics not loaded%
    }{See the gnuplot documentation for explanation.%
    }{The gnuplot epslatex terminal needs graphicx.sty or graphics.sty.}%
    \renewcommand\includegraphics[2][]{}%
  }%
  \providecommand\rotatebox[2]{#2}%
  \@ifundefined{ifGPcolor}{%
    \newif\ifGPcolor
    \GPcolortrue
  }{}%
  \@ifundefined{ifGPblacktext}{%
    \newif\ifGPblacktext
    \GPblacktextfalse
  }{}%
  % define a \g@addto@macro without @ in the name:
  \let\gplgaddtomacro\g@addto@macro
  % define empty templates for all commands taking text:
  \gdef\gplbacktext{}%
  \gdef\gplfronttext{}%
  \makeatother
  \ifGPblacktext
    % no textcolor at all
    \def\colorrgb#1{}%
    \def\colorgray#1{}%
  \else
    % gray or color?
    \ifGPcolor
      \def\colorrgb#1{\color[rgb]{#1}}%
      \def\colorgray#1{\color[gray]{#1}}%
      \expandafter\def\csname LTw\endcsname{\color{white}}%
      \expandafter\def\csname LTb\endcsname{\color{black}}%
      \expandafter\def\csname LTa\endcsname{\color{black}}%
      \expandafter\def\csname LT0\endcsname{\color[rgb]{1,0,0}}%
      \expandafter\def\csname LT1\endcsname{\color[rgb]{0,1,0}}%
      \expandafter\def\csname LT2\endcsname{\color[rgb]{0,0,1}}%
      \expandafter\def\csname LT3\endcsname{\color[rgb]{1,0,1}}%
      \expandafter\def\csname LT4\endcsname{\color[rgb]{0,1,1}}%
      \expandafter\def\csname LT5\endcsname{\color[rgb]{1,1,0}}%
      \expandafter\def\csname LT6\endcsname{\color[rgb]{0,0,0}}%
      \expandafter\def\csname LT7\endcsname{\color[rgb]{1,0.3,0}}%
      \expandafter\def\csname LT8\endcsname{\color[rgb]{0.5,0.5,0.5}}%
    \else
      % gray
      \def\colorrgb#1{\color{black}}%
      \def\colorgray#1{\color[gray]{#1}}%
      \expandafter\def\csname LTw\endcsname{\color{white}}%
      \expandafter\def\csname LTb\endcsname{\color{black}}%
      \expandafter\def\csname LTa\endcsname{\color{black}}%
      \expandafter\def\csname LT0\endcsname{\color{black}}%
      \expandafter\def\csname LT1\endcsname{\color{black}}%
      \expandafter\def\csname LT2\endcsname{\color{black}}%
      \expandafter\def\csname LT3\endcsname{\color{black}}%
      \expandafter\def\csname LT4\endcsname{\color{black}}%
      \expandafter\def\csname LT5\endcsname{\color{black}}%
      \expandafter\def\csname LT6\endcsname{\color{black}}%
      \expandafter\def\csname LT7\endcsname{\color{black}}%
      \expandafter\def\csname LT8\endcsname{\color{black}}%
    \fi
  \fi
  \setlength{\unitlength}{0.0500bp}%
  \begin{picture}(9360.00,4320.00)%
    \gplgaddtomacro\gplbacktext{%
      \csname LTb\endcsname%
      \put(0,660){\makebox(0,0)[r]{\strut{}20}}%
      \put(0,1035){\makebox(0,0)[r]{\strut{}22}}%
      \put(0,1410){\makebox(0,0)[r]{\strut{}24}}%
      \put(0,1785){\makebox(0,0)[r]{\strut{}26}}%
      \put(0,2160){\makebox(0,0)[r]{\strut{}28}}%
      \put(0,2534){\makebox(0,0)[r]{\strut{}30}}%
      \put(0,2909){\makebox(0,0)[r]{\strut{}32}}%
      \put(0,3284){\makebox(0,0)[r]{\strut{}34}}%
      \put(0,3659){\makebox(0,0)[r]{\strut{}36}}%
      \put(446,440){\makebox(0,0){\strut{}-3}}%
      \put(918,440){\makebox(0,0){\strut{} 0}}%
      \put(1390,440){\makebox(0,0){\strut{} 3}}%
      \put(1861,440){\makebox(0,0){\strut{} 6}}%
      \put(2333,440){\makebox(0,0){\strut{} 9}}%
      \put(2805,440){\makebox(0,0){\strut{} 12}}%
      \put(-374,2159){\rotatebox{-270}{\makebox(0,0){\strut{}$\log E L$ [erg s$^{-1}$]}}}%
      \put(1625,220){\makebox(0,0){\strut{}$\log E$ [eV]}}%
      \put(761,3472){\makebox(0,0){\large $t = 0.1$ Myr}}%
    }%
    \gplgaddtomacro\gplfronttext{%
      \csname LTb\endcsname%
      \put(2509,2237){\makebox(0,0)[r]{\strut{}CR}}%
      \csname LTb\endcsname%
      \put(2509,2017){\makebox(0,0)[r]{\strut{}$\star$ 1}}%
      \csname LTb\endcsname%
      \put(2509,1797){\makebox(0,0)[r]{\strut{}$\star$ 2}}%
      \csname LTb\endcsname%
      \put(2509,1577){\makebox(0,0)[r]{\strut{}$\star$ 3}}%
      \csname LTb\endcsname%
      \put(2509,1357){\makebox(0,0)[r]{\strut{}$\star$ 4}}%
      \csname LTb\endcsname%
      \put(2509,1137){\makebox(0,0)[r]{\strut{}$\star$ 5}}%
      \csname LTb\endcsname%
      \put(2509,917){\makebox(0,0)[r]{\strut{}$\star$ 6}}%
    }%
    \gplgaddtomacro\gplbacktext{%
      \csname LTb\endcsname%
      \put(3120,660){\makebox(0,0)[r]{\strut{}}}%
      \put(3120,1035){\makebox(0,0)[r]{\strut{}}}%
      \put(3120,1410){\makebox(0,0)[r]{\strut{}}}%
      \put(3120,1785){\makebox(0,0)[r]{\strut{}}}%
      \put(3120,2160){\makebox(0,0)[r]{\strut{}}}%
      \put(3120,2534){\makebox(0,0)[r]{\strut{}}}%
      \put(3120,2909){\makebox(0,0)[r]{\strut{}}}%
      \put(3120,3284){\makebox(0,0)[r]{\strut{}}}%
      \put(3120,3659){\makebox(0,0)[r]{\strut{}}}%
      \put(3566,440){\makebox(0,0){\strut{}-3}}%
      \put(4038,440){\makebox(0,0){\strut{} 0}}%
      \put(4510,440){\makebox(0,0){\strut{} 3}}%
      \put(4981,440){\makebox(0,0){\strut{} 6}}%
      \put(5453,440){\makebox(0,0){\strut{} 9}}%
      \put(5925,440){\makebox(0,0){\strut{} 12}}%
      \put(4745,220){\makebox(0,0){\strut{}$\log E$ [eV]}}%
      \put(4745,3879){\makebox(0,0){\Large $R = 50$ pc}}%
      \put(3881,3472){\makebox(0,0){\large $t = 0.6$ Myr}}%
    }%
    \gplgaddtomacro\gplfronttext{%
    }%
    \gplgaddtomacro\gplbacktext{%
      \csname LTb\endcsname%
      \put(6240,660){\makebox(0,0)[r]{\strut{}}}%
      \put(6240,1035){\makebox(0,0)[r]{\strut{}}}%
      \put(6240,1410){\makebox(0,0)[r]{\strut{}}}%
      \put(6240,1785){\makebox(0,0)[r]{\strut{}}}%
      \put(6240,2160){\makebox(0,0)[r]{\strut{}}}%
      \put(6240,2534){\makebox(0,0)[r]{\strut{}}}%
      \put(6240,2909){\makebox(0,0)[r]{\strut{}}}%
      \put(6240,3284){\makebox(0,0)[r]{\strut{}}}%
      \put(6240,3659){\makebox(0,0)[r]{\strut{}}}%
      \put(6686,440){\makebox(0,0){\strut{}-3}}%
      \put(7158,440){\makebox(0,0){\strut{} 0}}%
      \put(7630,440){\makebox(0,0){\strut{} 3}}%
      \put(8101,440){\makebox(0,0){\strut{} 6}}%
      \put(8573,440){\makebox(0,0){\strut{} 9}}%
      \put(9045,440){\makebox(0,0){\strut{} 12}}%
      \put(7865,220){\makebox(0,0){\strut{}$\log E$ [eV]}}%
      \put(7001,3472){\makebox(0,0){\large $t = 1$ Myr}}%
    }%
    \gplgaddtomacro\gplfronttext{%
    }%
    \gplbacktext
    \put(0,0){\includegraphics{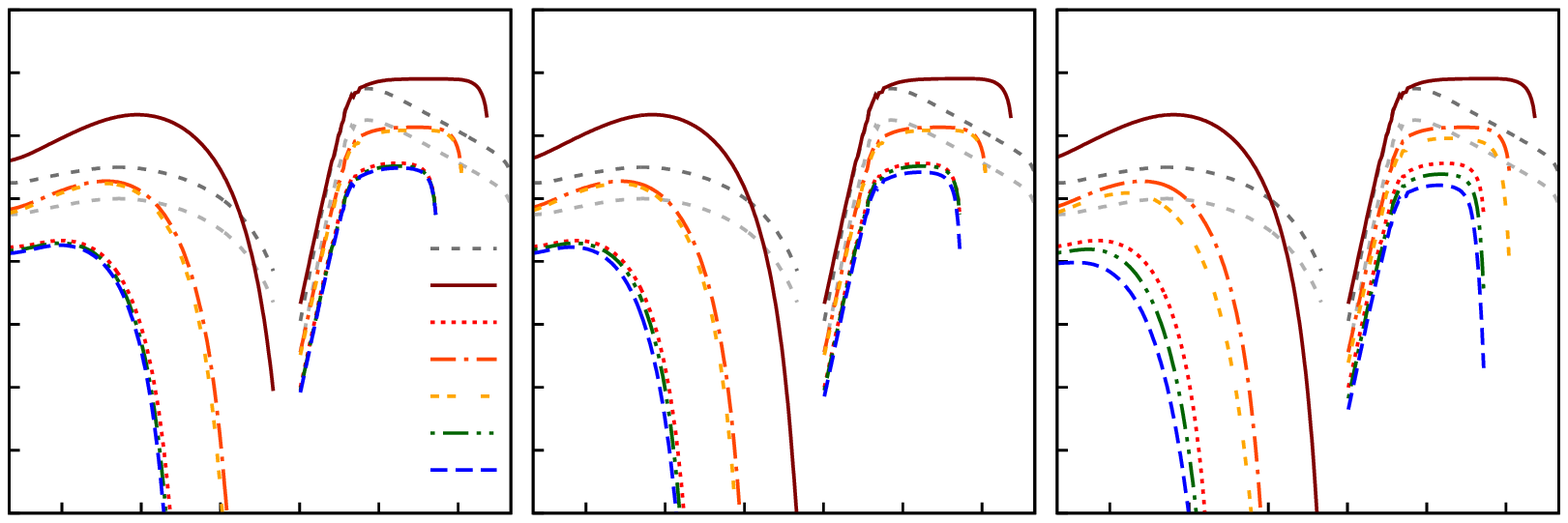}}%
    \gplfronttext
  \end{picture}%
\endgroup

%% file: chi0p001_SED_CORE.tex
% GNUPLOT: LaTeX picture with Postscript
\begingroup
  \makeatletter
  \providecommand\color[2][]{%
    \GenericError{(gnuplot) \space\space\space\@spaces}{%
      Package color not loaded in conjunction with
      terminal option `colourtext'%
    }{See the gnuplot documentation for explanation.%
    }{Either use 'blacktext' in gnuplot or load the package
      color.sty in LaTeX.}%
    \renewcommand\color[2][]{}%
  }%
  \providecommand\includegraphics[2][]{%
    \GenericError{(gnuplot) \space\space\space\@spaces}{%
      Package graphicx or graphics not loaded%
    }{See the gnuplot documentation for explanation.%
    }{The gnuplot epslatex terminal needs graphicx.sty or graphics.sty.}%
    \renewcommand\includegraphics[2][]{}%
  }%
  \providecommand\rotatebox[2]{#2}%
  \@ifundefined{ifGPcolor}{%
    \newif\ifGPcolor
    \GPcolortrue
  }{}%
  \@ifundefined{ifGPblacktext}{%
    \newif\ifGPblacktext
    \GPblacktextfalse
  }{}%
  % define a \g@addto@macro without @ in the name:
  \let\gplgaddtomacro\g@addto@macro
  % define empty templates for all commands taking text:
  \gdef\gplbacktext{}%
  \gdef\gplfronttext{}%
  \makeatother
  \ifGPblacktext
    % no textcolor at all
    \def\colorrgb#1{}%
    \def\colorgray#1{}%
  \else
    % gray or color?
    \ifGPcolor
      \def\colorrgb#1{\color[rgb]{#1}}%
      \def\colorgray#1{\color[gray]{#1}}%
      \expandafter\def\csname LTw\endcsname{\color{white}}%
      \expandafter\def\csname LTb\endcsname{\color{black}}%
      \expandafter\def\csname LTa\endcsname{\color{black}}%
      \expandafter\def\csname LT0\endcsname{\color[rgb]{1,0,0}}%
      \expandafter\def\csname LT1\endcsname{\color[rgb]{0,1,0}}%
      \expandafter\def\csname LT2\endcsname{\color[rgb]{0,0,1}}%
      \expandafter\def\csname LT3\endcsname{\color[rgb]{1,0,1}}%
      \expandafter\def\csname LT4\endcsname{\color[rgb]{0,1,1}}%
      \expandafter\def\csname LT5\endcsname{\color[rgb]{1,1,0}}%
      \expandafter\def\csname LT6\endcsname{\color[rgb]{0,0,0}}%
      \expandafter\def\csname LT7\endcsname{\color[rgb]{1,0.3,0}}%
      \expandafter\def\csname LT8\endcsname{\color[rgb]{0.5,0.5,0.5}}%
    \else
      % gray
      \def\colorrgb#1{\color{black}}%
      \def\colorgray#1{\color[gray]{#1}}%
      \expandafter\def\csname LTw\endcsname{\color{white}}%
      \expandafter\def\csname LTb\endcsname{\color{black}}%
      \expandafter\def\csname LTa\endcsname{\color{black}}%
      \expandafter\def\csname LT0\endcsname{\color{black}}%
      \expandafter\def\csname LT1\endcsname{\color{black}}%
      \expandafter\def\csname LT2\endcsname{\color{black}}%
      \expandafter\def\csname LT3\endcsname{\color{black}}%
      \expandafter\def\csname LT4\endcsname{\color{black}}%
      \expandafter\def\csname LT5\endcsname{\color{black}}%
      \expandafter\def\csname LT6\endcsname{\color{black}}%
      \expandafter\def\csname LT7\endcsname{\color{black}}%
      \expandafter\def\csname LT8\endcsname{\color{black}}%
    \fi
  \fi
  \setlength{\unitlength}{0.0500bp}%
  \begin{picture}(9360.00,4320.00)%
    \gplgaddtomacro\gplbacktext{%
      \csname LTb\endcsname%
      \put(0,660){\makebox(0,0)[r]{\strut{}20}}%
      \put(0,1035){\makebox(0,0)[r]{\strut{}22}}%
      \put(0,1410){\makebox(0,0)[r]{\strut{}24}}%
      \put(0,1785){\makebox(0,0)[r]{\strut{}26}}%
      \put(0,2160){\makebox(0,0)[r]{\strut{}28}}%
      \put(0,2534){\makebox(0,0)[r]{\strut{}30}}%
      \put(0,2909){\makebox(0,0)[r]{\strut{}32}}%
      \put(0,3284){\makebox(0,0)[r]{\strut{}34}}%
      \put(0,3659){\makebox(0,0)[r]{\strut{}36}}%
      \put(446,440){\makebox(0,0){\strut{}-3}}%
      \put(918,440){\makebox(0,0){\strut{} 0}}%
      \put(1390,440){\makebox(0,0){\strut{} 3}}%
      \put(1861,440){\makebox(0,0){\strut{} 6}}%
      \put(2333,440){\makebox(0,0){\strut{} 9}}%
      \put(2805,440){\makebox(0,0){\strut{} 12}}%
      \put(-374,2159){\rotatebox{-270}{\makebox(0,0){\strut{}$\log E L$ [erg s$^{-1}$]}}}%
      \put(1625,220){\makebox(0,0){\strut{}$\log E$ [eV]}}%
      \put(761,3472){\makebox(0,0){\large $t = 0.1$ Myr}}%
    }%
    \gplgaddtomacro\gplfronttext{%
    }%
    \gplgaddtomacro\gplbacktext{%
      \csname LTb\endcsname%
      \put(3120,660){\makebox(0,0)[r]{\strut{}}}%
      \put(3120,1035){\makebox(0,0)[r]{\strut{}}}%
      \put(3120,1410){\makebox(0,0)[r]{\strut{}}}%
      \put(3120,1785){\makebox(0,0)[r]{\strut{}}}%
      \put(3120,2160){\makebox(0,0)[r]{\strut{}}}%
      \put(3120,2534){\makebox(0,0)[r]{\strut{}}}%
      \put(3120,2909){\makebox(0,0)[r]{\strut{}}}%
      \put(3120,3284){\makebox(0,0)[r]{\strut{}}}%
      \put(3120,3659){\makebox(0,0)[r]{\strut{}}}%
      \put(3566,440){\makebox(0,0){\strut{}-3}}%
      \put(4038,440){\makebox(0,0){\strut{} 0}}%
      \put(4510,440){\makebox(0,0){\strut{} 3}}%
      \put(4981,440){\makebox(0,0){\strut{} 6}}%
      \put(5453,440){\makebox(0,0){\strut{} 9}}%
      \put(5925,440){\makebox(0,0){\strut{} 12}}%
      \put(4745,220){\makebox(0,0){\strut{}$\log E$ [eV]}}%
      \put(4745,3879){\makebox(0,0){\Large $R = 1.5$ pc}}%
      \put(3881,3472){\makebox(0,0){\large $t = 0.6$ Myr}}%
    }%
    \gplgaddtomacro\gplfronttext{%
    }%
    \gplgaddtomacro\gplbacktext{%
      \csname LTb\endcsname%
      \put(6240,660){\makebox(0,0)[r]{\strut{}}}%
      \put(6240,1035){\makebox(0,0)[r]{\strut{}}}%
      \put(6240,1410){\makebox(0,0)[r]{\strut{}}}%
      \put(6240,1785){\makebox(0,0)[r]{\strut{}}}%
      \put(6240,2160){\makebox(0,0)[r]{\strut{}}}%
      \put(6240,2534){\makebox(0,0)[r]{\strut{}}}%
      \put(6240,2909){\makebox(0,0)[r]{\strut{}}}%
      \put(6240,3284){\makebox(0,0)[r]{\strut{}}}%
      \put(6240,3659){\makebox(0,0)[r]{\strut{}}}%
      \put(6686,440){\makebox(0,0){\strut{}-3}}%
      \put(7158,440){\makebox(0,0){\strut{} 0}}%
      \put(7630,440){\makebox(0,0){\strut{} 3}}%
      \put(8101,440){\makebox(0,0){\strut{} 6}}%
      \put(8573,440){\makebox(0,0){\strut{} 9}}%
      \put(9045,440){\makebox(0,0){\strut{} 12}}%
      \put(7865,220){\makebox(0,0){\strut{}$\log E$ [eV]}}%
      \put(7001,3472){\makebox(0,0){\large $t = 1$ Myr}}%
    }%
    \gplgaddtomacro\gplfronttext{%
    }%
    \gplbacktext
    \put(0,0){\includegraphics{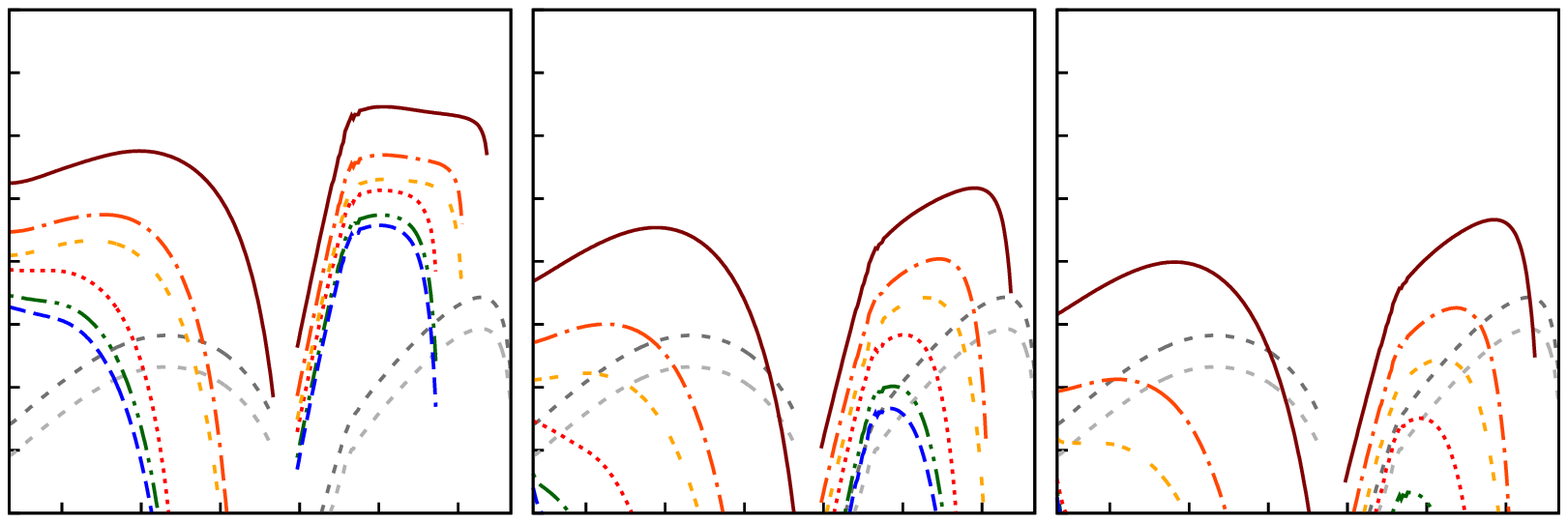}}%
    \gplfronttext
  \end{picture}%
\endgroup

%% file: chi0p1_en_prot_time.tex
% GNUPLOT: LaTeX picture with Postscript
\begingroup
  \makeatletter
  \providecommand\color[2][]{%
    \GenericError{(gnuplot) \space\space\space\@spaces}{%
      Package color not loaded in conjunction with
      terminal option `colourtext'%
    }{See the gnuplot documentation for explanation.%
    }{Either use 'blacktext' in gnuplot or load the package
      color.sty in LaTeX.}%
    \renewcommand\color[2][]{}%
  }%
  \providecommand\includegraphics[2][]{%
    \GenericError{(gnuplot) \space\space\space\@spaces}{%
      Package graphicx or graphics not loaded%
    }{See the gnuplot documentation for explanation.%
    }{The gnuplot epslatex terminal needs graphicx.sty or graphics.sty.}%
    \renewcommand\includegraphics[2][]{}%
  }%
  \providecommand\rotatebox[2]{#2}%
  \@ifundefined{ifGPcolor}{%
    \newif\ifGPcolor
    \GPcolortrue
  }{}%
  \@ifundefined{ifGPblacktext}{%
    \newif\ifGPblacktext
    \GPblacktextfalse
  }{}%
  % define a \g@addto@macro without @ in the name:
  \let\gplgaddtomacro\g@addto@macro
  % define empty templates for all commands taking text:
  \gdef\gplbacktext{}%
  \gdef\gplfronttext{}%
  \makeatother
  \ifGPblacktext
    % no textcolor at all
    \def\colorrgb#1{}%
    \def\colorgray#1{}%
  \else
    % gray or color?
    \ifGPcolor
      \def\colorrgb#1{\color[rgb]{#1}}%
      \def\colorgray#1{\color[gray]{#1}}%
      \expandafter\def\csname LTw\endcsname{\color{white}}%
      \expandafter\def\csname LTb\endcsname{\color{black}}%
      \expandafter\def\csname LTa\endcsname{\color{black}}%
      \expandafter\def\csname LT0\endcsname{\color[rgb]{1,0,0}}%
      \expandafter\def\csname LT1\endcsname{\color[rgb]{0,1,0}}%
      \expandafter\def\csname LT2\endcsname{\color[rgb]{0,0,1}}%
      \expandafter\def\csname LT3\endcsname{\color[rgb]{1,0,1}}%
      \expandafter\def\csname LT4\endcsname{\color[rgb]{0,1,1}}%
      \expandafter\def\csname LT5\endcsname{\color[rgb]{1,1,0}}%
      \expandafter\def\csname LT6\endcsname{\color[rgb]{0,0,0}}%
      \expandafter\def\csname LT7\endcsname{\color[rgb]{1,0.3,0}}%
      \expandafter\def\csname LT8\endcsname{\color[rgb]{0.5,0.5,0.5}}%
    \else
      % gray
      \def\colorrgb#1{\color{black}}%
      \def\colorgray#1{\color[gray]{#1}}%
      \expandafter\def\csname LTw\endcsname{\color{white}}%
      \expandafter\def\csname LTb\endcsname{\color{black}}%
      \expandafter\def\csname LTa\endcsname{\color{black}}%
      \expandafter\def\csname LT0\endcsname{\color{black}}%
      \expandafter\def\csname LT1\endcsname{\color{black}}%
      \expandafter\def\csname LT2\endcsname{\color{black}}%
      \expandafter\def\csname LT3\endcsname{\color{black}}%
      \expandafter\def\csname LT4\endcsname{\color{black}}%
      \expandafter\def\csname LT5\endcsname{\color{black}}%
      \expandafter\def\csname LT6\endcsname{\color{black}}%
      \expandafter\def\csname LT7\endcsname{\color{black}}%
      \expandafter\def\csname LT8\endcsname{\color{black}}%
    \fi
  \fi
  \setlength{\unitlength}{0.0500bp}%
  \begin{picture}(4750.00,4320.00)%
    \gplgaddtomacro\gplbacktext{%
      \csname LTb\endcsname%
      \put(462,660){\makebox(0,0)[r]{\strut{}$44$}}%
      \put(462,1465){\makebox(0,0)[r]{\strut{}$46$}}%
      \put(462,2270){\makebox(0,0)[r]{\strut{}$48$}}%
      \put(462,3074){\makebox(0,0)[r]{\strut{}$50$}}%
      \put(462,3879){\makebox(0,0)[r]{\strut{}$52$}}%
      \put(594,440){\makebox(0,0){\strut{}$0.0$}}%
      \put(1399,440){\makebox(0,0){\strut{}$0.2$}}%
      \put(2203,440){\makebox(0,0){\strut{}$0.4$}}%
      \put(3008,440){\makebox(0,0){\strut{}$0.6$}}%
      \put(3812,440){\makebox(0,0){\strut{}$0.8$}}%
      \put(4617,440){\makebox(0,0){\strut{}$1.0$}}%
      \put(88,2269){\rotatebox{-270}{\makebox(0,0){\strut{}$\log E$ [erg]}}}%
      \put(2605,220){\makebox(0,0){\strut{}$t$ [Myr]}}%
      \put(2605,4099){\makebox(0,0){\strut{}Total protons energy}}%
    }%
    \gplgaddtomacro\gplfronttext{%
    }%
    \gplbacktext
    \put(0,0){\includegraphics{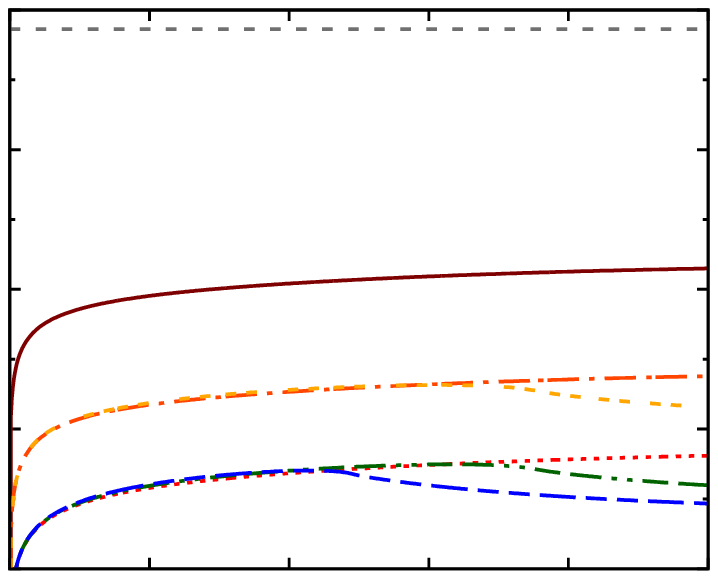}}%
    \gplfronttext
  \end{picture}%
\endgroup

%% file: chi0p1_en_elec_time.tex
% GNUPLOT: LaTeX picture with Postscript
\begingroup
  \makeatletter
  \providecommand\color[2][]{%
    \GenericError{(gnuplot) \space\space\space\@spaces}{%
      Package color not loaded in conjunction with
      terminal option `colourtext'%
    }{See the gnuplot documentation for explanation.%
    }{Either use 'blacktext' in gnuplot or load the package
      color.sty in LaTeX.}%
    \renewcommand\color[2][]{}%
  }%
  \providecommand\includegraphics[2][]{%
    \GenericError{(gnuplot) \space\space\space\@spaces}{%
      Package graphicx or graphics not loaded%
    }{See the gnuplot documentation for explanation.%
    }{The gnuplot epslatex terminal needs graphicx.sty or graphics.sty.}%
    \renewcommand\includegraphics[2][]{}%
  }%
  \providecommand\rotatebox[2]{#2}%
  \@ifundefined{ifGPcolor}{%
    \newif\ifGPcolor
    \GPcolortrue
  }{}%
  \@ifundefined{ifGPblacktext}{%
    \newif\ifGPblacktext
    \GPblacktextfalse
  }{}%
  % define a \g@addto@macro without @ in the name:
  \let\gplgaddtomacro\g@addto@macro
  % define empty templates for all commands taking text:
  \gdef\gplbacktext{}%
  \gdef\gplfronttext{}%
  \makeatother
  \ifGPblacktext
    % no textcolor at all
    \def\colorrgb#1{}%
    \def\colorgray#1{}%
  \else
    % gray or color?
    \ifGPcolor
      \def\colorrgb#1{\color[rgb]{#1}}%
      \def\colorgray#1{\color[gray]{#1}}%
      \expandafter\def\csname LTw\endcsname{\color{white}}%
      \expandafter\def\csname LTb\endcsname{\color{black}}%
      \expandafter\def\csname LTa\endcsname{\color{black}}%
      \expandafter\def\csname LT0\endcsname{\color[rgb]{1,0,0}}%
      \expandafter\def\csname LT1\endcsname{\color[rgb]{0,1,0}}%
      \expandafter\def\csname LT2\endcsname{\color[rgb]{0,0,1}}%
      \expandafter\def\csname LT3\endcsname{\color[rgb]{1,0,1}}%
      \expandafter\def\csname LT4\endcsname{\color[rgb]{0,1,1}}%
      \expandafter\def\csname LT5\endcsname{\color[rgb]{1,1,0}}%
      \expandafter\def\csname LT6\endcsname{\color[rgb]{0,0,0}}%
      \expandafter\def\csname LT7\endcsname{\color[rgb]{1,0.3,0}}%
      \expandafter\def\csname LT8\endcsname{\color[rgb]{0.5,0.5,0.5}}%
    \else
      % gray
      \def\colorrgb#1{\color{black}}%
      \def\colorgray#1{\color[gray]{#1}}%
      \expandafter\def\csname LTw\endcsname{\color{white}}%
      \expandafter\def\csname LTb\endcsname{\color{black}}%
      \expandafter\def\csname LTa\endcsname{\color{black}}%
      \expandafter\def\csname LT0\endcsname{\color{black}}%
      \expandafter\def\csname LT1\endcsname{\color{black}}%
      \expandafter\def\csname LT2\endcsname{\color{black}}%
      \expandafter\def\csname LT3\endcsname{\color{black}}%
      \expandafter\def\csname LT4\endcsname{\color{black}}%
      \expandafter\def\csname LT5\endcsname{\color{black}}%
      \expandafter\def\csname LT6\endcsname{\color{black}}%
      \expandafter\def\csname LT7\endcsname{\color{black}}%
      \expandafter\def\csname LT8\endcsname{\color{black}}%
    \fi
  \fi
  \setlength{\unitlength}{0.0500bp}%
  \begin{picture}(4750.00,4320.00)%
    \gplgaddtomacro\gplbacktext{%
      \csname LTb\endcsname%
      \put(462,660){\makebox(0,0)[r]{\strut{}$44$}}%
      \put(462,1465){\makebox(0,0)[r]{\strut{}$46$}}%
      \put(462,2270){\makebox(0,0)[r]{\strut{}$48$}}%
      \put(462,3074){\makebox(0,0)[r]{\strut{}$50$}}%
      \put(462,3879){\makebox(0,0)[r]{\strut{}$52$}}%
      \put(594,440){\makebox(0,0){\strut{}$0.0$}}%
      \put(1399,440){\makebox(0,0){\strut{}$0.2$}}%
      \put(2203,440){\makebox(0,0){\strut{}$0.4$}}%
      \put(3008,440){\makebox(0,0){\strut{}$0.6$}}%
      \put(3812,440){\makebox(0,0){\strut{}$0.8$}}%
      \put(4617,440){\makebox(0,0){\strut{}$1.0$}}%
      \put(88,2269){\rotatebox{-270}{\makebox(0,0){\strut{}$\log E$ [erg]}}}%
      \put(2605,220){\makebox(0,0){\strut{}$t$ [Myr]}}%
      \put(2605,4099){\makebox(0,0){\strut{}Total electrons energy}}%
    }%
    \gplgaddtomacro\gplfronttext{%
      \csname LTb\endcsname%
      \put(1128,3706){\makebox(0,0)[r]{\strut{}$\star$ 1}}%
      \csname LTb\endcsname%
      \put(1128,3486){\makebox(0,0)[r]{\strut{}$\star$ 2}}%
      \csname LTb\endcsname%
      \put(1128,3266){\makebox(0,0)[r]{\strut{}$\star$ 3}}%
      \csname LTb\endcsname%
      \put(2379,3706){\makebox(0,0)[r]{\strut{}$\star$ 4}}%
      \csname LTb\endcsname%
      \put(2379,3486){\makebox(0,0)[r]{\strut{}$\star$ 5}}%
      \csname LTb\endcsname%
      \put(2379,3266){\makebox(0,0)[r]{\strut{}$\star$ 6}}%
      \csname LTb\endcsname%
      \put(3630,3706){\makebox(0,0)[r]{\strut{}CR}}%
    }%
    \gplbacktext
    \put(0,0){\includegraphics{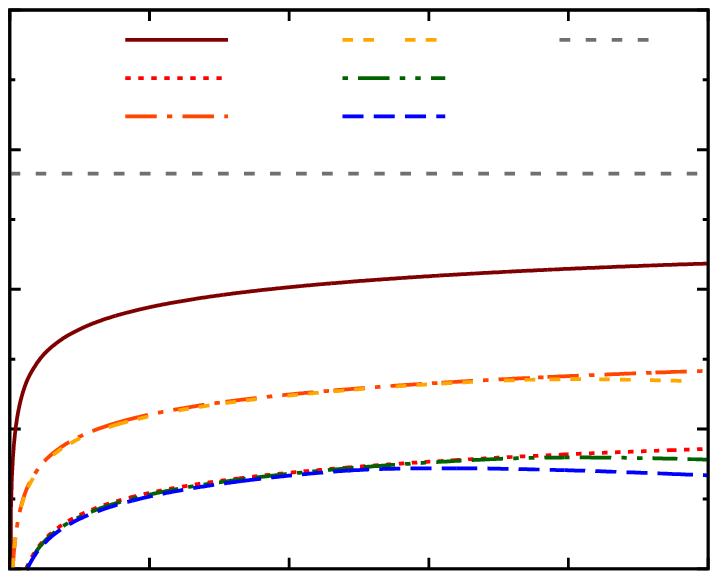}}%
    \gplfronttext
  \end{picture}%
\endgroup

%% file: chi0p1_L_gamm_time.tex
% GNUPLOT: LaTeX picture with Postscript
\begingroup
  \makeatletter
  \providecommand\color[2][]{%
    \GenericError{(gnuplot) \space\space\space\@spaces}{%
      Package color not loaded in conjunction with
      terminal option `colourtext'%
    }{See the gnuplot documentation for explanation.%
    }{Either use 'blacktext' in gnuplot or load the package
      color.sty in LaTeX.}%
    \renewcommand\color[2][]{}%
  }%
  \providecommand\includegraphics[2][]{%
    \GenericError{(gnuplot) \space\space\space\@spaces}{%
      Package graphicx or graphics not loaded%
    }{See the gnuplot documentation for explanation.%
    }{The gnuplot epslatex terminal needs graphicx.sty or graphics.sty.}%
    \renewcommand\includegraphics[2][]{}%
  }%
  \providecommand\rotatebox[2]{#2}%
  \@ifundefined{ifGPcolor}{%
    \newif\ifGPcolor
    \GPcolortrue
  }{}%
  \@ifundefined{ifGPblacktext}{%
    \newif\ifGPblacktext
    \GPblacktextfalse
  }{}%
  % define a \g@addto@macro without @ in the name:
  \let\gplgaddtomacro\g@addto@macro
  % define empty templates for all commands taking text:
  \gdef\gplbacktext{}%
  \gdef\gplfronttext{}%
  \makeatother
  \ifGPblacktext
    % no textcolor at all
    \def\colorrgb#1{}%
    \def\colorgray#1{}%
  \else
    % gray or color?
    \ifGPcolor
      \def\colorrgb#1{\color[rgb]{#1}}%
      \def\colorgray#1{\color[gray]{#1}}%
      \expandafter\def\csname LTw\endcsname{\color{white}}%
      \expandafter\def\csname LTb\endcsname{\color{black}}%
      \expandafter\def\csname LTa\endcsname{\color{black}}%
      \expandafter\def\csname LT0\endcsname{\color[rgb]{1,0,0}}%
      \expandafter\def\csname LT1\endcsname{\color[rgb]{0,1,0}}%
      \expandafter\def\csname LT2\endcsname{\color[rgb]{0,0,1}}%
      \expandafter\def\csname LT3\endcsname{\color[rgb]{1,0,1}}%
      \expandafter\def\csname LT4\endcsname{\color[rgb]{0,1,1}}%
      \expandafter\def\csname LT5\endcsname{\color[rgb]{1,1,0}}%
      \expandafter\def\csname LT6\endcsname{\color[rgb]{0,0,0}}%
      \expandafter\def\csname LT7\endcsname{\color[rgb]{1,0.3,0}}%
      \expandafter\def\csname LT8\endcsname{\color[rgb]{0.5,0.5,0.5}}%
    \else
      % gray
      \def\colorrgb#1{\color{black}}%
      \def\colorgray#1{\color[gray]{#1}}%
      \expandafter\def\csname LTw\endcsname{\color{white}}%
      \expandafter\def\csname LTb\endcsname{\color{black}}%
      \expandafter\def\csname LTa\endcsname{\color{black}}%
      \expandafter\def\csname LT0\endcsname{\color{black}}%
      \expandafter\def\csname LT1\endcsname{\color{black}}%
      \expandafter\def\csname LT2\endcsname{\color{black}}%
      \expandafter\def\csname LT3\endcsname{\color{black}}%
      \expandafter\def\csname LT4\endcsname{\color{black}}%
      \expandafter\def\csname LT5\endcsname{\color{black}}%
      \expandafter\def\csname LT6\endcsname{\color{black}}%
      \expandafter\def\csname LT7\endcsname{\color{black}}%
      \expandafter\def\csname LT8\endcsname{\color{black}}%
    \fi
  \fi
  \setlength{\unitlength}{0.0500bp}%
  \begin{picture}(4750.00,4320.00)%
    \gplgaddtomacro\gplbacktext{%
      \csname LTb\endcsname%
      \put(462,660){\makebox(0,0)[r]{\strut{}$26$}}%
      \put(462,1304){\makebox(0,0)[r]{\strut{}$28$}}%
      \put(462,1948){\makebox(0,0)[r]{\strut{}$30$}}%
      \put(462,2591){\makebox(0,0)[r]{\strut{}$32$}}%
      \put(462,3235){\makebox(0,0)[r]{\strut{}$34$}}%
      \put(462,3879){\makebox(0,0)[r]{\strut{}$36$}}%
      \put(594,440){\makebox(0,0){\strut{}$0.0$}}%
      \put(1399,440){\makebox(0,0){\strut{}$0.2$}}%
      \put(2203,440){\makebox(0,0){\strut{}$0.4$}}%
      \put(3008,440){\makebox(0,0){\strut{}$0.6$}}%
      \put(3812,440){\makebox(0,0){\strut{}$0.8$}}%
      \put(4617,440){\makebox(0,0){\strut{}$1.0$}}%
      \put(88,2269){\rotatebox{-270}{\makebox(0,0){\strut{}$\log L$ [erg s$^{-1}$]}}}%
      \put(2605,220){\makebox(0,0){\strut{}$t$ [Myr]}}%
      \put(2605,4099){\makebox(0,0){\strut{}Total gamma luminosity}}%
    }%
    \gplgaddtomacro\gplfronttext{%
    }%
    \gplbacktext
    \put(0,0){\includegraphics{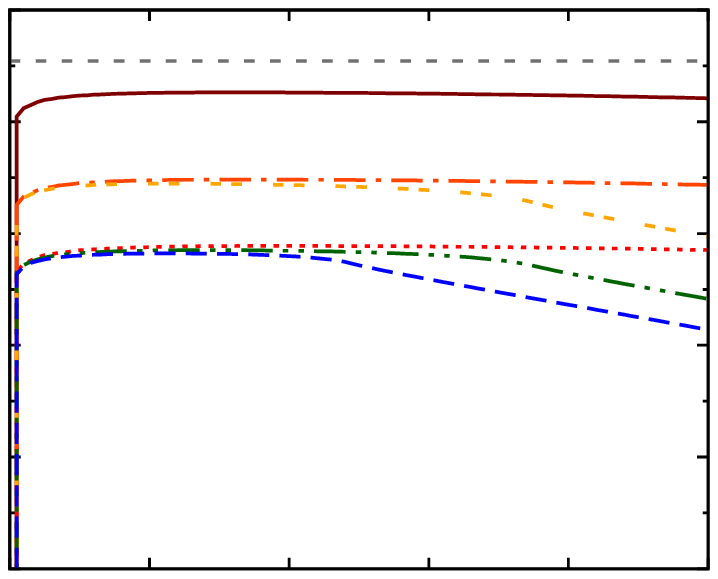}}%
    \gplfronttext
  \end{picture}%
\endgroup

%% file: chi0p1_L_sync_time.tex
% GNUPLOT: LaTeX picture with Postscript
\begingroup
  \makeatletter
  \providecommand\color[2][]{%
    \GenericError{(gnuplot) \space\space\space\@spaces}{%
      Package color not loaded in conjunction with
      terminal option `colourtext'%
    }{See the gnuplot documentation for explanation.%
    }{Either use 'blacktext' in gnuplot or load the package
      color.sty in LaTeX.}%
    \renewcommand\color[2][]{}%
  }%
  \providecommand\includegraphics[2][]{%
    \GenericError{(gnuplot) \space\space\space\@spaces}{%
      Package graphicx or graphics not loaded%
    }{See the gnuplot documentation for explanation.%
    }{The gnuplot epslatex terminal needs graphicx.sty or graphics.sty.}%
    \renewcommand\includegraphics[2][]{}%
  }%
  \providecommand\rotatebox[2]{#2}%
  \@ifundefined{ifGPcolor}{%
    \newif\ifGPcolor
    \GPcolortrue
  }{}%
  \@ifundefined{ifGPblacktext}{%
    \newif\ifGPblacktext
    \GPblacktextfalse
  }{}%
  % define a \g@addto@macro without @ in the name:
  \let\gplgaddtomacro\g@addto@macro
  % define empty templates for all commands taking text:
  \gdef\gplbacktext{}%
  \gdef\gplfronttext{}%
  \makeatother
  \ifGPblacktext
    % no textcolor at all
    \def\colorrgb#1{}%
    \def\colorgray#1{}%
  \else
    % gray or color?
    \ifGPcolor
      \def\colorrgb#1{\color[rgb]{#1}}%
      \def\colorgray#1{\color[gray]{#1}}%
      \expandafter\def\csname LTw\endcsname{\color{white}}%
      \expandafter\def\csname LTb\endcsname{\color{black}}%
      \expandafter\def\csname LTa\endcsname{\color{black}}%
      \expandafter\def\csname LT0\endcsname{\color[rgb]{1,0,0}}%
      \expandafter\def\csname LT1\endcsname{\color[rgb]{0,1,0}}%
      \expandafter\def\csname LT2\endcsname{\color[rgb]{0,0,1}}%
      \expandafter\def\csname LT3\endcsname{\color[rgb]{1,0,1}}%
      \expandafter\def\csname LT4\endcsname{\color[rgb]{0,1,1}}%
      \expandafter\def\csname LT5\endcsname{\color[rgb]{1,1,0}}%
      \expandafter\def\csname LT6\endcsname{\color[rgb]{0,0,0}}%
      \expandafter\def\csname LT7\endcsname{\color[rgb]{1,0.3,0}}%
      \expandafter\def\csname LT8\endcsname{\color[rgb]{0.5,0.5,0.5}}%
    \else
      % gray
      \def\colorrgb#1{\color{black}}%
      \def\colorgray#1{\color[gray]{#1}}%
      \expandafter\def\csname LTw\endcsname{\color{white}}%
      \expandafter\def\csname LTb\endcsname{\color{black}}%
      \expandafter\def\csname LTa\endcsname{\color{black}}%
      \expandafter\def\csname LT0\endcsname{\color{black}}%
      \expandafter\def\csname LT1\endcsname{\color{black}}%
      \expandafter\def\csname LT2\endcsname{\color{black}}%
      \expandafter\def\csname LT3\endcsname{\color{black}}%
      \expandafter\def\csname LT4\endcsname{\color{black}}%
      \expandafter\def\csname LT5\endcsname{\color{black}}%
      \expandafter\def\csname LT6\endcsname{\color{black}}%
      \expandafter\def\csname LT7\endcsname{\color{black}}%
      \expandafter\def\csname LT8\endcsname{\color{black}}%
    \fi
  \fi
  \setlength{\unitlength}{0.0500bp}%
  \begin{picture}(4750.00,4320.00)%
    \gplgaddtomacro\gplbacktext{%
      \csname LTb\endcsname%
      \put(462,660){\makebox(0,0)[r]{\strut{}$26$}}%
      \put(462,1304){\makebox(0,0)[r]{\strut{}$28$}}%
      \put(462,1948){\makebox(0,0)[r]{\strut{}$30$}}%
      \put(462,2591){\makebox(0,0)[r]{\strut{}$32$}}%
      \put(462,3235){\makebox(0,0)[r]{\strut{}$34$}}%
      \put(462,3879){\makebox(0,0)[r]{\strut{}$36$}}%
      \put(594,440){\makebox(0,0){\strut{}$0.0$}}%
      \put(1399,440){\makebox(0,0){\strut{}$0.2$}}%
      \put(2203,440){\makebox(0,0){\strut{}$0.4$}}%
      \put(3008,440){\makebox(0,0){\strut{}$0.6$}}%
      \put(3812,440){\makebox(0,0){\strut{}$0.8$}}%
      \put(4617,440){\makebox(0,0){\strut{}$1.0$}}%
      \put(88,2269){\rotatebox{-270}{\makebox(0,0){\strut{}$\log L$ [erg s$^{-1}$]}}}%
      \put(2605,220){\makebox(0,0){\strut{}$t$ [Myr]}}%
      \put(2605,4099){\makebox(0,0){\strut{}Total synchrotron luminosity}}%
    }%
    \gplgaddtomacro\gplfronttext{%
      \csname LTb\endcsname%
      \put(1128,3706){\makebox(0,0)[r]{\strut{}$\star$ 1}}%
      \csname LTb\endcsname%
      \put(1128,3486){\makebox(0,0)[r]{\strut{}$\star$ 2}}%
      \csname LTb\endcsname%
      \put(1128,3266){\makebox(0,0)[r]{\strut{}$\star$ 3}}%
      \csname LTb\endcsname%
      \put(2379,3706){\makebox(0,0)[r]{\strut{}$\star$ 4}}%
      \csname LTb\endcsname%
      \put(2379,3486){\makebox(0,0)[r]{\strut{}$\star$ 5}}%
      \csname LTb\endcsname%
      \put(2379,3266){\makebox(0,0)[r]{\strut{}$\star$ 6}}%
      \csname LTb\endcsname%
      \put(3630,3706){\makebox(0,0)[r]{\strut{}CR}}%
    }%
    \gplbacktext
    \put(0,0){\includegraphics{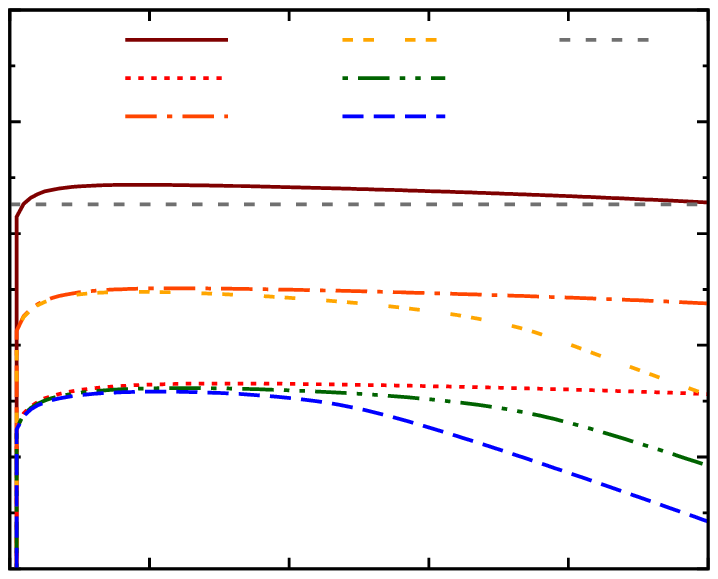}}%
    \gplfronttext
  \end{picture}%
\endgroup